\documentclass[12pt]{article}
\usepackage{latexsym}
\usepackage{amsmath}
\usepackage{amssymb}
\usepackage{amssymb}
\catcode`\@=11
\newcount\hour
\newcount\minute
\newtoks\amorpm \hour=\time\divide\hour by 60\minute
=\time{\multiply\hour by 60 \global\advance\minute by-\hour}
\edef\standardtime{{\ifnum\hour<12 \global\amorpm={am}%
        \else\global\amorpm={pm}\advance\hour by-12 \fi
        \ifnum\hour=0 \hour=12 \fi
        \number\hour:\ifnum\minute<10
        0\fi\number\minute\the\amorpm}}
\edef\militarytime{\number\hour:\ifnum\minute<10
0\fi\number\minute}
\def\draftlabel#1{{\@bsphack\if@filesw {\let\thepage\relax
   \xdef\@gtempa{\write\@auxout{\string
      \newlabel{#1}{{\@currentlabel}{\thepage}}}}}\@gtempa
   \if@nobreak \ifvmode\nobreak\fi\fi\fi\@esphack}
        \gdef\@eqnlabel{#1}}
\def\@eqnlabel{}
\def\@vacuum{}
\def\marginnote#1{}
\def\draftmarginnote#1{\marginpar{\raggedright\scriptsize\tt#1}}
\def\draft{
        \pagestyle{plain}
        \overfullrule=2pt
        \oddsidemargin -.1truein
        \def\@oddhead{\sl \phantom{\today\quad\militarytime} \hfil
        \smash{\Large\sl DRAFT} \hfil \today\quad\militarytime}
        \let\@evenhead\@oddhead
        \let\label=\draftlabel
        \let\marginnote=\draftmarginnote
        \def\ps@empty{\let\@mkboth\@gobbletwo
        \def\@oddfoot{\hfil \smash{\Large\sl DRAFT} \hfil}
        \let\@evenfoot\@oddhead}
        \def\@eqnnum{(\theequation)\rlap{\kern\marginparsep\tt\@eqnlabel}%
        \global\let\@eqnlabel\@vacuum}  }

\renewcommand{\theequation}{\thesection.\arabic{equation}}
\renewcommand{\thefootnote}{\fnsymbol{footnote}}
\newcommand{\newsection}{    
\setcounter{equation}{0}\section}
\def\appendix#1{\addtocounter{section}{1}\setcounter{equation}{0}
\renewcommand{\thesection}{\Alph{section}}
\section*{Appendix \thesection\protect\indent \parbox[t]{11.15cm}{#1}}
\addcontentsline{toc}{section}{Appendix \thesection\ \ \ #1}}

\def \bi{\bibitem}
\def \la {\label}

\def \b {\beta}

\def \d {\partial}

\def\be{\begin{equation}}
\def\ee{\end{equation}}

\newcommand{\cont}[1]{{}_{#1}{}^{#1}}

\catcode`\@=11

\def\bea{\begin{eqnarray}}
\def\eea{\end{eqnarray}}
\def\beann{\begin{eqnarray*}}
\def\eeann{\end{eqnarray*}}
\def\beq{\begin{equation}}
\def\eeq{\end{equation}}
\def\ba{\begin{array}}
\def\ea{\end{array}}
\def\ben{\begin{enumerate}}
\def\een{\end{enumerate}}
 \def \l {\lambda}

 \def \la {\label}
 \def\be{\begin{equation}}
\def\ee{\end{equation}}

\def \la {\label}

\font\mybb=msbm10 at 11pt

\def\bb#1{\hbox{\mybb#1}}

\def\bR {\bb{R}}

\def\bH {\bb{H}}
\def\bC {\bb{C}}

\def\e  {\epsilon}

\def \k {\kappa}

\def \ee {\epsilon}

\def \g {\gamma}
\def \bi{\bibitem}
\def\a{\alpha }

\def \d {\delta}

\def \l {\lambda}

\def \g {\gamma}

\def \b {\beta}

\def\be{\begin{equation}}
\def\ee{\end{equation}}

\def \bi {\bibitem}
\def \la{\label}

\begin{document}
\date{November 2002}
\begin{titlepage}
\begin{center}
\hfill UB-ECM-PF-07-04 \\
\hfill hep-th/yymmnnn \\

\vspace{3.0cm} {\Large \bf Geometry of all supersymmetric type I backgrounds }
\\[.2cm]

\vspace{1.5cm}
 {\large  U.~Gran$^1$,
 G.~Papadopoulos$^2$, D.~Roest$^3$ and P.~Sloane$^2$
 }

\vspace{0.5cm}

${}^1$ Fundamental Physics\\
Chalmers University of Technology\\
SE-412 96 G\"oteborg, Sweden\\

\vspace{0.5cm}
${}^2$ Department of Mathematics\\
King's College London\\
Strand,
London WC2R 2LS, UK\\

\vspace{0.5cm}
${}^3$ Departament Estructura i Constituents de la Materia \\
    Facultat de F\'{i}sica, Universitat de Barcelona \\
    Diagonal 647, 08028 Barcelona, Spain \\

\end{center}

\vskip 1.5 cm
\begin{abstract}

We find the geometry  of all supersymmetric type I backgrounds by solving the gravitino and dilatino
 Killing spinor equations, using the spinorial geometry technique, in all cases.
 The solutions of the gravitino Killing spinor equation are characterized by their
 isotropy group  in $Spin(9,1)$, while the solutions of the dilatino Killing spinor equation are characterized by
 their isotropy group in the subgroup $\Sigma({\cal P})$ of $Spin(9,1)$ which preserves the space of parallel spinors
 ${\cal P}$.
Given a solution of the gravitino Killing spinor equation with $L$
parallel spinors, $L=1,2,3,4,5,6, 8$,
  the dilatino Killing spinor equation allows for solutions with $N$ supersymmetries
 for any $0<N\leq L$. Moreover   for $L=16$, we confirm that $N=8, 10, 12, 14, 16$.
 We find that in most cases the Bianchi identities and the field equations of type I backgrounds
  imply a further reduction of the holonomy of the supercovariant connection. In addition,  we show that
in some cases   if the holonomy group of the supercovariant connection is precisely the isotropy group
 of the parallel spinors, then  all parallel
 spinors are Killing and so there are no backgrounds with $N<L$ supersymmetries.

\end{abstract}
\end{titlepage}
\newpage
\setcounter{page}{1}
\renewcommand{\thefootnote}{\arabic{footnote}}
\setcounter{footnote}{0}

\setcounter{tocdepth}{1}
\tableofcontents

\setcounter{section}{0}
\setcounter{subsection}{0}
\newsection{Introduction}

The last twenty years, supersymmetric solutions of the type I supergravities and their geometries  have been the focus of intensive
investigation  because of their  applications in type I and heterotic superstrings,
see e.g.~\cite{strominger}-\cite{becker}.
Type I supergravities
have three types of Killing spinor equations associated with the vanishing of  the supersymmetry variations
of the gravitino, dilatino and gaugino.
The gravitino Killing spinor equation is a parallel transport equation of a metric connection with skew-symmetric torsion,
$\hat \nabla$, where the torsion is the NS$\otimes$NS  or R$\otimes$R three-form field strength in the heterotic\footnote{
Similar geometries appear in the context of  ($1+1$)- and (1+0)-dimensional supersymmetric sigma models,
 see e.g.~\cite{hull, howe, howegp, coles}.}
or type I superstrings, respectively.
So the holonomy of $\hat\nabla$, ${\rm hol}(\hat\nabla)$,  is contained in $Spin(9,1)$.
The existence of parallel spinors requires that ${\rm hol}(\hat\nabla)$ must be a subgroup of their
isotropy group in $Spin(9,1)$.
Therefore either the Killing spinors have a non-trivial (proper) stability Lie subgroup
in $Spin(9,1)$ or the stability subgroup is $\{1\}$ and the curvature $\hat R$ of $\hat\nabla$ vanishes, $\hat R=0$.
The isotropy or stability subgroups, up to a discrete identification, of  Majorana-Weyl spinors in $Spin(9,1)$  are
 \bea
&&Spin(7)\ltimes\bR^8 ~(1)\supset SU(4)\ltimes\bR^8~(2)\supset Sp(2)\ltimes\bR^8~(3)\supset (SU(2)\times SU(2))\ltimes\bR^8~(4)
\cr
&&~~~~~~~~~~~~~~~~~~~~~~~\supset SU(2)\ltimes \bR^8~(5)\supset U(1)\ltimes\bR^8~(6)\supset \bR^8~(8)\,,
\cr
&&Spin(7)\ltimes\bR^8 ~(1)\supset G_2~(2)\supset SU(3)~(4)\supset  SU(2)~(8)\supset \{1\}~(16)~,
\la{stabspin}
\eea
where in parenthesis we have denoted the number of linearly independent  invariant spinors. The maximal compact subgroups
of (\ref{stabspin}) have appeared before, see  \cite{bj}, in the context of supersymmetric  M-brane  configurations.
 Lists of isotropy groups
of $Spin(9,1)$ and $Spin(10,1)$
spinors in various representations\footnote{ The isotropy groups of spinors are representation sensitive. There are many
more isotropy groups that appear for $Spin(9,1)$ Majorana  spinors, and the  $Spin(10,1)$ Majorana spinors
have different isotropy groups.} can be found in \cite{figueroab}.  Most of above groups
have also appeared in \cite{het}. To our knowledge the concise
 list of isotropy groups of $Spin(9,1)$ Majorana-Weyl spinors
has been given for the first time in this paper  and
a proof that (\ref{stabspin}) is complete can be found  in
appendix B.
As can easily be  seen,
 there are two classes of
stability subgroups characterized by their topology.
Moreover the  isotropy group of 9  or more spinors is $\{1\}$. Therefore backgrounds with more than 8 parallel spinors
necessarily have $\hat R=0$.

The dilatino Killing spinor equation is not amenable to such a straightforward Lie
algebraic interpretation. This has been one of the obstacles to find the geometry of all supersymmetric type I backgrounds.
Nevertheless much progress has been made to systematically understand the geometry
  of supersymmetric type I backgrounds. In  \cite{het}, the Killing spinor
equations of type I supergravities have been solved, using the spinorial geometry method of \cite{ggp},
 under the assumption that all the $\hat\nabla$-parallel spinors are Killing, i.e.~all  solutions of the gravitino Killing spinor equation are also solutions  the dilatino one, see also \cite{common}
for an application to the common sector.

The supersymmetric backgrounds with $\hat R=0$ have been examined in \cite{jfofhet}. In particular,
$\hat R=0$ and $dH=0$ imply that the spacetime is a Lorentzian  metric Lie group. These  groups have
been classified in \cite{kawano, jfofhet},
based on
some earlier work on Lorentzian Lie groups \cite{medina}.
So, the class of supersymmetric backgrounds
 that remains to be examined is that for which some of the $\hat\nabla$-parallel spinors do not solve the dilatino
 Killing spinor equation and $\hat R\not=0$.

In this paper, we classify  the geometry of all supersymmetric type I backgrounds.
This is done by completing the program, i.e.~by solving the Killing spinor equations for
those backgrounds for which only some of  the $\hat\nabla$-parallel spinors solve the dilatino Killing spinor
equation. We shall find that the Killing spinor equations allow for  backgrounds for any $N\leq 8$.
We have carried out the classification using a  combination of
the spinorial geometry method of \cite{ggp} and   its recent adaptation to nearly maximally
supersymmetric backgrounds proposed in  \cite{iibpreons, Mpreons}. The first part of the task is to find the $\hat\nabla$-parallel
spinors and to solve the gravitino Killing spinor equation. This has been done in \cite{het} and the parallel
spinors have been identified in most cases. We give the parallel spinors of the $SU(2)\ltimes\bR^8$ and $U(1)\ltimes\bR^8$
cases that have not been included in \cite{het}.

Next it remains to identify the Killing spinors of a supersymmetric background, i.e.~those
 $\hat\nabla$-parallel spinors that solve the dilatino Killing spinor equation as well. Clearly for a background with $N$ Killing spinors
 and $L$ $\hat\nabla$-parallel spinors, $1\leq N\leq L$. The backgrounds with $N<L$ are referred as ``{\it descendants}''.
 The $N$ Killing spinors
 of a supersymmetric background
can span any $N$-plane in the $L$-dimensional vector space ${\cal
P}$ of $\hat\nabla$-parallel spinors. Generically there are infinitely
many choices of $N$-planes in an $L$-dimensional vector space, so at
first sight it appears that the program cannot be carried out.
However, according to the spinorial geometry method of \cite{ggp},
the Killing spinors should be identified up to a gauge
transformation of the Killing spinor equations. So not all choices
of $N$-planes give rise to different spacetime geometries. In
particular any two $N$-planes that are related  by a $Spin(9,1)$
transformation which {\sl preserves the space of parallel spinors}
give rise to the same spacetime geometry and fluxes up to a Lorentz
transformation. Given that  the solutions $(\e_1,\dots, \e_L)$ of
the gravitino Killing spinor equation span ${\cal P}$,
$\hat\nabla\e_i=0$, we shall identify the Killing spinors up to
transformations of the group \bea \Sigma({\cal P})={\rm Stab}({\cal
P})/{\rm Stab}(\e_1,\dots, \e_L)~, \eea where ${\rm Stab}({\cal P})$
is the subgroup of $Spin(9,1)$ which preserves the $L$-dimensional
vector space ${\cal P}$ and ${\rm Stab}(\e_1,\dots, \e_L)$ is the
subgroup of $Spin(9,1)$ which preserves each $(\e_1,\dots, \e_L)$
individually. We shall see that $\Sigma({\cal P})$ is the product of
a $Spin$ group and an $R$-symmetry group of an appropriate
lower-dimensional supergravity. It turns out that after an
appropriate identification using $\Sigma({\cal P})$, there is a
finite number of distinct  $N$-planes, for $N\leq L/2$, and so the
classification can be completed. In particular the Killing spinors
of all the descendants can be identified.  For $N>L/2$, one can use
a similar argument to specify the normals to the Killing spinors. As
in \cite{iibpreons},
 these can be used to determine the Killing spinors.   The dilatino Killing spinor equation can again  be solved.

The Killing spinors of all descendants can be characterized by two
groups. One is the isotropy group of the parallel spinors we have
already mentioned. The other is the isotropy group of the Killing
spinors ${\rm Stab}_{\Sigma}(\e_1, \dots, \e_N)$ in $\Sigma({\cal
P})$. In the description of ${\rm Stab}_\Sigma$, it is sufficient to
consider $N\leq L/2$. This is because for $N>L/2$, it is more
convenient to consider the analogous groups for the normals to the
Killing spinors. However, these coincide with those of the Killing
spinors for $N\leq L/2$.

A special case are the descendants  of backgrounds with $L=16$
parallel spinors. These  backgrounds are parallelizable, $\hat
R=0$. In this case, our method reproduces
 the results
of \cite{jfofhet}. A summary of the geometric properties of all cases can be found at the conclusions.

We also investigate the conditions on the descendants imposed by
the Bianchi identities and field equations of the type I
supergravities. For ${\rm Stab}(\e_1,\dots, \e_L)$ non-compact,
$L\leq 8$, it turns out that $dH=0$ and the field equations of
type I supergravities imply that the descendants exist if and only
if the holonomy of $\hat\nabla$, ${\rm hol}(\hat\nabla)$, reduces
to a proper subgroup of ${\rm Stab}(\e_1,\dots, \e_L)$, i.e.~${\rm
hol}(\hat\nabla)\subset {\rm Stab}(\e_1,\dots, \e_L)$. In
particular, if one insists that ${\rm hol}(\hat\nabla)= {\rm
Stab}(\e_1,\dots, \e_L)$, then under the same conditions, the
gravitino Killing spinor equation implies the dilatino one and so
the only backgrounds that  occur are those for which $N=L$, i.e.~those investigated in \cite{het}. For ${\rm Stab}(\e_1,\dots,
\e_L)$ compact, there are descendants for which ${\rm
hol}(\hat\nabla)= {\rm Stab}(\e_1,\dots, \e_L)$. Moreover, the
gravitino Killing spinor equation implies the dilatino one provided
some conditions are satisfied in addition to those implied by
$dH=0$, the field equations and  ${\rm hol}(\hat\nabla)= {\rm
Stab}(\e_1,\dots, \e_L)$.

The gaugino Killing spinor equation, $F\e=0$,  can also be understood in a similar way  to that of the gravitino Killing spinor
equation. In particular, the spacetime indices of the gauge field strength  $F$ can be interpreted as taking values
in $\mathfrak{spin}(9,1)$, so either the spinors $\e$ have a non-trivial stability subgroup in $Spin(9,1)$ or the
gauge connection is flat, $F=0$. We shall not present a detailed analysis of the conditions on $F$ implied by the
dilatino Killing spinor equation. This is because the geometry of spacetime is not affected by the solutions
of the gaugino Killing spinor equation\footnote{The gauge field may contribute in the
modification of the Bianchi identity of $H$ due to the anomaly cancellation mechanism,
and so it  affects the spacetime geometry only in the case that
$dH\not=0$. However to lowest order in $\a'$, $dH=0$. If the anomaly correction is included, then
the sigma model two-loop contribution to the field equations should be taken
into account, see e.g.~\cite{tsimpis}. In any case, most of our analysis is independent of such assumptions on $dH$.}. Of course one can consider the possibility that some
of the solutions of the gravitino and dilatino Killing spinor equations solve the gaugino one as well.
However, it is more usual to take that either all parallel spinors solve the gaugino Killing spinor equation or
that all solutions of the gravitino and dilatino Killing spinor equations also solve the
gaugino one. In all cases the solutions of the gaugino Killing spinor equation
 can be deduced from those of the gravitino Killing spinor equation.

This paper is organized as follows: In section two, we describe how the gauge symmetry of the
Killing spinors can be used to identify the Killing spinors of all supersymmetric type I backgrounds.
In sections three to five, we investigate the descendants of $SU(4)\ltimes\bR^8$-, $Sp(2)\ltimes\bR^8$- and
$(SU(2)\times SU(2))\ltimes \bR^8$-invariant parallel spinors
and compare their geometry to that of backgrounds for which all parallel spinors are Killing in each case. In sections six and seven,
we solve the Killing spinor equations, and those of their descendants, of backgrounds
with $SU(2)\ltimes\bR^8$- and  $U(1)\ltimes\bR^8$-invariant parallel spinors. In section eight, we examine the Killing spinor equations
of the descendants of $\bR^8$-parallel spinors. In section nine, we use the Bianchi identities
and the field equations to investigate the conditions under which the holonomy of $\hat\nabla$ reduces
to a subgroup of the isotropy group of the parallel spinors. We also present some applications.
In sections ten, eleven and twelve, we solve the Killing spinor equations of the descendants
of backgrounds with $G_2$-, $SU(3)$- and $SU(2)$-invariant parallel spinors, respectively. We also
investigate the reduction of the holonomy and its consequences in each case. In section thirteen,
we investigate the parallelizable backgrounds using the methods developed in this paper and confirm
the results of \cite{kawano, jfofhet},
and in section fifteen we give our conclusions.
In appendix A, we summarize some aspects of the geometry of manifolds which admits $\hat\nabla$-parallel spinors
and outline some of their geometric properties. In appendix B, we show that the list presented in
(\ref{stabspin}) is complete, and in appendix C, we summarize some results on a group representation that we have used
to investigate an $N=4$ descendant of $SU(2)$-invariant parallel spinors. In appendix D, we give the additional parallel spinor bi-linears
for the $SU(2)\ltimes\bR^8$ and $U(1)\ltimes \bR^8$ cases.

\newsection{Preliminaries}

The Killing spinor equations of type I and heterotic supergravities are
\bea
{\cal D}(e,H)_A\e=\hat\nabla_A \epsilon=0~,~~~{\cal A}(e,H,\Phi)\e=(\Gamma^A \partial_A\Phi-{1\over12} H_{ABC} \Gamma^{ABC})\e=0
\eea
where $e$ is a frame, $\Phi$ is the dilaton, $H$ is the NS$\otimes$NS three-form field strength and
\bea
\hat\nabla_B Y^A=\nabla_B Y^A+{1\over2} H^A{}_{BC} Y^C~,
\eea
is a metric connection with torsion $H$.  The spinors $\e$ are in the positive chirality Majorana-Weyl representation $S^+$
of $Spin(9,1)$ which in the conventions of \cite{het} are represented by even-degree forms. (We use the
conventions of \cite{het} throughout this paper.)

The Lie subgroups of $Spin(9,1)$ that   leave  some spinors
invariant have been listed in (\ref{stabspin}).  We collectively
denote them with ${\rm Stab}(\e_1,\dots,\e_L)$ for
$L=1,2,3,4,5,6,8$ and $16$. These stability subgroups have the
property that they leave every individual spinor invariant. Since the
holonomy of $\hat\nabla$ is contained in $Spin(9,1)$, the
gravitino Killing spinor equation has solutions provided that
\bea {\rm hol}(\hat\nabla)\subseteq {\rm Stab}(\e_1,\dots,\e_L)~.
\eea If ${\rm Stab}(\e)=\{1\}$, then the curvature of $\hat\nabla$
vanishes, $\hat R=0$. This together with the closure of $H$,
$dH=0$, imply that the spacetime is a Lorentzian metric Lie group.

To solve the dilatino Killing spinor equation, we first assume
that we have a solution of the gravitino Killing spinor equation,
i.e.~we have a given number of parallel spinors spanning a
subspace ${\cal P}$ in the space of spinors $S^+$.  Then we try to
find the conditions for which some of the parallel spinors solve the
dilatino Killing spinor equation as well. If $N$ is the number of
Killing spinors, then necessarily they are at most as many as the
parallel $\hat\nabla$-spinors, so \bea N\leq {\rm dim}~{\cal
P}=L~. \eea

Let $\{\eta_i\}$ be a basis in the space of parallel spinors,
$\hat\nabla\eta_i=0$, ${\cal P}=\bR<\eta_1,\dots, \eta_L>$. The
Killing spinors can now be written as \bea \epsilon_r=\sum^L_{i=1}
f_{ri}\eta_i~, \eea where $f$ is a matrix of spacetime functions of
rank $N$. Since $\epsilon_r$ must remain $\hat\nabla$-parallel and
$\{\eta_i\}$ is a basis, it is easy to show that in fact $f$ is a
constant matrix. Let ${\cal K}$ be the $N$-plane in ${\cal P}$
spanned by the Killing spinors, ${\cal K}=\bR<\e_1,\dots\e_N>$.

Next suppose that $\ell\in Spin(9,1)$ and that it
preserves\footnote{The assumption $\ell{\cal P}\subseteq {\cal P}$
can be relaxed but it is more convenient to consider only those
$\ell$ that preserve ${\cal P}$.} ${\cal P}$, $\ell{\cal P}\subseteq
{\cal P}$.
 Then consider $\ell\e_r$ and observe that
\bea {\cal D}(e^{\ell^{-1}}, H^{\ell^{-1}}) \ell\e_r=\ell {\cal D}(e,
H) \e_r=0~, \cr {\cal A}(e^{\ell^{-1}}, H^{\ell^{-1}},
\Phi^{\ell^{-1}})\ell\e_r= \ell {\cal A}(e,H,\Phi) \e_r=0~, \eea
where
$e^{\ell^{-1}}, H^{\ell^{-1}}, \Phi^{\ell^{-1}}$ are the Lorentz
transformed frame, $H$ and dilaton with respect to the inverse
$\ell^{-1}$ Lorentz transformation\footnote{We denote with the same
symbol the element $\ell\in Spin(9,1)$ and its projection on the
Lorentz group.} associated with $\ell\in Spin(9,1)$. Therefore the
spinors $\ell\e_r$ are also solutions of the Killing spinor equations
up to a Lorentz rotation of the frame and the fluxes. Since we
identify backgrounds related by frame Lorentz transformations, one
concludes that the $N$-planes ${\cal K}$ and $\ell{\cal K}$ give rise
to the same spacetime geometry and fluxes. Thus to classify the
$N$-supersymmetric backgrounds, it is sufficient to find all
$N$-planes in ${\cal P}$ up to transformations in $Spin(9,1)$ that
preserve ${\cal P}$.

To continue we have to identify the subgroup $\Sigma({\cal
P})\subseteq Spin(9,1)$ which preserves ${\cal P}$, where ${\cal
P}=\bR<\eta_1,\dots, \eta_L>$. As we have mentioned in the
introduction, first define the stability subgroup of ${\cal P}$ as
\bea
{\rm Stab}({\cal P})=\{\ell\in Spin(9,1)~{\rm s}.{\rm t.}~~
\ell{\cal P}\subset {\cal P}\}
\eea
Clearly ${\rm
Stab}(\eta_1,\dots\eta_L)\subseteq {\rm Stab}({\cal P})$. In fact
${\rm Stab}(\eta_1,\dots\eta_L)$
 is a normal subgroup. Then we define
\bea \Sigma({\cal P})={\rm Stab}({\cal P})/{\rm
Stab}(\eta_1,\dots\eta_L)~. \eea $\Sigma({\cal P})$ may act
non-trivially on the space of parallel spinors preserving the
subspace  spanned by them and takes the r\^ole of the gauge group
in the context of the spinorial geometry approach to solving the
Killing spinor equations. The groups $\Sigma({\cal P})$ are
summarized in table 1. It can be easily seen that they are
products of the type $\Sigma({\cal P})=Spin(d,1)\times R$. Such
groups  are reminiscent of the gauge groups of
$(d+1)$-supergravities, where $R$ is the R-symmetry group. It may be
possible to make this correspondence more precise by considering
compactifications of type I supergravity on  supersymmetric
backgrounds with an appropriate $\Sigma({\cal P})$ group.

\begin{table}[ht]
 \begin{center}
\begin{tabular}{|c|c|c|}\hline
$L$ & ${\mathrm Stab}(\e_1,\dots,\e_L)$ & $\Sigma({\cal P})$
 \\ \hline \hline
$1$ & $Spin(7)\ltimes\bR^8$& $Spin(1,1)$ \\
\hline
$2$ &  $SU(4)\ltimes\bR^8$&$Spin(1,1)\times U(1)$
\\ \hline
$3$ & $Sp(2)\ltimes\bR^8$&$Spin(1,1)\times SU(2)$
\\ \hline
$4$ & $(SU(2)\times SU(2))\ltimes\bR^8$&$Spin(1,1)\times Sp(1)\times Sp(1)$
\\ \hline
$5$ & $SU(2)\ltimes\bR^8$&$Spin(1,1)\times Sp(2)$
\\ \hline
$6$ & $U(1)\ltimes\bR^8$&$Spin(1,1)\times SU(4)$
\\ \hline
$8$ & $\bR^8$&$Spin(1,1)\times Spin(8)$
\\ \hline \hline
$2$ & $G_2$&$Spin(2,1)$
\\ \hline
$4$ & $SU(3)$&$Spin(3,1)\times U(1)$
\\ \hline
$8$ & $SU(2)$&$Spin(5,1)\times SU(2)$
\\ \hline
$16$ & $\{1\}$&$Spin(9,1)$
\\ \hline
\end{tabular}
\end{center}
\caption{In the  columns are the numbers of parallel spinors, their isotropy groups  and the $\Sigma({\cal P})$ groups, respectively. The $\Sigma({\cal P})$
groups are a product of a $Spin$ group and an R-symmetry group.}
\end{table}

To see how $\Sigma({\cal P})$ is used, let us first choose ${\cal
P}$ and suppose that only one of the $\hat\nabla$-parallel spinors
also solves the dilatino Killing spinor equation, say $\e$ and so
$N=1$. The spinor $\e$ can be expressed as a linear combination of
a basis of parallel spinors $\e=f_i \eta_i$. As we have explained,
$\e$ and $\ell\e$, $\ell\in \Sigma({\cal P})$, give rise to the same
spacetime geometry. So the Killing spinors which may lead to
different spacetime geometries are labeled by the orbits, ${\cal
O}_{\Sigma({\cal P})}({\cal P})$, of $\Sigma({\cal P})$ in
${\cal
P}$. Hence, to find all $N=1$ backgrounds with $L$
$\hat\nabla$-parallel spinors, it suffices to choose a single
representative from each orbit of  $\Sigma({\cal P})$ in ${\cal
P}$. In general these representatives depend on as many different
parameters as the number of deformations that preserve the orbit.
In particular the representatives of generic orbits, i.e.~orbits
of maximal co-dimension, the number of parameters is equal to the
co-dimension of the orbit in ${\cal P}$. The Killing spinor
equations  are linear, so the Killing spinor is specified up to an
overall scale. As a result, the number of independent parameters
that the  Killing spinor depends on is at most the dimension of the
deformations that preserve the associated orbit. The number of
parameters of representatives of generic orbits is either ${\rm
codim}\,{\cal O}_{\Sigma({\cal P})}({\cal P})$ or ${\rm
codim}\,{\cal O}_{\Sigma({\cal P})}({\cal P})-1$   depending on
whether $\Sigma({\cal P})$ contains a scale generator. In most
generic cases, it turns out that ${\rm codim}\,{\cal O}$ is either
zero or one  and so we have to specify a single direction.

To continue, we proceed inductively. Let ${\cal K}$ be the $N$-plane in ${\cal P}$ spanned by the
first $N$ Killing spinors,
\bea
0\rightarrow {\cal K}\rightarrow {\cal P}\rightarrow {\cal P}/{\cal K}\rightarrow 0~,
\eea
${\cal K}=\bR<\e_1,\dots,\e_N>$.
To choose the $(N+1)$-th Killing spinor, we first consider ${\rm Stab}({\cal K})\subseteq \Sigma({\cal P})$ that preserves
 ${\cal K}$, i.e.
\bea
{\rm Stab}({\cal K})=\{\ell\in \Sigma({\cal P})~,~~~\ell{\cal K}\subseteq {\cal K}\}~.
\eea
The strategy we adopt is to use  ${\rm Stab}({\cal K})$ to choose the $(N+1)$-th Killing spinor. For this, first choose
a spinor $\e_{N+1}$ which is linearly independent from those in ${\cal K}$. Since the Killing spinor equations are linear,
it suffices to choose $\e_{N+1}$ up to elements in ${\cal K}$. Thus $\e_{N+1}$ can be thought of as an element
in ${\cal P}/{\cal K}$. Moreover ${\rm Stab}({\cal K})$ acts on ${\cal P}/{\cal K}$ preserving the plane of the first $N$ Killing spinors.
Using again the identification of supersymmetric backgrounds under frame Lorentz rotations, the $(N+1)$-th Killing spinor $\e_{N+1}$
can be chosen to be a representative of the orbits ${\cal O}_{{\rm Stab}({\cal K})}({\cal P}/{\cal K})$ of
${\rm Stab}({\cal K})$ in ${\cal P}/{\cal K}$. Again the number of independent parameters that $\e_{N+1}$ has depends on the type of orbit
it represents. If it is a generic orbit, the number of parameters is
either ${\rm codim} \, {\cal O}_{{\rm Stab}({\cal K})}({\cal P}/{\cal K})$ or
${\rm codim} \, {\cal O}_{{\rm Stab}({\cal K})}({\cal P}/{\cal K})-1$ depending on whether ${\rm Stab}({\cal K})$ acts
with or without a scale transformation on ${\cal P}/{\cal K}$.

The above described procedure works well for all $1\leq N\leq
L/2$. For $N>L/2$ in some cases it is more convenient,
instead of determining the Killing spinors up to $Spin(9,1)$
transformations, to specify their normals. For this first recall
that we have chosen ${\cal P}\subseteq S^+$, where $S^+$ is the positive
chirality Majorana-Weyl representation of $Spin(9,1)$. The dual of
$(S^+)^\star$ is identified with $S^-$, the negative chirality
Majorana-Weyl spinors, via the Majorana $Pin$-invariant inner
product $B$, i.e.~$S^-=B((S^+)^\star)$, see \cite{het, iibpreons}
for details. Define ${\cal Q}=B({\cal P}^\star)$. Next, $Spin(9,1)$
acts on $S^-$ and so as before define $\Sigma({\cal Q})$. The
hyperplane of Killing spinors of $N=L-1$ supersymmetric
backgrounds has a unique normal in ${\cal Q}$. Using the
identification of supersymmetric backgrounds under frame Lorentz
transformations and an argument as above, backgrounds with
distinct geometries are labeled by the orbits ${\cal
O}_{\Sigma({\cal Q})}({\cal Q})$, i.e.~by the choice of the normal
$\nu$ up to gauge transformations that preserve ${\cal Q}$. The
normal spinor is specified up to an overall scale, i.e.~we need to
specify only the normal direction. Thus the number of
parameters that the normal spinor has depends on the type of orbit it
represents. For generic orbits, the number of parameters is either
${\rm codim}{\cal O}_{\Sigma({\cal Q})}({\cal Q})$
 or  ${\rm codim}{\cal O}_{\Sigma({\cal Q})}({\cal Q})-1$ depending on the way that $\Sigma({\cal Q})$ acts on ${\cal Q}$.
Then the hyperplane of the  Killing spinors is specified by the
orthogonality condition \bea B(\nu, {\cal K})=0~. \eea To
continue, one proceeds inductively. First define ${\cal N}$ as the
$(L-N)$-plane in ${\cal Q}$ spanned by the first $L-N$ normal
spinors. To specify an additional Killing spinor up to a
$Spin(9,1)$ transformation, define the subgroup ${\rm Stab}({\cal
N})\subseteq Spin(9,1)$ that preserves ${\cal N}$ and consider the
sequence \bea 0\rightarrow {\cal N}\rightarrow {\cal Q}\rightarrow
{\cal Q}/{\cal N}\rightarrow 0~. \eea The $(L-N+1)$ normal spinor
$\nu_{L-N+1}$ is chosen to be linearly independent from the first
$L-N$ normal spinors and it is specified up to elements in ${\cal
N}$. This is because the $(N-1)$-Killing spinors will span a
hyperplane in ${\cal K}$ and so they will be always orthogonal to
${\cal N}$. Thus $\nu_{L-N+1}$ can be thought of as an element in
${\cal Q}/{\cal N}$. Using the identification of supersymmetric
backgrounds under frame Lorentz transformations again, the
additional normal spinor can be chosen as a representative of the
orbits ${\cal O}_{{\rm Stab}({\cal N})}({\cal Q}/{\cal N})$. The number of
independent parameters of the new normal depend on the type of
orbit it represents. For generic orbits, the number of parameters
is either  ${\rm codim} \, {\cal O}_{{\rm Stab}({\cal N})}({\cal
Q}/{\cal N})$ or
 ${\rm codim} \, {\cal O}_{{\rm Stab}({\cal N})}({\cal Q}/{\cal N})-1$.
In turn the Killing spinors are determined by the orthogonality
condition \bea B({\cal N}, {\cal K})=0~, \eea where now ${\cal N}$
is spanned by all $N-L+1$  normal spinors. As we have mentioned in
the introduction, we refer to the backgrounds with $N$
supersymmetries, $N<L$, that arise from a given set of ${\rm
Stab}(\e_1,\dots,\e_L)$-invariant parallel spinors as
``descendants'' of ${\rm Stab}(\e_1,\dots,\e_L)$.

\newsection{The descendants of  $SU(4)\ltimes \bR^8$}

A (complex) basis\footnote{The associated real basis is $\big(1+e_{1234}, i(1-e_{1234})\big)$. This can be easily found by taking the
real and imaginary parts of the complex spinor $1$ with respect to a reality condition that defines the
Majorana-Weyl representation of $Spin(9,1)$, see \cite{het}.}
in the space of parallel spinors can be chosen as
\bea
1~.
\eea
Observe that $\Sigma({\cal P})=Spin(1,1)\times U(1)$,
where the generator of $Spin(1,1)$ is $\Gamma^{+-}$ and the generator of $U(1)$ can be chosen as $i \Gamma^{1\bar1}$.
Observe that $Spin(1,1)=\bR^*$.
There is a single descendant background with $N=1$ supersymmetry.
The dilatino Killing spinor equation
 can be written as
\bea
{\cal A}(1+e_{1234})=0~.
\eea
We use the conventions of \cite{het} to denote the spinors and forms that arise in the analysis that follows.
Note that the stability subgroup of the Killing spinor in $\Sigma({\cal P})$ is ${\rm Stab}_{\Sigma}(1+e_{1234})=\{1\}$.

The strategy we adopt to organize the solutions of the Killing
spinor equations in all cases is to first solve the gravitino
Killing spinor equation. The conditions that arise are the same
for all descendants. Then the dilatino Killing spinor equation is
solved for each descendant and the solutions are expressed  in
representations of ${\rm Stab}(\e_1,\dots\e_L)$, i.e.~the
isotropy group of the parallel spinors.

\subsection{Geometry of the gravitino Killing spinor equation}

The solution of the gravitino Killing spinor equation can be read off from the results of \cite{het}.
The conditions that this imposes on the geometry is that
${\rm hol}(\hat\nabla)\subseteq SU(4)\ltimes \bR^8$.
This is equivalent to requiring that the spacetime admits the $\hat\nabla$-parallel forms
\bea
e^-,~~~e^-\wedge \omega_I~,~~~e^-\wedge {\rm Re}\,\chi~,~~~e^-\wedge {\rm Im}\,\chi~,
\eea
where
\bea
\omega_I=-(e^1\wedge e^6+e^2\wedge e^7+e^3\wedge e^8+e^4\wedge e^9)~,~~\chi=(e^1+i e^6)\wedge \dots \wedge(e^4+i e^9)~,
\eea
and $I$ is an endomorphism constructed by the metric and the two form $\omega_I$.
 In particular $I$ can be thought of as an ``almost complex''
structure in the ``transverse space'' to the light-cone directions. In the Hermitian light-cone frame
$e^+, e^-, e^\a, e^{\bar\a}$,   it has components  $I^\a{}_\b=i \delta^\a{}_\b$, $\a,\b=1,2,3,4$.

To continue, the metric and three-form can be written as in appendix A. In this case,
$\mathfrak{k}=\mathfrak{su}(4)$. So, $\mathfrak{su}(4)^\perp$ is spanned by the
(2,0)- and (0,2)-forms, and $\omega_I$ in $\Lambda^2(\bR^8)\otimes\bC$.
As we have explained in appendix A, the components of $H^{\mathfrak{su}(4)^\perp}_{-ij}$
 are determined by the geometry. In particular,
one finds that
\bea
H^{2,0+0,2}_{-ij}&=&-{1\over2}[i_I(\nabla_-\omega)]_{ij}={1\over2\cdot 3!} [(\nabla_-{\rm Re}\,\chi)_{ik_1k_2k_3}
{\rm Re}\,\chi_j{}^{k_1k_2k_3}]^{2,0+0,2}
\cr
&=&{1\over2\cdot 3!} [(\nabla_-{\rm Im}\,\chi)_{ik_1k_2k_3}
{\rm Im}\,\chi_j{}^{k_1k_2k_3}]^{2,0+0,2}~,
\cr
H_{-ij} \omega_I^{ij}&=&{1\over2\cdot 4!} (\nabla_-{\rm Re}\,\chi)_{k_1k_2k_3k_4} {\rm Im}\chi^{k_1k_2k_3k_4}~, ~~i,j, \dots=1,2,3,4,6,7,8,9.
\eea
Furthermore, the conditions along the transverse directions give
\bea
H^{\rm rest}&=&{1\over3!} H_{ijk} e^i\wedge e^j\wedge e^k
\cr
&=&-i_I \tilde d\omega_I- 2{\cal N}(I)=\star(\tilde d\omega_I\wedge \omega_I)
-{1\over2}\star
(\theta_{\omega_I}\wedge \omega_I\wedge \omega_I)+{\cal N}(I),~~
\eea
where $\tilde d$ denotes the exterior derivative projected along the eight directions transverse to the light-cone
 and the Hodge duality $\star$  operation\footnote{Note that $\star \psi_{i_1\dots i_{n-k}}={1\over k!} \psi_{j_1\dots j_k} \e^{j_1\dots j_k}{}_{i_1\dots i_{n-k}}$.}
is taken with volume form $d{\rm vol}=e^1\wedge\dots \wedge e^4\wedge e^6\wedge\dots\wedge e^9$.
For a similar expression for $H^{\rm rest}$ in the context of Riemannian geometry see \cite{stefang2}.
In addition
$\theta_{\omega_I}=-\star(\star\tilde d\omega_I\wedge \omega_I)$ is the Lee form of $\omega_I$, and ${\cal N}(I)$
is a (3,0) and (0,3) tensor, the Nijenhuis tensor of the endomorphism $I$, ${\cal N}(I)_{\a\b\g}=4H_{\a\b\g}$.
It remains to find the conditions on the geometry. It turns out that
\bea
(de_+)^{2,0+0,2}_{ij}&=&-{1\over2}[i_I(\nabla_+\omega)]_{ij}~,
\cr
(de_+)_{ij} \omega^{ij}&=&{1\over2\cdot 4!} (\nabla_+{\rm Re}\,\chi)_{k_1k_2k_3k_4} {\rm Im}\chi^{k_1k_2k_3k_4}~,
\cr
W_2&=&0~,~~~\theta_\omega=\theta_{{\rm Re}\,\chi}~,
\eea
where $de^-=\eta^{-+} de_+$ and $\theta_{{\rm Re}\,\chi}=-{1\over4} \star(\star \tilde d {\rm Re}\,\chi\wedge {\rm Re}\,\chi)$
is the Lee form of ${\rm Re}\,\chi$. The first two conditions are required for the compatibility of determining
$H_{+ij}^{\mathfrak{k}^\perp}$ in terms of both the Lie derivative of $\omega$ and $\chi$ along the parallel vector field.
The last two conditions, are required for the existence of $H^{\rm rest}$. $W_2$ is one of the Gray-Hervella classes
for determining $U(n)$ structures \cite{grayhervella}. The vanishing of $W_2$ implies that that the Nijenhuis tensor
is skew-symmetric in all three indices.  The non-vanishing of the Nijenhuis tensor indicates that
the endomorphism $I$ is not integrable. The equality of the Lee forms $\theta_\omega$ and $\theta_{{\rm Re}\chi}$ can also
be expressed as a condition on $SU(4)$ classes by saying that $W_4=W_5$.

\subsection{Geometry of $N=1$ supersymmetric backgrounds}

The solution of the dilatino Killing spinor equation is
\bea
&&\partial_+\Phi=0~,~~~H_{+\a}{}^\a=0~,~~~-H_{+\bar\a_1\bar\a_2}+{1\over2} H_{+\b_1\b_2} \e^{\b_1\b_2}{}_{\bar\a_1\bar\a_2}=0~,
\cr
&&\partial_{\bar\a}\Phi+{1\over6} H_{\b_1\b_2\b_3}\e^{\b_1\b_2\b_3}{}_{\bar\a}-{1\over2} H_{\bar\a \b}{}^\b-{1\over2} H_{-+\bar\a}=0~.
\la{dn=1}
\eea
This is the same as that which has been found in \cite{het} for $N=1$ supersymmetric $Spin(7)\ltimes\bR^8$
backgrounds.
The above conditions are in addition to those we have stated in the previous section for the existence of a solution
to the gravitino Killing spinor equation. In particular, (\ref{dn=1}) can be rewritten as
\bea
&&\partial_+\Phi=0~,~~~de^-\in \mathfrak{spin}(7)\oplus_s\bR^8~,
\cr
&&(d\Phi)_i+{1\over8\cdot 3!} {\cal N}_{k_1k_2k_3} ({\rm Re}\,\chi)^{k_1k_2k_3}{}_i-{1\over2} (\theta_{\omega_I})_i-{1\over2} H_{-+i}=0~,
\la{dn=1sp2}
\eea
The space of spacetime two-forms decomposes under the action of $Spin(7)\ltimes\bR^8$
into irreducible representations. The condition $de^-\in \mathfrak{spin}(7)\oplus_s\bR^8$ means that the two-form $de^-$ takes values
in the  $\mathfrak{spin}(7)\oplus_s\bR^8$ subspace. This is equivalent to writing $de^-=\a+e^-\wedge\b$, where $\a$ is a two-form  with
values in the ${\bf 21}$ irreducible representation in  the decomposition of the transverse  two-forms in $Spin(7)$ representations
and $\b$ is a transverse one-form.
Alternatively, this condition can be written as
\bea
(de^-)_{ij} \omega^{ij}=0~,~~~(de^-)_{ij}={1\over4} (de^-)_{kl}\, {\rm Re\, \chi}_{ij}{}^{kl}~.
\eea
Both these conditions can be thought of  as additional conditions on the geometry of spacetime. The components
of $de^-$ that lie in $\mathfrak{su}^\perp(4)\subset \mathfrak{spin}(7)$ are not required to vanish.
The components of $de^-$ along $\mathfrak{su}(4)\oplus_s\bR^8$ are not restricted by the Killing spinor equations.
The last condition in (\ref{dn=1sp2}) is a generalization of the conformal balance condition that it is well-known
for some supersymmetric type I backgrounds, see e.g.~\cite{ivanovgp}. Variations of this condition
appear in all solutions of dilatino Killing spinor equation for all descendants.

\subsection{Comparison with N=2}
It is instructive to compare the conditions we have found  for the
$N=1$ backgrounds  with the results of \cite{het} for the $N=L=2$
backgrounds. The solution of the dilatino Killing spinor equation
is \bea
&&\partial_+\Phi=0~,~~~H_{+\a}{}^\a=0~,~~~H_{+\bar\a_1\bar\a_2}=H_{\b_1\b_2\b_3}=0~,
\cr &&\partial_{\bar\a}\Phi-{1\over2} H_{\bar\a \b}{}^\b-{1\over2}
H_{-+\bar\a}=0~. \la{dn=2} \eea This can also be rewritten as \bea
&&\partial_+\Phi=0~,~~~{\cal N}(I)_{ijk}=0~,~~~de^-\in
\mathfrak{su}(4)\oplus_s\bR^8\subset
\mathfrak{spin}(7)\oplus_s\bR^8~, \cr &&(d\Phi)_i-{1\over2}
(\theta_{\omega_I})_i-{1\over2} H_{-+i}=0~, \la{dn=2sp2} \eea i.e.~the endomorphism $I$ is integrable and both $e^-\wedge \omega_I$
and $e^-\wedge \chi$ are invariant under the action of the
$\hat\nabla$-parallel vector field $e_+$, i.e. \bea {\cal
L}_{e_+}(e^-\wedge \omega_I)={\cal L}_{e_+}(e^-\wedge \chi)=0~.
\eea
 The conditions that arise from the gravitino Killing spinor equation are the same.
The differences  of $N=1$ and $N=2$ backgrounds are summarized in table 2. The isotropy group ${\rm Stab}_{\Sigma}$ of the $N=1$
Killing spinor in
$\Sigma({\cal P})$ is also tabulated.

 \begin{table}[ht]
 \begin{center}
\begin{tabular}{|c|c|c|c|}\hline
$SU(4)\ltimes \bR^8$ & $de^-$& ${\cal N}$&${\rm Stab}_{\Sigma}$
 \\ \hline \hline
$N=1$& $\mathfrak{spin}(7)\oplus_s\bR^8$& ${\cal N}(I)\not=0 $ & \{1\}\\
\hline
 $N=2$&$\mathfrak{su}(4)\oplus_s\bR^8$ &${\cal N}(I)=0 $&
\\ \hline
\end{tabular}
\end{center}
\caption{The differences in the geometry of $N=1$ and $N=2$ backgrounds are in the non-vanishing
components of $de^-$ and ${\cal N}(I)$. It is understood that the remaining conditions
of the dilatino Killing spinor equation for $N=1$ backgrounds are valid.}
\end{table}

\newsection{The descendants of $Sp(2)\ltimes\bR^8$}
A basis in the space of parallel spinors can be chosen as
\bea
1+e_{1234}~, ~~~i(1-e_{1234})~,~~~ i(e_{12}+e_{34})~,
\eea
i.e.~${\cal P}=\bR<1+e_{1234}, i(1-e_{1234}), i(e_{12}+e_{34})>$.
It is easy to see that in this case  $\Sigma({\cal P})=Spin(1,1)\times SU(2)$, where $SU(2)$ acts on ${\cal P}$ with the
three-dimensional representation. In particular, in the basis given above   $\mathfrak{su}(2)$ is spanned by
$\Gamma^{\bar 1\bar 2}-\Gamma^{34}, \Gamma^{1 2}-\Gamma^{\bar 3\bar 4}, {i\over2} (\Gamma^{1\bar1}+\Gamma^{2\bar2}+
\Gamma^{3\bar3}+\Gamma^{4\bar4})$, and the generator of $Spin(1,1)$ is $\Gamma^{+-}$. {}From these,
it is straightforward to find the $N=1$ and $N=2$ descendants.

As in the previous case, we first solve the gravitino Killing spinor equation. The conditions that arise
from the analysis of both the gravitino and dilatino Killing spinor equations in all cases can be most efficiently
organized as conditions on two endomorphisms $I$ and $J$. It turns out that this is a generic feature
of all cases that have parallel spinors with non-compact isotropy groups. In every new case, we shall introduce an appropriate  new
endomorphism.

\subsection{Geometry of the gravitino Killing spinor equation}

The geometry of the gravitino Killing spinor equations can be investigated as in the $SU(4)\ltimes\bR^8$ case.
The difference is that there are three $\hat\nabla$-parallel three-forms, instead of one, associated with the
Hermitian forms of an almost hyper-complex structure. Let $\{I_r, r=1,2,3\}=\{I,J,K\}$ be endomorphisms such that
$I_r I_s=-\delta_{rs}{\bf 1}_{8\times8}+\epsilon_{rst} I_t$. Then if the forms
\bea
&&e^-~,~~~e^-\wedge \omega_I~,~~~e^-\wedge\omega_J~,~~~
\eea
are $\hat\nabla$-parallel, then ${\rm hol}(\hat\nabla)\subseteq Sp(2)\ltimes\bR^8$,
where $\omega_I, \omega_J$ and $\omega_K$ are the associated
Hermitian forms. One can easily show that $e^-\wedge \omega_K$ is parallel as well.
In particular $\omega_I$ can be chosen as in the $SU(4)\ltimes\bR^8$ case while
$\omega_J=2 {\rm Re}(e^1\wedge e^2+  e^3\wedge e^4)$.

The conditions on the geometry can be described as two
copies of those of the $SU(4)\ltimes\bR^8$ case, with each copy associated with one of the endomorphisms $I, J$, that have to be valid simultaneously.
To proceed, we have to identify the directions that lie in $\mathfrak{sp}(2)^\perp$,
$\Lambda^2(\bR^8)=\mathfrak{sp}(2)\oplus\mathfrak{sp}(2)^\perp$. For this first observe that $\mathfrak{sp}(2)$ is spanned
by the (1,1)-forms in $\Lambda^2(\bR^8)\otimes\bC$ with respect to both $I$ and $J$. Thus $\mathfrak{sp}(2)^\perp$ is
spanned by the (2,0)- and (0,2)-forms with respect to $I$, and
those (1,1)-forms with respect to $I$ that are  (2,0) and (0,2) with respect to $J$. So if one sets
$H^{\mathfrak{sp}(2)^\perp}_-=(H^{2,0+0,2}_-,  \check H^{1,1}_{-})$, then
one can write
\bea
H^{2,0+0,2}_{-ij}&=&-{1\over2}[i_I(\nabla_-\omega_I)]_{ij}~,~~~
\cr
 \check H^{1,1}_{-ij}&=&-{1\over2}[i_J(\nabla_-\omega_J)]_{ij}^{1,1}~,
\eea
where the projections (2,0), (0,2) and (1,1) have been taken with respect to the $I$ endomorphism.
In addition, we get the geometric conditions
\bea
(de_+)^{2,0+0,2}_{ij}&=&-{1\over2}[i_I(\nabla_+\omega_I)]_{ij}~,~~~
\cr
{\check {(de_+)}}^{1,1}_{ij}&=&-{1\over2}[i_J(\nabla_+\omega_J)]_{ij}^{1,1}~.
\la{psp2con}
\eea

Furthermore, one finds that
\bea
H^{\rm rest}&=&-i_I \tilde d\omega_I- 2{\cal N}(I)=\star(\tilde d\omega_I\wedge \omega_I)-{1\over2}\star
(\theta_{\omega_I}\wedge \omega_I\wedge \omega_I)+{\cal N}(I)
\cr
&=&-i_J \tilde d\omega_J- 2{\cal N}(J)=\star(\tilde d\omega_J\wedge \omega_J)-{1\over2}\star
(\theta_{\omega_J}\wedge \omega_J\wedge \omega_J)+{\cal N}(J)~.
\eea
The equality involving the $I$ and $J$ expression should be interpreted as a condition on the geometry.
Moreover $W_2(I)=W_2(J)=0$ which is equivalent to the condition that the Nijenhuis tensor of both
$I$ and $J$ is skew-symmetric.

\subsection{N=1}
The dilatino Killing spinor equation is
\bea
{\cal A}(1+e_{1234})=0~,
\eea
The solution has been given in  (\ref{dn=1}) or equivalently (\ref{dn=1sp2}).

\subsection{N=2}
The dilatino Killing spinor equation is
\bea
{\cal A}1=0~.
\eea
The solution to the dilatino Killing spinor equation has already been given in either (\ref{dn=2}) or equivalently (\ref{dn=2sp2}).

\subsection{Comparison with N=3}

The conditions that arise form the dilatino Killing spinor equation in this case have been computed
in \cite{het} and  can be summarized as
\bea
&&\partial_+\Phi=0~,~~~de^-\in \mathfrak{sp}(2)\oplus_s\bR^8~,~~~{\cal N}(I)_{ijk}={\cal N}(J)_{ijk}=0~,
\cr
&&2\partial_i\Phi - H_{-+i}=(\theta_{\omega_I})_i=(\theta_{\omega_J})_i~.
\la{dn=3sp2}
\eea

The conditions on the geometry that arise from the gravitino Killing spinor equation are the same in all cases. The differences
arise in the
solution to the  dilatino Killing spinor equation  and have been summarized in table 3. We also give ${\rm Stab}_\Sigma$.

\begin{table}[ht]
 \begin{center}
\begin{tabular}{|c|c|c|c|c|}\hline
$Sp(2)\ltimes \bR^8$ & $de^-$& ${\cal N}$&$\theta$&${\rm Stab}_{\Sigma}$
 \\ \hline \hline
$N=1$& $\mathfrak{spin}(7)\oplus_s\bR^8$& ${\cal N}(I), {\cal N}(J)\not=0$ &$-$ &$U(1)$\\
\hline
 $N=2$&$\mathfrak{su}(4)\oplus_s\bR^8$ &${\cal N}(I)=0, {\cal N}(J)\not=0$&$-$&
\\ \hline
 $N=3$&$\mathfrak{sp}(2)\oplus_s\bR^8$ &${\cal N}(I)={\cal N}(J)=0$&$\theta_{\omega_I}=\theta_{\omega_J}$&
\\ \hline
\end{tabular}
\end{center}
\caption{The differences in the geometry of $N=1$, $N=2$ and $N=3$ backgrounds are in the non-vanishing
components of $de^-$, and ${\cal N}(I)$ and ${\cal N}(J)$, and the relation between the Lee forms. $-$
indicates that there is no relation between the Lee forms. It is understood that the remaining conditions
of the dilatino Killing spinor equation for  $N=1$ backgrounds are valid.}
\end{table}

\newsection{The descendants of $(SU(2)\times SU(2))\ltimes \bR^8$}

A complex basis in the space of $(SU(2)\times SU(2))\ltimes \bR^8$-invariant spinors is
\bea
1~,~~~e_{12}~.
\eea
It is easy to see that in this case $\Sigma({\cal P})=Spin(1,1)\times Sp(1)_L\times Sp(1)_R$. Identifying
${\cal P}=\bH$, $Sp(1)_L\times Sp(1)_R$ acts as
\bea
x\rightarrow ax \bar b~,~~~x\in \bH~,~~~ a\in Sp(1)_L~,~~~b\in Sp(1)_R~.
\eea
In addition, $Spin(1,1)$ has the generator $\Gamma^{+-}$.
There is a single type of orbit in ${\cal P}$ with stability subgroup $Sp(1)$ acting with the three-dimensional
representation on the remaining space. {}From this, one can easily determine the Killing spinors for all cases.
The ${\rm Stab}_{\Sigma}$ groups are given in table 4.

 \begin{table}[ht]
 \begin{center}
\begin{tabular}{|c|c|}\hline
$N$&${\rm Stab}_{\Sigma}$\\ \hline \hline
$1$&$SU(2)$\\ \hline
 $2$&$U(1)$ \\ \hline
\end{tabular}
\end{center}
\caption{The first column denotes the number of supersymmetries and the second column the stability
subgroups of  Killing spinors for $N\leq 2$ in $\Sigma({\cal P})=Spin(1,1)\times Sp(1)_L\times Sp(1)_R$.}
\end{table}

\subsection{Geometry of the gravitino Killing spinor equation}

The gravitino Killing spinor equation implies that ${\rm hol}(\hat\nabla)\subseteq (SU(2)\times SU(2))\ltimes \bR^8$.
In turn this is equivalent to requiring \cite{het} that the  forms
\bea
e^-~,~~~e^-\wedge \omega_1~,~~~e^-\wedge \omega_2~,~~~e^-\wedge \chi_1~,~~~e^-\wedge\chi_2~,
\eea
are $\hat\nabla$-parallel,
where $\omega_1=-i(e^1\wedge e^{\bar 1}+e^2\wedge e^{\bar 2})$, $\omega_2=-i(e^3\wedge e^{\bar 3}+e^4\wedge e^{\bar 4})$,
$\chi_1=2 e^1\wedge e^2$, and $\chi_2=2 e^3\wedge e^4$.
In this case $\mathfrak{k}^\perp$ is spanned by the (2,0) and (0,2) forms of the endomorphisms $I$,  $J$ and $L$, where
 $\omega_I=\omega_1+\omega_2$,    $\omega_J={\rm Re} (\chi_1+\chi_2)$  and $\omega_L=\omega_1-\omega_2$.
 The endomorphisms
satisfy the algebra
\bea
I^2=J^2=L^2=-1_{8\times 8}~,~~~IJ=-JI~,~~~IL=LI~,~~~JL=-LJ~.
\eea
In addition $(I,J, K=IJ)$ and $(L, M=ILJ, N=-IJ)$ are almost hyper-complex structures, and $ P=IL$ is an almost product structure.
 The geometric conditions that arise from the gravitino Killing spinor equation
are those that arise from three $U(4)\ltimes\bR^8$ structures each associated with $I,J$ and $L$, respectively.
  Applying the results of appendix A, we find the
geometric conditions
\bea
&&(de_+)_{\a\b}=-{1\over2} (i_{I_1} \nabla_+\omega_1)_{\a\b}~,~~~
(de_+)_{pq}=-{1\over2} (i_{I_2} \nabla_+\omega_2)_{pq}
\cr
&&(de_+)_{p \a }=-2(i_{I_1}\nabla_+\omega_1)_{p\a}~,~~~
(de_+)_{\bar p\a}=-2(i_{I_1}\nabla_+\omega_1)_{\bar p\a}
\cr
&& (de_+)_{ij}\, \omega^{ij}_1=(\nabla_+{\rm Re}\chi_1)_{ij}\, {\rm Im} \chi_1^{ij}~,~~~(\nabla_+\omega_1)_{pq}=0~,
\cr
&&(de_+)_{ij}\, \omega^{ij}_2=(\nabla_+{\rm Re}\chi_2)_{ij}\, {\rm Im} \chi_2^{ij}~,~~~(\nabla_+\omega_2)_{\a\b}=0~,
\eea
where $\a,\b=1,2$ and $p,q=3,4$.
{}From these it is also straightforward to express the components of $H_-^{\mathfrak{k}^\perp}$ in terms of the geometry.
In particular, one has
that
\bea
&&H_{-\a\b}=-{1\over2} (i_{I_1} \nabla_-\omega_1)_{\a\b}=(\nabla_-{\rm Re} \chi_1)_{\a i}\, {\rm Re}(\chi_1)_{\b}{}^i
+(\nabla_-{\rm Im}\, \chi_1)_{\a i}\, {\rm Im}(\chi_1)_{\b}{}^i
\cr
&&H_{-pq}=-{1\over2} (i_{I_2} \nabla_-\omega_2)_{pq}=(\nabla_-{\rm Re} \chi_2)_{p i}\, {\rm Re}(\chi_2)_{q}{}^i
+(\nabla_-{\rm Im}\, \chi_2)_{p i}\, {\rm Im}(\chi_2)_{q}{}^i
\cr
&&H_{-p \a }=-2(i_{I_1}\nabla_-\omega_1)_{p\a}=-2(i_{I_2}\nabla_-\omega_2)_{p\a}
=2 (\nabla_-{\rm Re} \chi_1)_{pi} \,
({\rm Re}\, \chi_1)_\a{}^i
\cr
&&~~~~~~~~~~~~~~~~~~~~~~~~
=-2(\nabla_-{\rm Re} \chi_2)_{\a i} \,
({\rm Re}\, \chi_2)_{p}{}^i~,
\cr
&&H_{-\bar p\a}=-2(i_{I_1}\nabla_-\omega_1)_{\bar p\a}=-2(i_{I_2}\nabla_-\omega_2)_{\bar p\a}
=2 (\nabla_-{\rm Re} \chi_1)_{\bar pi} \,
({\rm Re}\, \chi_1)_\a{}^i~,
\cr
&&~~~~~~~~~~~~~~~~~~~~~~~~
=-2(\nabla_-{\rm Re} \chi_2)_{\a i} \,
({\rm Re}\, \chi_2)_{\bar p}{}^i~,
\cr
&& H_{-ij}\, \omega^{ij}_1=(\nabla_-{\rm Re}\chi_1)_{ij}\, {\rm Im} \chi_1^{ij}~,~~~(\nabla_-\omega_1)_{pq}=0~,
\cr
&&H_{-ij}\, \omega^{ij}_2=(\nabla_-{\rm Re}\chi_2)_{ij}\, {\rm Im} \chi_2^{ij}~,~~~(\nabla_-\omega_2)_{\a\b}=0~.
\eea
This concludes the analysis of the conditions along the light-cone directions.

Next consider  the parallel transport equations along the transverse directions. It turns out that
\bea
H^{\rm rest}&=&-i_I\tilde d\omega_I- 2{\cal N}(I)=\star(\tilde d\omega_I\wedge \omega_I)-{1\over2}\star
(\theta_{\omega_I}\wedge \omega_I\wedge \omega_I)+{\cal N}(I)
\cr
&=&-i_J \tilde d\omega_J- 2{\cal N}(J)=\star(\tilde d\omega_J\wedge \omega_J)-{1\over2}\star
(\theta_{\omega_J}\wedge \omega_J\wedge \omega_J)+{\cal N}(J)
\cr
&=&-i_L \tilde d\omega_L- 2{\cal N}(L)=\star(\tilde d\omega_L\wedge \omega_L)-{1\over2}\star
(\theta_{\omega_L}\wedge \omega_L\wedge \omega_L)+{\cal N}(L)~.
\la{transv}
\eea
In addition, the $W_2$ Gray-Hervella classes for each endomorphism should also vanish, i.e.
\bea
W_2(I)=W_2(J)=W_2(L)=0~.
\la{w2}
\eea
This in turn implies that the Nijenhuis tensors of all endomorphisms are skew-symmetric.

\subsection{N=1}
As in previous $N=1$ cases the dilatino Killing spinor equation is
\bea
{\cal A}(1+e_{1234})=0~.
\eea
The solution has been given in either (\ref{dn=1}) or equivalently (\ref{dn=1sp2}). It can be easily
decomposed in $SU(2)\times SU(2)$ representations but the way it is stated in (\ref{dn=1sp2}) suffices
for our purpose.

\subsection{N=2}

The dilatino Killing spinor equation is
\bea
{\cal A}1=0~.
\eea
The solution has been given in either (\ref{dn=2}) or equivalently (\ref{dn=2sp2}).
Again the solution can be easily decomposed in $SU(2)\times SU(2)$ representations
but the way it has been expressed in (\ref{dn=2sp2}) will suffice.

\subsection{N=3}

 The dilatino Killing spinor equation is
\bea
{\cal A} 1=0~,~~~{\cal A}(e_{12}-e_{34})=0~.
\eea
The solution of the dilatino Killing spinor equation is the same as for the $N=3$ supersymmetric backgrounds
with $Sp(2)\ltimes\bR^8$-invariant parallel spinors. These conditions have already been stated in (\ref{dn=3sp2}).

\subsection{Comparison with N=4}

The solution of the dilatino Killing spinor equation has been given in \cite{het}. This is summarized
as
\bea
&&\partial_+\Phi=0~,~~~de^-\in (\mathfrak{su}(2)\oplus \mathfrak{su}(2))\oplus_s\bR^8~,
\cr
&&{\cal N}(I)_{ijk}={\cal N}(J)_{ijk}={\cal N}(L)_{ijk}=0~,
\cr
&&
2\partial_i\Phi- H_{-+i}=(\theta_{\omega_I})_i=(\theta_{\omega_J})_i=(\theta_{\omega_L})_i~.
\la{dn=4su22}
\eea
The conditions that arise from the gravitino Killing spinor equation are the same in all cases. The differences
in the geometry of the descendants that arise from the dilatino Killing spinor equation are summarized in  table 5.

\begin{table}[ht]
 \begin{center}
\begin{tabular}{|c|c|c|c|}\hline
$\times^2 SU(2)\ltimes \bR^8$ & $de^-$& ${\cal N}$&$\theta$
 \\ \hline \hline
$N=1$& $\mathfrak{spin}(7)\oplus_s\bR^8$& ${\cal N}(I), {\cal N}(J), {\cal N}(L)\not=0$ &$-$\\
\hline
 $N=2$&$\mathfrak{su}(4)\oplus_s\bR^8$ &${\cal N}(I)=0, {\cal N}(J), {\cal N}(J)\not=0$&$-$
\\ \hline
 $N=3$&$\mathfrak{sp}(2)\oplus_s\bR^8$ &${\cal N}(I)={\cal N}(J)=0, {\cal N}(L)\not=0$&$\theta_{\omega_I}=\theta_{\omega_J}$
\\ \hline
$N=4$&$(\mathfrak{su}(2)\oplus \mathfrak{su}(2))\oplus_s\bR^8$ &${\cal N}(I)={\cal N}(J)={\cal N}(L)=0$&$\theta_{\omega_I}=\theta_{\omega_J}=
\theta_{\omega_L}$
\\ \hline
\end{tabular}
\end{center}
\caption{The differences in the geometry of descendants  are in the non-vanishing
components of $de^-$, and ${\cal N}(I)$,  ${\cal N}(J)$ and ${\cal N}(L)$, and the relation between the Lee forms. $-$
indicates that there is no relation between the Lee forms. It is understood that the remaining conditions
of the dilatino Killing spinor equation of $N=1$ supersymmetric
backgrounds are valid.  }
\end{table}

\newsection{$SU(2)\ltimes \bR^8$ and its descendants}

A basis in the space of $\hat\nabla$-parallel spinors  ${\cal P}$ is
\bea
1~,~~~ e_{12}~,~~~e_{13}+e_{24}~.
\eea
It can be easily verified by a direct computation that the above spinors are invariant under  $SU(2)\ltimes\bR^8$,
where ${\mathfrak{su}}(2)$ is generated by
\bea
i ( \Gamma^{1 \bar 1} - \Gamma^{2 \bar 2}
     - \Gamma^{3 \bar 3} + \Gamma^{4 \bar 4})~,~~~
i( \Gamma^{1 \bar 2} + \Gamma^{2 \bar 1}
   - \Gamma^{3 \bar 4} - \Gamma^{4 \bar 3})~,~~~
  \Gamma^{1 \bar 2} - \Gamma^{2 \bar 1}
     + \Gamma^{3 \bar 4} - \Gamma^{4 \bar 3}~.
  \eea
 Alternatively  observe that chiral Majorana-Weyl representation of $Spin(8)$, $\Delta^+_{\bf 8}$, decomposes
under $SU(2)\times SU(2)$ as $\Delta^+_{\bf 8}=\oplus^4\bR\oplus \bH$, where
the first four  directions are spanned by  the  $SU(2)\times SU(2)$-invariant spinors. Moreover $SU(2)\times SU(2)$
acts on $\bH$ by left and right quaternionic multiplication. Consequently, the
diagonal $SU(2)$ subgroup leaves invariant an additional spinor, and so $SU(2)\ltimes\bR^8$ leaves invariant five spinors.

To investigate the descendants of $SU(2)\ltimes \bR^8$ first observe that $\Sigma({\cal P})=Spin(1,1)\times Sp(2)$, where
$Sp(2)$ acts on ${\cal P}$ with the five-dimensional vector representation, $Sp(2)=Spin(5)$. This can be  verified either
by a direct computation or by observing that any three linearly independent spinors in $\Delta^+_{\bf 8}$ have stability
subgroup $Sp(2)\subset Spin(8)$. Again $Spin(1,1)$ is generated by $\Gamma^{+-}$.
The group $Sp(2)$ acts transitively on the $ S^4\subset {\cal P}$ with stability subgroup $SU(2)\times SU(2)$. Using this,
it is easy to construct all the descendants. The ${\rm Stab}_{\Sigma}$ groups are collected in table 6.
\begin{table}[ht]
 \begin{center}
\begin{tabular}{|c|c|}\hline
$N$&${\rm Stab}_{\Sigma}$\\ \hline \hline
$1$&$SU(2)\times SU(2)$\\ \hline
 $2$&$SU(2)$ \\ \hline
\end{tabular}
\end{center}
\caption{The first column denotes the number of supersymmetries and the second column the stability
subgroups of  Killing spinors for $N\leq 2$ in $\Sigma({\cal P})=Spin(1,1)\times Sp(2)$.}
\end{table}

\subsection{The geometry of the gravitino Killing spinor equation}

The gravitino Killing spinor equation is
\bea
\hat\nabla 1=\hat\nabla e_{12}=\hat\nabla (e_{13}+e_{24})=0~,
\eea
which implies  that ${\rm hol}(\hat\nabla)\subseteq SU(2)\ltimes\bR^8$. In turn this  is equivalent to requiring that
the forms
\bea
e^-~,~~~e^-\wedge \omega_I~,~~~e^-\wedge\omega_J~,~~~e^-\wedge \omega_L~,~~~e^-\wedge \omega_Q
\eea
are $\hat\nabla$-parallel, where the first four forms are defined as in the $(SU(2)\ltimes SU(2))\ltimes \bR^8$ case and
\bea
\omega_Q= e^1\wedge e^3+ e^2\wedge e^4+e^{\bar 1}\wedge e^{\bar 3}+ e^{\bar 2}\wedge e^{\bar 4}~.
\eea
The form spinor bilinears are given in appendix D.
The new endomorphism $Q$ satisfies the algebraic conditions
\bea
IQ=-QI~,~~JQ=QJ~,~~~QL=LQ~,~~~Q^2=-{\bf 1}_{8\times 8}~.
\eea
It is clear that  $Q$ is on the same footing as the other three. The $\mathfrak{su}(2)^\perp$ is spanned
by those forms in $\Lambda^2(\bR^8)$ which are  (2,0) and (0,2) with respect to all endomorphisms.
So the conditions on the geometry
are four copies of those that we have found for $SU(4)\ltimes\bR^8$.
In particular, the gravitino Killing spinor equation along the directions transverse to the light-cone gives
conditions like (\ref{transv}) and (\ref{w2}) but now for all endomorphisms $I$, $J$, $L$ and $Q$, see also the general
analysis of appendix A. The results are tabulated in table 7.

\subsection{N=1}
The dilatino Killing spinor equation is
\bea
{\cal A}(1+e_{1234})=0
\eea
The solution has been given in (\ref{dn=1sp2}).
\subsection{N=2}
The dilatino Killing spinor equations are
\bea
{\cal A}1=0
\eea
The solution has been given in (\ref{dn=2sp2}).
\subsection{N=3}
The dilatino Killing spinor equations are
\bea
{\cal A}1={\cal A}(e_{12}-e_{34})=0
\eea
The solution has been given in (\ref{dn=3sp2}).
\subsection{N=4}
The dilatino Killing spinor equations are
\bea
{\cal A}1={\cal A}e_{12}=0
\eea
The solution has been given in (\ref{dn=4su22}).
\subsection{N=5 and comparison with the descendants}
The dilatino Killing spinor equations are
\bea
{\cal A}1={\cal A}e_{12}={\cal A}(e_{13}+e_{24})=0~.
\eea
The solution to the dilatino Killing spinor equation is
\bea
&&\partial_+\Phi=0~,~~~de^-\in \mathfrak{su}(2)\oplus_s\bR^8~,
\cr
&&{\cal N}(I)_{ijk}={\cal N}(J)_{ijk}={\cal N}(L)_{ijk}={\cal N}(Q)_{ijk}=0~,
\cr
&& 2\partial_i\Phi-H_{-+i}=(\theta_{\omega_I})_i=(\theta_{\omega_J})_i=(\theta_{\omega_L})_i=(\theta_{\omega_Q})_i~.
\la{dn=5su2}
\eea

The conditions that arise from the gravitino Killing spinor equation are the same in all cases. The $N=5$ case and the descendants
differ in the conditions that arise from the dilatino Killing spinor equation. The differences are summarized in table 7.

\begin{table}[ht]
 \begin{center}
\begin{tabular}{|c|c|c|c|}\hline
$SU(2)\ltimes \bR^8$ & $de^-$& ${\cal N}$&$\theta$
 \\ \hline \hline
$N=1$& $\mathfrak{spin}(7)\oplus_s\bR^8$& ${\cal N}(I), {\cal N}(J), $ &$-$
\\
&&${\cal N}(L), {\cal N}(Q)\not=0$&
\\
\hline
 $N=2$&$\mathfrak{su}(4)\oplus_s\bR^8$ &${\cal N}(I)=0$&$-$
 \\
&&${\cal N}(J), {\cal N}(J), {\cal N}(Q)\not=0$&
\\ \hline
 $N=3$&$\mathfrak{sp}(2)\oplus_s\bR^8$ &${\cal N}(I)={\cal N}(J)=0, $&$\theta_{\omega_I}=\theta_{\omega_J}$
 \\
 &&${\cal N}(L), {\cal N}(Q)\not=0$&
\\ \hline
$N=4$&$(\mathfrak{su}(2)\oplus \mathfrak{su}(2))\oplus_s\bR^8$
&${\cal N}(I)={\cal N}(J)={\cal N}(L)=0 $&$\theta_{\omega_I}=\theta_{\omega_J}=
\theta_{\omega_L}$
\\
&&${\cal N}(Q)\not=0$&
\\ \hline
$N=5$&$\mathfrak{su}(2)\oplus_s\bR^8$ &${\cal N}(I)={\cal N}(J)=$&$\theta_{\omega_I}=\theta_{\omega_J}=
\theta_{\omega_L}=\theta_{\omega_Q}$
\\
&&${\cal N}(L)={\cal N}(Q)=0$&
\\ \hline
\end{tabular}
\end{center}
\caption{As in previous cases, the differences in the geometry of descendants  are in the non-vanishing
components of $de^-$, and ${\cal N}(I)$,  ${\cal N}(J)$,  ${\cal N}(L)$, and ${\cal N}(Q)$ and the relation between the Lee forms. $-$
indicates that there is no relation between the Lee forms. It is understood that the remaining conditions
of the dilatino Killing spinor equation of $N=1$ supersymmetric
backgrounds are valid. }
\end{table}

\newsection{ $U(1)\ltimes \bR^8$ and its descendants}

A complex basis in the space of parallel spinors  ${\cal P}$ is
\bea
1,~~~ e_{12}~,~~~e_{13}~.
\eea
The presence of backgrounds with six parallel spinors
is a direct consequence of the previous $SU(2)\ltimes \bR^8$ case. To see this, we decompose $\Delta_{\bf 8}^+$ under $SU(2)$ as
$\Delta_{\bf 8}^+=\oplus^5\bR\oplus \bR^3$, where the first five singlets span the five  $SU(2)\ltimes \bR^8$-invariant spinors.
Since $SU(2)$ acts with the vector representation
on $\bR^3$, there is an additional invariant spinor with stability subgroup
$U(1)\ltimes \bR^8$. In the basis chosen above,  $\mathfrak{u}(1)$ is generated by
\bea
i ( \Gamma^{1 \bar 1} - \Gamma^{2 \bar 2}
     - \Gamma^{3 \bar 3} + \Gamma^{4 \bar 4})~.
     \eea

To investigate the descendants of $U(1)\ltimes \bR^8$ first observe that $\Sigma({\cal P})=Spin(1,1)\times SU(4)$, where
$SU(4)$ acts on ${\cal P}$ with the real six-dimensional vector representation, $SU(4)=Spin(6)$. This can easily be seen
from previous results by a direct commutation. Alternatively, it is a consequence of the fact that the stability subgroup in $Spin(8)$
of two linearly independent
spinors in $\Delta_{\bf 8}^+$ is $SU(4)$, and that $SU(4)$ acts on the remaining spinors with the six-dimensional representation.
The descendants can be easily found using group theory and the observation that $SU(4)$ acts transitively on the
$S^5$ in ${\cal P}$ with stability subgroup $Sp(2)$. The ${\rm Stab}_{\Sigma}$ groups have been collected in table 8.

\begin{table}[ht]
 \begin{center}
\begin{tabular}{|c|c|}\hline
$N$&${\rm Stab}_{\Sigma}$\\ \hline \hline
$1$&$Sp(2)$ \\ \hline
$2$&$SU(2)\times SU(2)$\\ \hline
 $3$&$SU(2)$ \\ \hline
\end{tabular}
\end{center}
\caption{The first column denotes the number of supersymmetries and the second column the stability
subgroups of  Killing spinors for $N\leq 3$ in $\Sigma({\cal P})=Spin(1,1)\times SU(4)$.}
\end{table}

\subsection{The geometry of the gravitino Killing spinor equation}
The gravitino Killing spinor equation is
\bea
\hat\nabla 1=\hat\nabla e_{12}=\hat\nabla e_{13}=0~.
\eea
This is equivalent to requiring that ${\rm hol}(\hat\nabla)\subseteq U(1)\ltimes\bR^8$. Explicitly, the solution is
 \begin{align}
  & \hat{\Omega}_{A,B+} = \hat{\Omega}_{A,\a\b} = \hat{\Omega}_{A, \a
  \bar \b} = 0 \,, \quad (\a \neq \b) \,, \notag \\
  & \hat{\Omega}_{A,1 \bar 1} = - \hat{\Omega}_{A,2 \bar 2} = - \hat{\Omega}_{A,3 \bar 3}
  = \hat{\Omega}_{A,4 \bar 4} \,.
 \end{align}
The condition that ${\rm hol}(\hat\nabla)\subseteq U(1)\ltimes\bR^8$ is also equivalent to requiring that
the forms
\bea
e^-~,~~~e^-\wedge \omega_I~,~~~e^-\wedge\omega_J~,~~~e^-\wedge \omega_L~,~~~e^-\wedge \omega_Q~,~~~e^-\wedge \omega_T~,
\eea
are $\hat\nabla$-parallel, where the first five forms are defined as in the $SU(2)\ltimes \bR^8$ case and
\bea
\omega_T=-i (e^1\wedge e^{\bar 1}- e^2\wedge e^{\bar 2}+e^{3}\wedge e^{\bar 3}- e^{4}\wedge e^{\bar 4})~.
\eea
The form spinor bilinears are given in appendix D.
The new endomorphism obeys the algebraic conditions
\bea
IT=TI~,~~~JT=TJ~,~~~TL=LT~,~~~TQ=-QT~,~~~T^2=-{\bf 1}_{8\times 8}~.
\eea
It is clear that the endomorphism $T$ is on the same footing as the other four. So the conditions on the geometry
are five copies of those that we have found for $SU(4)\ltimes\bR^8$ case. In particular,
 the gravitino Killing spinor equation along the directions transverse to the light-cone gives
conditions like (\ref{transv}) and (\ref{w2}) but now for all endomorphisms $I$, $J$, $L$, $Q$ and $T$, see also
appendix A. The results are tabulated in table 9.

\subsection{N=1}
The dilatino Killing spinor equation is
\bea
{\cal A}(1+e_{1234})=0~.
\eea
The solution has been given in (\ref{dn=1sp2}).

\subsection{N=2}
The dilatino Killing spinor equation is
\bea
{\cal A}1=0~.
\eea
The solution has been given in (\ref{dn=2sp2}).

\subsection{N=3}
The dilatino Killing spinor equations are
\bea
{\cal A}1={\cal A}(e_{12}+e_{34})=0~.
\eea
The solution has been given in (\ref{dn=3sp2}).
\subsection{N=4}
The dilatino Killing spinor equations are
\bea
{\cal A}1={\cal A}e_{12}=0
\eea
The solution has been given in (\ref{dn=4su22}).
\subsection{N=5}
 The dilatino Killing spinor equations are
\bea
{\cal A}1={\cal A}e_{12}={\cal A}(e_{13}+e_{24})=0
\eea
The solution has been given in (\ref{dn=5su2}).
\subsection{N=6 and comparison with the descendants}
The dilatino Killing spinor equations are
\bea
{\cal A}1={\cal A}e_{12}={\cal A}e_{13}=0~.
\eea
The solution can be written as
\bea
&&\partial_+\Phi=0~,~~~de^-\in \mathfrak{u}(1)\oplus_s\bR^8~,
\cr
&&{\cal N}(I)_{ijk}={\cal N}(J)_{ijk}={\cal N}(L)_{ijk}={\cal N}(Q)_{ijk}={\cal N}(T)_{ijk}=0~,
\cr
&& 2\partial_i\Phi-H_{-+i}=(\theta_{\omega_I})_i=(\theta_{\omega_J})_i=(\theta_{\omega_L})_i=(\theta_{\omega_Q})_i=(\theta_{\omega_T})_i~.
\la{dn=6u1}
\eea
Again, the conditions that arise from the gravitino Killing spinor equation are common to all cases. So the differences
arise from the conditions implied by the dilatino Killing spinor equation. These have been summarized in table 9.

\begin{table}[ht]
 \begin{center}
\begin{tabular}{|c|c|c|c|}\hline
$U(1)\ltimes \bR^8$ & $de^-$& ${\cal N}$&$\theta$
 \\ \hline \hline
$N=1$& $\mathfrak{spin}(7)\oplus_s\bR^8$& ${\cal N}(I), {\cal N}(J), $ &$-$
\\
&&${\cal N}(L), {\cal N}(Q), {\cal N}(T)\not=0$&
\\
\hline
 $N=2$&$\mathfrak{su}(4)\oplus_s\bR^8$ &${\cal N}(I)=0$&$-$
 \\
&&${\cal N}(J), {\cal N}(J), {\cal N}(Q), {\cal N}(T)\not=0$&
\\ \hline
 $N=3$&$\mathfrak{sp}(2)\oplus_s\bR^8$ &${\cal N}(I)={\cal N}(J)=0, $&$\theta_{\omega_I}=\theta_{\omega_J}$
 \\
 &&${\cal N}(L), {\cal N}(Q), {\cal N}(T)\not=0$&
\\ \hline
$N=4$&$(\mathfrak{su}(2)\oplus \mathfrak{su}(2))\oplus_s\bR^8$
&${\cal N}(I)={\cal N}(J)={\cal N}(L)=0 $&$\theta_{\omega_I}=\theta_{\omega_J}=
\theta_{\omega_L}$
\\
&&${\cal N}(Q), {\cal N}(T)\not=0$&
\\ \hline
$N=5$&$\mathfrak{su}(2)\oplus_s\bR^8$ &${\cal N}(I)={\cal N}(J)=$&$\theta_{\omega_I}=\theta_{\omega_J}=$
\\
&&${\cal N}(L)={\cal N}(Q)=0, {\cal N}(T)\not=0$& $\theta_{\omega_L}=\theta_{\omega_Q}$
\\ \hline
$N=6$&$\mathfrak{u}(1)\oplus_s\bR^8$ &${\cal N}(I)={\cal N}(J)=$&$\theta_{\omega_I}=\theta_{\omega_J}=
$
\\
&&${\cal N}(L)={\cal N}(Q)= {\cal N}(T)=0$&$\theta_{\omega_L}=\theta_{\omega_Q}=\theta_{\omega_T}$
\\ \hline
\end{tabular}
\end{center}
\caption{As in previous cases, the differences in the geometry of descendants  are in the non-vanishing
components of $de^-$, and ${\cal N}(I)$,  ${\cal N}(J)$,  ${\cal N}(L)$,  ${\cal N}(Q)$ and ${\cal N}(T)$,
and the relation between the Lee forms. $-$
indicates that there is no relation between the Lee forms. It is understood that the remaining conditions
of the dilatino Killing spinor equation of $N=1$ supersymmetric
backgrounds are valid. }
\end{table}

\newsection{The descendants of $\bR^8$}

A complex basis in the space of $\hat\nabla$-parallel spinors  ${\cal P}$  is
\bea
1~,~~~e_{ij}~,~~~i,j\leq 4.
\eea
It is clear that these spinors are invariant under $\bR^8$.  Direct inspection reveals that ${\cal P}$  can be identified  with the
 positive chirality Majorana-Weyl representation $\Delta_{\bf 8}^+$ of $Spin(8)$, ${\cal P}=\Delta_{\bf 8}^+$.
 Using this, we find that
$\Sigma({\cal P})=Spin(1,1)\times Spin(8)$, where the generator of $Spin(1,1)$ is $\Gamma^{+-}$.

In the investigation of the descendants with $N>4$ it is also necessary to consider the
normals to the parallel spinors. Using the definitions in section two,
 ${\cal Q}$ can also be identified with the
positive chirality  Majorana-Weyl representation  $\Delta_{\bf 8}^+$ of $Spin(8)$, ${\cal Q}=\Delta_{\bf 8}^+$.
The identification of  descendants of $\bR^8$ is the most involved so far. Because of this, we shall
describe each case in more detail. The descendants can be easily found using group theory and the observation that $Spin(7)$
acts transitively on the
$S^7$ in ${\cal P}$ with stability subgroup $Spin(7)$. The ${\rm Stab}_{\Sigma}$ groups have been collected in table 10.

\begin{table}[ht]
 \begin{center}
\begin{tabular}{|c|c|}\hline
$N$&${\rm Stab}_{\Sigma}$\\ \hline \hline
$1$&$Spin(7)$ \\ \hline
$2$&$SU(4)$ \\ \hline
$3$&$Sp(2)$ \\ \hline
$4$&$SU(2)\times SU(2)$\\ \hline
\end{tabular}
\end{center}
\caption{The first column denotes the number of supersymmetries and the second column the stability
subgroups of  Killing spinors for $N\leq 4$ in $\Sigma({\cal P})=Spin(1,1)\times Spin(8)$.}
\end{table}

\subsection{Geometry of the gravitino Killing spinor equation}

The condition that ${\rm hol}(\hat\nabla)\subseteq \bR^8$ is equivalent to requiring that
the forms
\bea
e^-~,~~~e^-\wedge e^i~,~~~i=1,2,3,4,6,7,8,9~.
\la{pppp}
\eea
are $\hat\nabla$-parallel. In this case, all the components of $H$ are determined in terms of the geometry. To see
this define the one-forms $(v_i)=\delta_{ij} e^j$. Then
\bea
H_{ijk}=-2 \nabla_i(v_j)_k~,~~~~H_{-ij} =-2 \nabla_-(v_i)_j~.
\eea
In addition, one also has the geometric conditions
\bea
\nabla_{i}(v_j)_k=\nabla_{[i}(v_j)_{k]}~,~~~(de_+)_{ij}=-2\nabla_+(v_i)_j~.
\eea
We can also describe the solution of the gravitino Killing spinor equation by choosing, $e^-, e^-\wedge \omega$, as $\hat\nabla$-parallel forms, where
$\omega$ is a shorthand for a basis in the space of two-forms.
This would have been more uniform with previous cases but the choice of the parallel forms in (\ref{pppp}), even though they are not associated
 with spinor bilinears,  leads to a simpler
description of the spacetime geometry.
As in previous cases, the dilatino Killing spinor equation imposes additional conditions on the fluxes and geometry.

\subsection{N=1}

As we have explained in section two, to choose the first Killing spinor in ${\cal P}$, it suffices to find the orbits
of  $\Sigma({\cal P})=Spin(1,1)\times Spin(8) $ in ${\cal P}= \Delta_{\bf 8}^+$. There is only one type of orbit
of co-dimension zero which has stability subgroup $Spin(7)$ in $\Sigma({\cal P})$. In particular, the
 dilatino Killing spinor equation is
\bea
{\cal A} (1+e_{1234})=0~.
\eea
The solution of this equation expressed in $Spin(7)$ representations is given in \cite{het} and reads
\bea
&&\partial_+\Phi=0~,~~~de^-\in \mathfrak{spin}(7)\oplus_s\bR^8~,
\cr
&&\partial_i\Phi-{1\over2}(\theta_\phi)_i-{1\over2}H_{-+i}=0~,
\eea
where $\theta_\phi=-{1\over6}\star(\star \tilde d\phi\wedge \phi)$ is the Lee form of the $Spin(7)$-invariant form $\phi$, and
the Hodge dual has been taken with respect to the volume form $d{\rm vol}=e^1\wedge e^2\wedge e^3\wedge e^4\wedge e^6\wedge e^7
\wedge e^8\wedge e^9$.

\subsection{N=2}

The first Killing spinor $\e$ is chosen as in the $N=1$ case above, $\e_1=\e$. To choose the direction of the second Killing spinor $\e_2$
observe that $\Delta^+_{\bf 8}$ decomposes under the stability subgroup of the first normal as $\Delta_{\bf 8}^+=\bR<1+e_{1234}>
\oplus \Lambda^1_{\bf 7}(\bR^7)$, i.e.~${\cal P}/{\cal K}=\Lambda^1_{\bf 7}(\bR^7)$. In addition,
 ${\rm Stab}({\cal K})=Spin(1,1)\times Spin(7)$, where $Spin(7)$ is the stability
subgroup of $\e_1$. Since $Spin(7)$ acts transitively on the sphere in the space of one-forms of $\bR^7$, ${\rm Stab}({\cal P})$
has a single orbit
in $\Lambda^1_{\bf 7}(\bR^7)$ of codimension zero with stability subgroup $SU(4)$. So we can choose
$\e_2=i(1-e_{1234})$.
The dilatino Killing spinor equation is
\bea
{\cal A} 1=0~.
\eea
The solution organized in $SU(4)$ representations is given in (\ref{dn=2sp2}).

\subsection{N=3}

Next consider
${\cal K}=\bR<\e_1,\e_2>$, where $e_1, \e_2$ are the Killing spinors of the $N=2$ case above,
 and observe that ${\rm Stab}({\cal P})= Spin(1,1)\times SU(4)\times U(1)$. This group\footnote{There may be discrete
 identifications in ${\rm Stab}({\cal P})$ that we do not take into account because the analysis is focused
 on the Lie algebra level, i.e.~one may have instead
 ${\rm Stab}({\cal P})= Spin(1,1)\times(SU(4)\cdot U(1))$.}  is constructed from the
stability subgroup $SU(4)$ of both spinors, a $U(1)$ generated by $i\Gamma^{1\bar1}$ which rotates $\e_1$ and $\e_2$ and
a boost (scaling)  generated by $\Gamma^{+-}$. Next observe that under ${\rm Stab}({\cal K})$, ${\cal P}$ decomposes
as ${\cal P}={\cal K}\oplus {\rm Re}\Lambda^2_{\bf 6}(\bC^4)$, and so ${\cal P}/{\cal K}={\rm Re}\Lambda^2_{\bf 6}(\bC^4)$.
For this, we have used the decomposition
$\Lambda^1_{\bf 7}(\bR^7)=\bR<i(1-e_{1234})>\oplus {\rm Re}\Lambda^2_{\bf 6}(\bC^4)$ under $SU(4)$, where $SU(4)=Spin(6)$
acts with the vector representation\footnote{The reality condition in $\Lambda^2(\bC^4)$ is defined by the anti-linear
map $\tau$ constructed from
complex conjugation followed by a duality map. Observe that this commutes with the $SU(4)$ action on $\Lambda^2(\bC^4)$.
So ${\rm Re} \Lambda^2(\bC^4)$
is defined as the fixed point set of $\tau$.}  on ${\rm Re}\Lambda^2_{\bf 6}(\bC^4)=\bR^6$.
Thus $Spin(1,1)\times SU(4)\times U(1)$ has one type of orbit in ${\rm Re}\Lambda^2_{\bf 6}(\bC^4)$ of codimension zero
with stability subgroup $Sp(2)\times U(1)$, where $Sp(2)=Spin(5)$.
Thus a representative can be chosen as
\bea
\e_3=i(e_{12}+e_{34})~.
\eea
The dilatino Killing spinor equation is
\bea
{\cal A}1={\cal A}  (e_{12}+e_{34})=0~.
\eea
The solution of this Killing spinor equation has been given in (\ref{dn=3sp2}).

\subsection{N=4}

The $N=4$ can  be investigated in two ways. One is to use the gauge symmetry either to specify the  Killing spinors
or to determine their normals. It is the ``self-dual'' case under the correspondence
\bea
N\longleftrightarrow 8-N~.
\eea
The two ways of examining $N=4$ are equivalent, so without loss of generality, we shall determine  the Killing spinors.
We begin by choosing the first three Killing spinors, $\e_1, \e_2, \e_3$ as in the $N=3$ case above.
 To determine the forth Killing spinor $\e_4$,  let ${\cal K}=\bR<\e_1,\e_2, \e_3>$ be the vector space spanned by the
 three Killing spinors.
 First observe that $\Sigma({\cal K})=Spin(1,1)\times Sp(2)\times SU(2)$, where $Sp(2)$ is the stability
 subgroup of the first three spinors and $SU(2)$ acts on them with the vector representation. It suffices to focus on $Sp(2)$.
 To determine ${\cal P}/{\cal K}$ recall the results from $N=3$ and   observe that
 ${\rm Re}\,\Lambda^2(\bC^4)= \bR<i(e_{12}+e_{34})>\oplus \Lambda^1_{\bf 5}(\bR^5)$
under $Sp(2)$, therefore ${\cal P}/{\cal K}=\Lambda^1_{\bf 5}(\bR^5)$. Moreover $Sp(2)=Spin(5)$ acts with the vector
representation on ${\cal P}/{\cal K}$ and so it has a unique type of  orbit $S^4$ with stability subgroup $Spin(4)=SU(2)\times SU(2)$.
In fact ${\rm Stab}({\cal K})$ has an orbit in ${\cal P}/{\cal K}$ of codimension zero, a representative can be chosen as
\bea
\e_4=i(e_{12}+e_{34})~.
\eea
The  dilatino Killing spinor equation for $N=4$ backgrounds  becomes
\bea
{\cal A}1={\cal A}e_{12}=0~.
\eea
The solution of this has been given in (\ref{dn=4su22}). In table 11, we summarize the groups $\Sigma$ that have been used
in the identification of the descendants.

\begin{table}[ht]
 \begin{center}
\begin{tabular}{|c|c|}\hline
${\mathrm N}$ & $\Sigma$
 \\ \hline \hline
$1$& $Spin(1,1)\times Spin(8)$ \\
\hline
 $2$&$Spin(1,1) \times Spin(7)$
\\ \hline
$3$&$Spin(1,1) \times SU(4)\times U(1)$
\\ \hline
$4$&$Spin(1,1) \times Sp(2)\times SU(2)$
\\ \hline
\end{tabular}
\end{center}
\caption{For $N>4$ the same $\Sigma$ groups are used to determine the normals
of the Killing spinors.}
\end{table}

\subsection{N=5}

The selection of Killing spinors for the remaining $N>4$ backgrounds is straightforward from the analysis we have presented for the $N<4$
cases and the correspondence $N\leftrightarrow 8-N$. So the
dilatino Killing spinor equations for $N>4$ will be written down without further explanation. The choice of representatives is such that
the Killing spinors of $N$-supersymmetric backgrounds are included in the $N+1$-supersymmetric ones.

The dilatino Killing spinor equation is
\bea
{\cal A} 1={\cal A}e_{12}={\cal A}(e_{13}+e_{24})=0~.
\eea
The solution has been given in (\ref{dn=5su2}).

\subsection{N=6}
The dilatino Killing spinor equation is
\bea
{\cal A} 1={\cal A}e_{12}={\cal A}e_{13}=0~.
\eea
The solution has been given in (\ref{dn=6u1}).

\subsection{N=7}
The dilatino Killing spinor equation is
\bea
{\cal A} 1={\cal A}e_{12}={\cal A}e_{13}={\cal A}(e_{23}-e_{14})=0~.
\eea
The solution is
\bea
&&\partial_+\Phi=0~,~~~de^-\in \bR^8~,
\cr
&&{\cal N}(I)_{ijk}={\cal N}(J)_{ijk}={\cal N}(L)_{ijk}={\cal N}(Q)_{ijk}={\cal N}(T)_{ijk}={\cal N}(U)_{ijk}=0~,
\cr
&& 2\partial_i\Phi-H_{-+i}=(\theta_{\omega_I})_i=(\theta_{\omega_J})_i=(\theta_{\omega_L})_i=(\theta_{\omega_Q})_i=(\theta_{\omega_T})_i=(\theta_{\omega_U})_i~,
\la{dn=7}
\eea
where the sixth endomorphism $U$ is defined via the Hermitian form
 \begin{align}
  \omega_U = e^1 \wedge e^4 + e^{\bar 1} \wedge e^{\bar 4} - e^2 \wedge e^3 -
  e^{\bar 2} \wedge e^{\bar 3} \,.
 \end{align}
The new endomorphism satisfies the algebraic conditions
\bea
UI=-IU~,~~UJ=JU~,~~UL=LU~,~~UQ=-QU~,~~UT=TU,~~U^2=-{\bf 1}_{8\times 8}~.
\eea
The dilatino Killing spinor equations imply that all the Lee forms of the endomorphisms are equal. However, this does not imply
that all components of $H^{{\rm rest}}$ vanish. In particular,  the non-vanishing components are
\bea
&& \tfrac12 H_{1 \bar 2 \bar 3} = + H_{\bar 4 1 \bar 1} = - H_{\bar 4 2 \bar 2} = - H_{\bar 4 3 \bar 3}~,
\cr
 && \tfrac12 H_{4 \bar 2 \bar 3} = - H_{\bar 1 4 \bar 4} =  H_{\bar 1 2 \bar 2} = H_{\bar 1 3 \bar 3}~,
 \cr
&&  \tfrac12 H_{2 \bar 1 \bar 4} = + H_{\bar 3 2 \bar 2} = - H_{\bar 3 1 \bar 1} = - H_{\bar 3 4 \bar 4}~,
\cr
&&  \tfrac12 H_{3 \bar 1 \bar 4} = - H_{\bar 2 3 \bar 3} = H_{\bar 2 1 \bar 1}
= H_{\bar 2 4 \bar 4}~. \label{N=7-comp}
\eea
\subsection{N=8 and comparison with the descendants}
 The solution of the dilatino Killing spinor equation of $N=8$ supersymmetric backgrounds \cite{het}
is
\bea
\partial_+\Phi=0~,~~~de^-\in \bR^8~,~~~H_{ijk}=0~,~~~2\partial_i\Phi-H_{-+i}=0~.
\eea
The conditions that arise from the gravitino Killing spinor equation are common in all cases. The differences
arise from those of the dilatino Killing spinor equation. We have summarize these in table 12.

\begin{table}[ht]
 \begin{center}
\begin{tabular}{|c|c|c|c|}\hline
$SU(2)\ltimes \bR^8$ & $de^-$& ${\cal N}$&$\theta$
 \\ \hline \hline
$N=1$& $\mathfrak{spin}(7)\oplus_s\bR^8$& ${\cal N}(I), {\cal N}(J), {\cal N}(L)$ &$-$
\\
&&${\cal N}(Q), {\cal N}(T), {\cal N}(U)\not=0$&
\\
\hline
 $N=2$&$\mathfrak{su}(4)\oplus_s\bR^8$ &${\cal N}(I)=0, {\cal N}(J), {\cal N}(J),$&$-$
 \\
&&$ {\cal N}(Q), {\cal N}(T), {\cal N}(U)\not=0$&
\\ \hline
 $N=3$&$\mathfrak{sp}(2)\oplus_s\bR^8$ &${\cal N}(I)={\cal N}(J)=0, {\cal N}(L),$&$\theta_{\omega_I}=\theta_{\omega_J}$
 \\
 &&$ {\cal N}(Q), {\cal N}(T), {\cal N}(U)\not=0$&
\\ \hline
$N=4$&$(\mathfrak{su}(2)\oplus \mathfrak{su}(2))\oplus_s\bR^8$
&${\cal N}(I)={\cal N}(J)={\cal N}(L)=0 $&$\theta_{\omega_I}=\theta_{\omega_J}=
\theta_{\omega_L}$
\\
&&${\cal N}(Q), {\cal N}(T), {\cal N}(U)\not=0$&
\\ \hline
$N=5$&$\mathfrak{su}(2)\oplus_s\bR^8$ &${\cal N}(I)={\cal N}(J)={\cal N}(L)=$&$\theta_{\omega_I}=\theta_{\omega_J}=$
\\
&&${\cal N}(Q)=0, {\cal N}(T), {\cal N}(U)\not=0$& $\theta_{\omega_L}=\theta_{\omega_Q}$
\\ \hline
$N=6$&$\mathfrak{u}(1)\oplus_s\bR^8$ &${\cal N}(I)={\cal N}(J)={\cal N}(L)=$&$\theta_{\omega_I}=\theta_{\omega_J}=
$
\\
&&${\cal N}(Q)= {\cal N}(T)=0, {\cal N}(U)\not=0$&$\theta_{\omega_L}=\theta_{\omega_Q}=\theta_{\omega_T}$
\\ \hline
$N=7$&$\bR^8$ &${\cal N}(I)={\cal N}(J)={\cal N}(L)=$&$\theta_{\omega_I}=\theta_{\omega_J}=\theta_{\omega_L}=
$
\\
&&${\cal N}(Q)= {\cal N}(T)={\cal N}(U)=0$&$\theta_{\omega_Q}=\theta_{\omega_T}=\theta_{\omega_U}$
\\ \hline
$N=8$&$\bR^8$ &$H_{ijk}=0$ &
\\ \hline
\end{tabular}
\end{center}
\caption{As in previous cases, the differences in the geometry of descendants  are in the non-vanishing
components of $de^-$, and ${\cal N}(I)$,  ${\cal N}(J)$,  ${\cal N}(L)$,  ${\cal N}(Q)$, ${\cal N}(T)$ and
${\cal N}(U)$
and the relation between the Lee forms. In the $N=8$ case, $H^{\rm rest}=0$ and so all the Nijenhuis tensors and
Lee forms vanish. $-$
indicates that there is no relation between the Lee forms. It is understood that the remaining conditions
of the dilatino Killing spinor equation of $N=1$ supersymmetric
backgrounds are valid.}
\end{table}

\newsection{Descendants and reduction of holonomy}

So far we have solved the Killing spinor equations for all
supersymmetric backgrounds for which the stability subgroup of the
parallel spinors is non-compact,  ${\rm
Stab}(\e_1,\dots,\e_L)=K\ltimes\bR^8$. The question that arises is
whether the Bianchi identity of $H$ and the field equations impose
additional conditions on the existence of the various descendants
we have found. We shall show that\footnote{This assumption is
sufficient.  What is required is that the term involving $dH$ in
the appropriate  Bianchi identity in appendix A does not
contribute in the calculations for the parallel forms.} if \bea
dH=0~, \eea and the field equations are satisfied, then for the
descendants ${\rm hol}(\hat\nabla)\subset K\ltimes\bR^8$. So the
holonomy of the $\hat\nabla$-connection  is a {\it proper}
subgroup of the stability group of the parallel spinors. Since the
holonomy of $\hat\nabla$ reduces,  the structure group of the
spacetime may reduce as well. Alternatively, if \bea dH=0~,~~~{\rm
hol}(\hat\nabla)= K\ltimes\bR^8 \eea and the field equations are
satisfied, then the gravitino Killing spinor equations imply the
dilatino ones, and all $\hat\nabla$-parallel spinors are Killing.
So there are no descendants and the only backgrounds that exist
are those investigated in \cite{het}.

To establish these, we shall  investigate in detail the $\hat\nabla$-parallel forms on the spacetime
that arise as a consequence of the gravitino Killing spinor equation, $dH=0$ and the field
equations of type I backgrounds. We shall focus first on the
 $SU(4)\ltimes \bR^8$ case. We shall find that the spacetime may  admit more
parallel forms than those that may have been expected from the $SU(4)\ltimes \bR^8$ isotropy group of the Killing spinors alone.
As a consequence, we shall
show the two statements mentioned above.

\subsection{Parallel forms of  $SU(4)\ltimes \bR^8$  backgrounds }

Suppose that $dH=0$ and ${\rm hol}\subseteq SU(4)\ltimes \bR^8$. To find  additional parallel forms, we
use the integrability condition of the gravitino Killing spinor equation
as well the Bianchi identities of the $\hat R$ curvature. These have been summarized in appendix A.
Since we have assumed $dH=0$, the Bianchi identity gives
\bea
\hat R_{A[B,CD]}=-{1\over3} \hat\nabla_A H_{BCD}~.
\la{bbb}
\eea

 To proceed, set $B=+, C=\a, D=\bar\b$ in (\ref{bbb}) and contract with $\delta^{\a\bar \b}$. Using that
${\rm hol}(\hat\nabla)\subseteq SU(4)\ltimes\bR^8$, e.g.~$\hat R_{AB,\a}{}^\a=0$, it is easy to see that (\ref{bbb}) implies that
\bea
\tau_1=iH_{+\a}{}^\a\, e^+
\eea
is $\hat\nabla$-parallel. Therefore if $\tau_1\not=0$, then ${\rm hol}(\hat\nabla)\subset SU(4)$. However,
if we insist that ${\rm hol}(\hat\nabla)= SU(4)\ltimes\bR^8$, then $\tau_1=0$.

To continue, set $B=+, C=\a, D=\b$ in (\ref{bbb}) and use that ${\rm hol}(\hat\nabla)\subseteq SU(4)\ltimes\bR^8$,
i.e.~$\hat R_{AB,+i}=\hat R_{AB,\a\b}=0$, then it is
easy to show that the three-form
\bea
\tau_2={1\over2} H_{+\a\b}\, e^+\wedge e^\a\wedge e^\b~,
\eea
is $\hat\nabla$-parallel,
\bea
\hat\nabla_A \tau_2=0~,
\eea
Since there is no such form invariant under $SU(4)\ltimes\bR^8$, one can only conclude that either the holonomy
of $\hat\nabla$ reduces to a proper subgroup of $SU(4)\ltimes\bR^8$ or $\tau_2=0$.

Next set $B=\a, C=\b, D=\g$ in (\ref{bbb}), and use that ${\rm hol}(\hat\nabla)\subseteq SU(4)\ltimes\bR^8$ and $\tau_2=0$, to show that
\bea
\tau_3={1\over3!} H_{\a\b\g} e^\a\wedge e^\b\wedge e^\g
\eea
is also $\hat\nabla$-parallel, i.e.
\bea
\hat\nabla_A\tau_3=0~.
\eea
Again there is no such form invariant under $SU(4)\ltimes\bR^8$. So   either the holonomy
of $\hat\nabla$ reduces to a proper subgroup of $SU(4)\ltimes\bR^8$ or $\tau_3=0$. Insisting that
${\rm hol}(\hat\nabla)= SU(4)\ltimes\bR^8$, we have to set $\tau_3=0$.
The observation that the Nijenhuis tensor of a Riemannian manifolds with a $U(n)$-structure compatible
with a connection with skew-symmetric torsion $H$, $dH=0$, is $\hat\nabla$-parallel has been made
in the context of supersymmetric sigma models in \cite{wit, howegpf, stefang2}.

There  are two additional parallel one-forms which can be found
 using the field equations
 \bea
\hat R_{A C,}{}^C{}_B-2 \hat\nabla_A \partial_B\Phi=0~.
\la{fieldeq}
 \eea
 Setting  $B=+$ and using ${\rm hol}(\hat\nabla)\subseteq SU(4)\ltimes\bR^8$, one can show
 that the one-form
 \bea
 \tau_4=\partial_+\Phi\, e^+~,
 \eea
 is $\hat\nabla$-parallel. Since there is no such one-form invariant under $SU(4)\ltimes \bR^8$, either
the holonomy of ${\rm hol}(\hat\nabla)$ reduces to a subgroup of $SU(4)\ltimes\bR^8$ or $\tau_4=0$.

Next set  $B=\a, C=\b, D=\bar\g$ in (\ref{bbb}), take the trace in $\b, \bar\g$, and use $\tau_1=\tau_2=\tau_4=0$
and ${\rm hol}(\hat\nabla)\subseteq SU(4)\ltimes\bR^8$ to find
\bea
\hat R_{A\b,}{}^\b{}_{\a}=-[\partial_A H_{\a\b}{}^\b -\hat\Omega_{A,}{}^\d{}_\a H_{\d\b}{}^\b-\hat\Omega_{A,}{}^+{}_\b H_{\a+}{}^\b]
\eea
Similarly set $B=+, C=-, D=\a$ in  (\ref{bbb}), to get that
\bea
\hat R_{A +,-\a}=-[\partial_A H_{+-\a}-\hat\Omega_{A,}{}^\d{}_{\a} H_{+-\d}-\hat\Omega_{A,}{}^{\bar \b}{}_- H_{+\bar\b\a}]~.
\eea
Substituting these into the field equations
\bea
\hat R_{A\b,}{}^\b{}_\a+\hat R_{A+,-\a}-2\hat\nabla_A \partial_\a\Phi=0~,
\eea
 we find that the one-form
\bea
\tau_5=(2\partial_i \Phi- \theta_i+H_{+-i}) e^i~,
\eea
is $\hat\nabla$-parallel. Again, since there is no such one-form invariant under $SU(4)\ltimes \bR^8$, either
the holonomy of ${\rm hol}(\hat\nabla)$ reduces to a subgroup of $SU(4)\ltimes\bR^8$ or $\tau_5=0$.

 {}For backgrounds to have precisely $N=1$ supersymmetry,
  neither $\tau_2$ nor  $\tau_3$ should vanish. As a consequence of the analysis above,
${\rm hol}(\hat\nabla)\subset SU(4)\ltimes\bR^8$ and so the holonomy reduces to a proper subgroup of the isotropy
group of the parallel spinors.

Another consequence of the analysis above is that if
 $dH=0$,  the field equations are satisfied and ${\rm hol}(\hat\nabla)= SU(4)\ltimes\bR^8$, then all $\hat\nabla$-parallel
 spinors are Killing. This is because in such a case
 $\tau_1=\tau_2=\tau_3=\tau_4=\tau_5=0$ which are precisely the conditions (\ref{dn=2sp2}) that arise from the dilatino
 Killing spinor equation of $N=2$ backgrounds.

\subsection{Parallel forms and descendants}

We shall now turn to show the two statements stated in the beginning of the
section.  We will treat all cases together apart from the $N=7$ descendant of $\bR^8$,
which will be discussed separately.
We begin by constructing the forms $\tau_1$, $\tau_2$ and $\tau_3$ with respect
all the endomorphisms $I$, $J$, $L$, and so on, available in each case.

If one of these is non-vanishing, and so a descendant exists, then
the holonomy of $\hat\nabla$ reduces. This is because the invariant forms of
non-compact isotropy groups $K\ltimes \bR^8$ are of the type
\bea
e^-\wedge \psi
\eea
where $\psi$ are forms in the ``transverse'' directions. Since $\tau_1$, $\tau_2$, $\tau_3$ and $\tau_4$
are not of this type, one concludes that either the  holonomy reduces or they should vanish.

Assuming that $\tau_1=\tau_2=\tau_3=\tau_4=0$ with respect to all endomorphisms, one can show,  using the argument we have presented above
to establish that $\tau_5$ is parallel in the $SU(4)\ltimes \bR^8$ case, that all the differences of Lee forms
\bea
\theta_{\omega_I}-\theta_{\omega_J},~~~\theta_{\omega_I}-\theta_{\omega_L}~,
\eea
and so on, are also $\hat\nabla$-parallel. Since again these forms are not invariant under
$K\ltimes \bR^8$,  either they vanish or the holonomy of $\hat\nabla$ reduces to a subgroup of
$K\ltimes \bR^8$.
If they do not vanish, the holonomy reduces and so we have established the first statement. If they do vanish,
and so ${\rm hol}(\hat\nabla)=K\ltimes \bR^8$, they imply the dilatino Killing spinor
equations for all parallel spinors. This establishes the second statement.

One can allow the holonomy to be reduced. The pattern of
 reductions depends on the choice of  parallel forms $\tau_1,\tau_2,\tau_3, \tau_5$ that will be allowed not to vanish.
For example if $\tau_1\not=0$, but the rest are zero, then the holonomy reduces from $SU(4)\ltimes \bR^8$ to $SU(4)$.
Similarly if $\tau_1, \tau_3\not=0$ but the rest vanish, then the holonomy reduces to $SU(3)$ and so on.

The  pattern of reductions of holonomy in the other cases is more involved. For example, consider the $Sp(2)\ltimes\bR^8$ case.
Suppose that $\tau_1\not=0$. Then the holonomy reduces to $Sp(2)$. The holonomy can remain $Sp(2)$
even if $\tau_2\not=0$. This is because one can  take $\tau_2=e^+\wedge \omega_J$, where $\omega_J$ is the hermitian
form of the $J$ endomorphism associated with this case. Therefore
if one allows appropriate reductions of the holonomy group, many descendants may exist.

Finally consider the $N=7$ descendant of $\bR^8$. The dilatino Killing spinor equations imply that
 $\tau_1=\tau_2=\tau_3=\tau_4=\tau_5=0$. So it may appear that for this descendant, the holonomy does not reduce.
 However, this is not the case because there are additional parallel forms which are the non-vanishing
 components of $H_{ijk}$. In particular using (\ref{bbb}), ${\rm hol}(\hat\nabla)\subseteq \bR^8$ and $H_{+ij}=(de_+)_{ij}=0$, it is easy to see
 that the three-form
\bea
H^{\rm rest}={1\over3 !} H_{ijk} e^i\wedge e^j\wedge e^k~,
\eea
constructed from the components \eqref{N=7-comp}   is $\hat\nabla$-parallel.
  A direct inspection of the integrability condition reveals that if ${\rm
 hol}(\hat\nabla) = \bR^8$, then $H^{\rm rest}=0$ and there is supersymmetry
 enhancement to $N=8$. If some components of \eqref{N=7-comp} are
 non-vanishing the holonomy reduces, i.e.~${\rm hol}(\hat\nabla)\subset
 \bR^8$. If it reduces to the identity the background preserves at least 8
 supersymmetries.
This arises as a consequence of the conditions $dH=\hat R=0$ and the dilatino
 Killing spinor equation \cite{kawano, jfofhet}. The argument is also reviewed in section \ref{flat}.

\newsection{The descendants of $G_2$}

A basis in the space of parallel spinors is
\bea
1+e_{1234}~,~~~e_{15}+e_{2345}~.
\eea
Moreover  $\Sigma({\cal P})=Spin(2,1)$ which acts with the Majorana representation on ${\cal P}$.
There is a single descendant background with $N=1$ supersymmetry. The dilatino Killing spinor equation
 can be written as
\bea
{\cal A}(1+e_{1234})=0~.
\eea
The stability subgroup of this spinor  is given in table 13.
\begin{table}[ht]
 \begin{center}
\begin{tabular}{|c|c|}\hline
$N$&${\rm Stab}_{\Sigma}$\\ \hline \hline
$1$&$\bR$ \\ \hline
\end{tabular}
\end{center}
\caption{The first column denotes the number of supersymmetries and the second column the stability
subgroup of  Killing spinor  in $\Sigma({\cal P})=Spin(2,1)$.}
\end{table}

\subsection{Geometry of the gravitino Killing spinor equation}
The condition that the gravitino Killing spinor equation imposes on the geometry is that ${\rm hol}(\hat\nabla)\subseteq G_2$, and has
been investigated in \cite{het}.
This is equivalent to requiring that the forms
\bea
e^+~,~~~ e^-~,~~~e^{\underline 1}~,~~~\varphi~,
\eea
are $\hat\nabla$-parallel,  where $\varphi={\rm Re}\,[(e^2+i e^7)\wedge (e^3+i e^8)\wedge (e^4+i e^9)]- e^6\wedge (e^2\wedge e^7+e^3\wedge e^8+e^4\wedge e^9)$
is the $G_2$ invariant three-form.
It is clear that in this case there are three $\hat\nabla$-parallel one-forms which we shall call collectively\footnote{We have
underlined one direction to emphasize that it is real.} $e^a$, $a=+,-,\underline 1$. As we have already explained in appendix
A, the associated vector fields $e_a$ are Killing and $i_a H=\eta_{ab} de^b$.

The geometric condition that arises from the compatibility of $e^a$ and $\varphi$ conditions, see appendix A, is that
\bea
[(de^a)_{ij}]^{{\bf 7}}={1\over 6} \eta^{ab} \nabla_b\varphi_{mn[i}\varphi^{mn}{}_{j]}~,~~~i,j,k,\dots=2,3,4,6,7,8,9~.
\la{g2geomcon}
\eea
{}For this, we have used the decomposition $\Lambda^2(\bR^7)=\Lambda^2_{{\bf 7}}\oplus \Lambda^2_{{\bf 14}}$, where
$\Lambda^2_{{\bf 14}}=\mathfrak{g}_2$.
The remaining components of $H$ are determined as
\bea
H^{\rm rest}=-{1\over6} (\tilde d\varphi, \star\varphi) \varphi+ \star \tilde d\varphi-\star(\theta\wedge \varphi)~,
\la{rrr}
\eea
where
\bea
\theta=-{1\over3}\star(\star \tilde d\varphi\wedge \varphi)~.~~~
\eea
Moreover, $\tilde d$ denotes the projection of the exterior derivative along the transverse directions and the
$\star$ operation has been taken with volume form $d{\rm vol}=e^2\wedge e^3\wedge e^4\wedge e^6\wedge \dots\wedge e^9$.
The geometry of Riemannian seven-dimensional manifolds with $G_2$-structure \cite{gray} compatible with a connection
with skew-symmetric torsion has been examined in detail in  \cite{stefang2}.
{}For use later, a straightforward computation reveals that
\bea
\theta_i=-{1\over 6} H_{kmn} \star\varphi^{kmn}{}_i~.
\eea
In addition, one also finds the geometric (integrability) condition
\bea
\tilde d\star\varphi=-\theta\wedge \star\varphi~.
\eea
This is equivalent to requiring that the $G_2$ class $X_2$ associated with the ${\bf 14}$ representation vanishes, $X_2=0$.
This is the only condition required for the existence of (\ref{rrr}).
This concludes the description of the geometry of the gravitino Killing spinor equation.

\subsection{Geometry of $N=1$ supersymmetric backgrounds}

The solution of the dilatino equation is that which one derives for the
$Spin(7)\ltimes\bR^8$ backgrounds \cite{het}. Organizing the
conditions in $G_2$
representations,
one has
\bea
&&\partial_+\Phi=0~,~~~H_{+\underline 1 i}+{1\over2} H_{+mn} \varphi^{mn}{}_i=0~,~~~
\cr
&&\partial_{\underline 1}\Phi-{1\over12} H_{ijk} \varphi^{ijk}-{1\over2} H_{-+\underline 1}=0~,
\cr
&&
\partial_i\Phi-{1\over12} H_{jkm} \star\varphi^{jkm}{}_i-{1\over4} H_{{\underline 1} jk} \varphi^{jk}{}_i-{1\over2} H_{-+i}=0~.
\eea
Using the relation between $H$ and $e^a$ established in appendix A, the above conditions can be rewritten as
\bea
&&\partial_+\Phi=0~,~~~[e_+, e_{\underline 1}]_i -{1\over2} (de_+)_{mn} \varphi^{mn}{}_i=0~,
\cr
&&\partial_{\underline 1}\Phi-{1\over12} H_{ijk} \varphi^{ijk}-{1\over2} H_{-+\underline 1}=0~,
\cr
&&
\partial_i\Phi-{1\over2} \theta_i-{1\over4} (de_{\underline 1})_{jk} \varphi^{jk}{}_i+{1\over2} [e_-, e_+]_i=0~.
\la{dn=1g2}
\eea
Note that $H_{-+\underline 1}$ can also be written in terms of the $\hat\nabla$-parallel vector fields
$e_+, e_-, e_{\underline 1}$ as $H_{-+\underline 1}=-g([e_-, e_+], e_{\underline 1})$, see appendix A, but it is more convenient
for simplicity of notation to leave it as it is in the equations.

There are various ways to interpret the above conditions. First observe that $\Phi$ is invariant only under the action
of one of the three Killing vector fields. The second condition is expected from the $N=1$ $Spin(7)\ltimes \bR^8$
results and  decomposition of $\mathfrak{spin}(7)=
\mathfrak{g}_2\oplus \Lambda^1_{\bf 7}$
in $\mathfrak{g}_2$ representations. The third condition relates the singlet in the decomposition of $H^{\rm rest}$ in $G_2$ representations
to the structure constants of $H_{-+\underline 1}$ and the derivative of $\Phi$ along $e_{\underline 1}$.
Finally, the last condition can be thought of as a generalization of the conformal balanced condition. The additional
terms involve the rotation of $e^{\underline 1}$ and  the commutator $[e_-, e_+]$.

Let $\mathfrak{h}=\bR<e_-, e_+, e_{\underline 1}>$. If $[\mathfrak{h}, \mathfrak{h}]\subseteq \mathfrak{h}$, i.e.~the algebra of three $\hat\nabla$-parallel vector fields  closes, then the conditions (\ref{dn=1g2}) can be written as
\bea
&&\partial_+\Phi=0~,~~~ (de_+)_{mn} \varphi^{mn}{}_i=0~,~~~
\cr
&&\partial_{\underline 1}\Phi-{1\over12} H_{ijk} \varphi^{ijk}-{1\over2} H_{-+\underline 1}=0~,
\cr
&&
\partial_i\Phi-{1\over2} \theta_i-{1\over4} (de_{\underline 1})_{jk} \varphi^{jk}{}_i=0~.
\la{g2alg}
\eea
In such a case, the spacetime is a principal bundle over a seven-dimensional base space, see \cite{het} and appendix A.
There are two cases to consider. If the isometry group is abelian, the curvature ${\cal F}^-$ of the principal bundle
is a $\mathfrak{g}_2$ instanton, and
${\cal F}^+$ and ${\cal F}^{\underline 1}$
take values in $\mathfrak{so}(7)$. Though in the two latter cases in  (\ref{g2geomcon}),  $({\cal F}^+)_{\bf 7}$ and $({\cal F}^{\underline 1})_{\bf 7}$
are related to the covariant derivative of $\varphi$. It is clear from these that both the dilaton $\Phi$ and the
three-form bilinear $\varphi$ may depend on the coordinates of the fiber and so they are not functions of the base space only of the
principal fibration.
If the dilaton is invariant under $e_{\underline 1}$, then the singlet
in the decomposition of $H$ vanishes.

A similar conclusion can also be reached in the case that the Lie algebra of isometries is $\mathfrak{sl}(2,\bR)$.
One of the differences
is that the singlet in the decomposition of $H$ does not vanish even if the dilaton is invariant.
In fact it is related to the structure constants
of $\mathfrak{h}$ as it can be seen in the second equation in (\ref{g2alg}).

\subsection{  $N=2$ }

The solution of the dilatino Killing spinor equation can be found\footnote{In \cite{het}, the solution has been organized
in this way only for the case that the
 algebra of isometries closes.} in  \cite{het}.
It turns out that it can be written as
\bea
&&\partial_a\Phi=0~,~~~\e_a{}^{bc}[e_b, e_c]_i-(de_a)_{mn} \varphi^{mn}{}_i=0~, ~~~\e_{+-\underline 1}=1~,
\cr
&&
{1\over6} H_{ijk} \varphi^{ijk}+ H_{-+\underline 1}=0~,~~~\partial_i\Phi-{1\over2} \theta_i=0~.
\la{dn=2g2}
\eea
The dilaton is invariant under all the three Killing vector fields. Moreover all $(de_{ij}^a)_{\bf 7}$ are related to the
commutator $\epsilon_{a}{}^{bc} ([X_b, X_c])_i$. In the case that the algebra of the three isometries closes, ${\cal F}^a$ takes values in
$\mathfrak{g}_2$
and so the principal bundle connection is a $\mathfrak{g}_2$ instanton. The geometry has been investigated in detail in
\cite{het} and we shall not explain this further here.

The $N=1$ and $N=2$ differ. It is clear that the conditions that arise
from the dilatino Killing spinor equations in the two cases are not the same.
The main differences lie in the conditions on $de^a_{ij}$ and whether the dilaton
is invariant under the isometries of the backgrounds.

\subsection{ Reduction of holonomy }

Reduction of the holonomy group can happen in backgrounds with both $N=1$ and $N=2$ supersymmetry. This is unlike
the non-compact case where we have shown that the Bianchi identities and the field equations force
a reduction of the holonomy only for the descendants.
We shall again use the Bianchi identities to find the additional parallel forms on the spacetime.

To begin, suppose that $dH=0$. It has been shown in \cite{het} that  either $[\mathfrak{h}, \mathfrak{h}]\subseteq \mathfrak{h}$,
where $\mathfrak{h}=\bR<e_a>$, $a=-,+,\underline{1}$,  or the holonomy of ${\rm hol}(\hat\nabla)\subset G_2$. This is because\footnote{In fact a necessary is for $H$ to be invariant.}
if $dH=0$, then the commutator of two $\hat\nabla$-parallel vector fields is $\hat\nabla$-parallel, see \cite{het}.
Thus either $[\mathfrak{h}, \mathfrak{h}]\subseteq \mathfrak{h}$ or there is an additional linearly independent vector field
which is $\hat\nabla$-parallel and so ${\rm hol}(\hat\nabla)\subset G_2$, i.e.~the holonomy reduces. This can also be shown using the Bianchi identity (\ref{bbb}), see
also appendix A.

Applying the Bianchi identity (\ref{bbb}) for $B=a, C=b, D=c$, we can show that $H_{abc}$ are constant.
In addition  contracting the Bianchi identity (\ref{bbb}) for  $B=i, C=i, D=k$ with the $\varphi$,
and using the condition that ${\rm hol}(\hat\nabla)\subseteq G_2$, i.e.
\bea
\hat R_{AB, a D}=\hat R_{AB, ij} \varphi^{ij}{}_k=0~,
\la{hoco}
\eea
one can also show that $H_{ijk} \varphi^{ijk}$ is constant as well.

Using (\ref{hoco}) and the Bianchi identity (\ref{bbb}) for $A=a, B=i, C=j$, one can show
that the Lie-algebra valued one-form
\bea
\tau^a_1={1\over2}\, de^a_{ij}\, \varphi^{ij}{}_k\, e^k~,
\eea
is $\hat\nabla$-parallel. Since $\tau^a_1$ are linearly independent from $e^a$,  either $\tau^a_1$ vanishes or the
 holonomy of  $\hat\nabla$ reduces to a subgroup of $G_2$.
Observe that $\tau^a_1$ is the 7-dimensional component of $\tilde de^a$ in the decomposition of two-forms in $G_2$
representations.

Substituting $B=a$ in the field equations (\ref{fieldeq}) and using ${\rm hol}(\hat\nabla)\subset G_2$, it is easy to see
that
\bea
\tau_2=\partial_a \Phi \, e^a
\eea
is $\hat\nabla$-parallel. Since $e^a$ are  $\hat\nabla$-parallel as well, this implies that $\partial_a \Phi=v_a$ are constant.

Next write the $G_2$ holonomy condition as
\bea
{1\over2} \hat R_{AB, kl} \star\varphi^{kl}{}_{ij}=\hat R_{AB, ij}~.
\eea
Setting $B=m$, contracting $m$ and $i$, using the field equations (\ref{fieldeq}) and the Bianchi identity (\ref{bbb}), we find that
\bea
\tau_3= (2\partial_i\Phi-\theta_i) e^i
\eea
is $\hat\nabla$-parallel. Since this one-form is linearly independent from $e^a$ either $\tau_3=0$ or ${\rm hol}(\hat\nabla)\subset G_2$
and so the holonomy reduces.

First consider the consequences of the above $\hat\nabla$-parallel
forms in the $N=2$ backgrounds. If one insist that ${\rm
hol}(\hat\nabla)= G_2$, then the field equations and $dH=0$ imply
all the conditions (\ref{dn=2g2}) that arise from the dilatino
Killing spinor equation apart from $\partial_a\Phi=v_a=0$ and
$H_{ijk} \varphi^{ijk}=0$ if $\mathfrak{h}$ is abelian, or
$H_{ijk} \varphi^{ijk}+6 H_{-+\underline 1}=0$ if $\mathfrak{h}$
is non-abelian, respectively. (In the non-abelian case a simple argument implies
that $v_a=0$.) This is   unlike the non-compact case, where under
the same assumptions the gravitino Killing spinor equation implies
all the conditions of the dilatino Killing spinor equation for the
$N=L$ backgrounds.

Next consider the applications of the additional  parallel forms in $N=1$ backgrounds. If either
$[\mathfrak{h}, \mathfrak{h}]\nsubseteq  \mathfrak{h}$ and/or $\tau^a_1\not=0$, then ${\rm hol}(\hat\nabla)\subset G_2$ and so
the holonomy reduces. However, unlike the non-compact cases, there may be backgrounds with $N=1$ supersymmetry
and  ${\rm hol}(\hat\nabla)= G_2$. For example   take  $\mathfrak{h}$ abelian,
$\tau^a_1=0$ and $12 v_{\underline{1}}- H_{ijk}\varphi^{ijk}=0$.
This is a linear  dilaton background.

\newsection{The descendants of $SU(3)$}

A complex basis in the space of parallel $SU(3)$-invariant spinors  ${\cal P}$ is
\bea
1~,~~~e_{15}~.
\eea
In this case $\Sigma({\cal P})=Spin(3,1)\times U(1)$, where $Spin(3,1)=SL(2,\bC)$ acts on
 ${\cal P}$ with the Majorana spinor representation
and $U(1)$ is generated by ${i\over2}(\Gamma^{2\bar2}+\Gamma^{3\bar3})$.
The generic orbit of $\Sigma({\cal P})$ on ${\cal P}$ is of co-dimension one. To see this observe that
 the generic orbit of $Spin(3,1)$ on ${\cal P}$ is of co-dimension two
and so one can choose
\bea
\e=\l_1(1+e_{1234})+i \l_2 (1-e_{1234})~,~~\l_1^2+\l_2^2=1~.
\eea
 Moreover $U(1)$ rotates the two spinors that appear in the expression above. So
\bea
\e=\l_1(1+e_{1234})~.
\eea
It is straightforward to choose the Killing spinors in all cases. For this observe that
$\Sigma(\bR<1+e_{1234}>)=(Spin(1,1)\times U(1))\ltimes \bR^2$. We simply state the results in the appropriate sections.
The stability groups of the Killing spinors are summarized in table 14.
\begin{table}[ht]
 \begin{center}
\begin{tabular}{|c|c|}\hline
$N$&${\rm Stab}_{\Sigma}$\\ \hline \hline
$1$&$U(1)\ltimes \bR^2$ \\ \hline
$2$&$\bR^2, \{1\}$ \\ \hline
\end{tabular}
\end{center}
\caption{The first column denotes the number of supersymmetries and the second column the stability
subgroup of  Killing spinor  in $\Sigma({\cal P})=Spin(3,1)\times U(1)$.}
\end{table}

\subsection{Geometry of the gravitino Killing spinor equation}

The gravitino Killing spinor equation implies that ${\rm hol}(\hat\nabla)\subseteq SU(3)$. This is equivalent
to requiring \cite{het} that the forms
\bea
&&e^a~,~~~\omega=\omega_I=-e^2\wedge e^7-e^2\wedge e^8-e^4\wedge e^9~,~~~
\cr
&&\chi=(e^2+ie^7)\wedge (e^3+ie^8)\wedge (e^4+ie^9)
\eea
are $\hat\nabla$-parallel, where $a=+,-,1, \bar 1$.
 So there are the four $\hat\nabla$-parallel vector fields and six transverse directions.
In this case, $\mathfrak{k}^\perp$ is spanned by
(2,0) and (0,2) forms with respect to $I$,   and $\omega$ in $\Lambda^2(\bR^6)$.
As we have explained in appendix A, $H_{aij}^{\mathfrak{k}^\perp}$, $i,j=2,3,4,7,8,9$,  is determined both by $de^a$ and the Levi-Civita
covariant derivative of the remaining parallel forms. The compatibility between the different ways of expressing $H$ leads
to the geometric conditions
\bea
(de_a)^{2,0+0,2}_{ij}&=&-{1\over2}[i_I(\nabla_a\omega)]_{ij}
\cr
(de_a)_{ij} \omega^{ij}&=&{1\over6} (\nabla_a{\rm Re}\,\chi)_{k_1k_2k_3} {\rm Im}\chi^{k_1k_2k_3}~.
\eea
Moreover
\bea
H^{\rm rest}=-i_I\tilde d\omega- 2{\cal N}=\star \tilde d\omega-\star(\theta_\omega\wedge \omega)+{\cal N}~,
\eea
where $\theta_\omega=-\star(\star \tilde d\omega\wedge\omega)$. One also finds the
additional geometric constraints
\bea
W_2=0~,~~~\theta_\omega=\theta_{{\rm Re}\chi}~,
\eea
where again the vanishing of  the Gray-Hervella class $W_2$  implies that the Nijenhuis
tensor ${\cal N}$ is skew symmetric, and
$\theta_{{\rm Re}\chi}=-{1\over2} \star(\star\tilde d{\rm Re}\chi\wedge {\rm Re}\chi)$ is the Lee form of ${\rm Re}\chi$.
The geometry of six-dimensional Riemannian manifolds with an $SU(3)$-structure \cite{chiossi}  and compatible
connection with skew-symmetric torsion have  been extensively investigated, see \cite{strominger, hullb,  ivanovgp, ivanovgpb,
tseytlin, gauntlett,
lustb,  goldstein, poonb, li, becker}. The equality of the two Lee forms can also be expressed in $SU(3)$ classes as $W_4=W_5$.

It has been explained in \cite{het} if  $[\mathfrak{h}, \mathfrak{h}]\subseteq \mathfrak{h}$, where $\mathfrak{h}=\bR<e_a>$,
then $\mathfrak{h}$ is
either abelian, $\bR\oplus^3\mathfrak{u}(1)$,
 $\bR\oplus \mathfrak{su}(2)$,  $\mathfrak{u}(1)\oplus \mathfrak{sl}(2, \bR)$, or
a pp-wave algebra.
This concludes the description of the geometry of the gravitino Killing spinor equation.

\subsection{N=1}

In this case the dilatino Killing spinor equation is
\bea
{\cal A}( 1+e_{1234})=0~.
\eea
The solution of the dilatino Killing spinor equation decomposed in $SU(3)$ representations is
\bea
&&\partial_+\Phi=0~,~~~H_{+1\bar 1}+H_{+ n}{}^n=0~,~~~-H_{+\bar 1\bar n}+{1\over2} H_{+pq} \e^{pq}{}_{\bar n}=0~,
\cr
&&
\partial_{\bar1}\Phi-{1\over6} H_{pqn}\e^{pqn}-{1\over2} H_{\bar1 n}{}^n-{1\over2} H_{-+\bar 1}=0~,
\cr
&&\partial_{\bar n}\Phi+{1\over2} H_{1pq}\e^{pq}{}_{\bar n}-{1\over2} H_{\bar n p}{}^p-{1\over2} H_{\bar n 1\bar1}-{1\over2} H_{-+\bar n}=0~,
\la{n=1su3con}
\eea
where $p,q,n=2,3,4$.
The dilaton is  invariant under the $e_+$ isometry of the spacetime but not necessarily the rest.
The remaining conditions can be interpreted in different ways.
For example observe that the above condition  implies that
\bea
de^-\in \mathfrak{spin}(7)\oplus_s\bR^8\subset \mathfrak{so}(8)\oplus_s\bR^8~.
\eea
Alternatively, they can be seen as relating the structure constants and commutators of the Killing vector fields to
the $\mathfrak{su}^(3)^\perp$ components of $de^-$. In particular  (\ref{n=1su3con}) can be rewritten as
\bea
&&\partial_+\Phi=0~,~~~H_{+1\bar 1}-{i\over2} (de_+)_{ij}\omega^{ij}=0~,~~~[e_+, e_{\bar1}]_{\bar n}+
{1\over2} (de_+)_{pq} \e^{pq}{}_{\bar n}=0~.
\cr
&&\partial_{\bar1}\Phi-{1\over24} {\cal N}_{pqn}\e^{pqn}+{i\over4}  (de_{\bar1})_{ij}\omega^{ij}-
{1\over2} H_{-+\bar 1}=0~,
\cr
&&\partial_{\bar n}\Phi+{1\over2}  (de_1)_{pq} \e^{pq}{}_{\bar n}-{1\over2} \theta_{\bar n}+{1\over2} [e_1, e_{\bar 1}]_{\bar n}
 +{1\over2} [e_-, e_+]_{\bar n} =0~.
\eea

If  $[\mathfrak{h}, \mathfrak{h}]\subseteq \mathfrak{h}$, then one finds that
\bea
&&\partial_+\Phi=0~,~~~H_{+1\bar 1}-{i\over2} (de_+)_{ij}\omega^{ij}=0~,~~~
 (de_+)^{2,0}=0~.
\cr
&&\partial_{\bar1}\Phi-{1\over24} {\cal N}_{pqn}\e^{pqn}+{i\over4}  (de_{\bar1})_{ij}\omega^{ij}-
{1\over2} H_{-+\bar 1}=0~,
\cr
&&\partial_{\bar n}\Phi+{1\over2}  (de_1)_{pq} \e^{pq}{}_{\bar n}-{1\over2} (\theta_\omega)_{\bar n}=0~.
\eea
It is clear from this that although ${\cal L}_+ \omega=0$ this is not the case for the rest of the parallel vector fields.
In addition in all cases ${\cal L}_a \chi\not=0$, unless $\mathfrak{h}$ is abelian in which case
${\cal L}_+\chi=0$. This is in agreement with results in the maximal $SU(3)$ case. Observe that even if one sets
${\cal N}=0$, the geometry of the Killing spinor equations is different from
that of the $N=4$ case in \cite{het}.

\subsection{N=2}

The dilatino Killing spinor equation is
either
\bea
{\cal A}1=0~,
\eea
with ${\rm Stab}_{\Sigma}(1)=\bR^2$
or
\bea
{\cal A}(1+e_{1234})=0~,~~~{\cal A}(e_{15}+e_{2345})=0~.
\eea
with ${\rm Stab}_{\Sigma}(1+e_{1234}, e_{15}+e_{2345})=\{1\}$.
So there are two cases to consider.

\subsubsection{${\cal A}1=0$}

The solution of the dilatino Killing spinor equation ${\cal A}1=0$  is
\bea
&&\partial_+\Phi=0~,~~~H_{+1\bar1}+H_{+n}{}^n=0~,~~~H_{+1n}=H_{+mn}=H_{1mn}=H_{mpq}=0~,
\cr
&&\partial_{\bar1}\Phi-{1\over2} H_{\bar1 n}{}^n-{1\over2} H_{-+\bar1}=0~,~~~\partial_{\bar n}\Phi-{1\over2} H_{\bar n p}{}^p-{1\over2} H_{\bar n 1\bar1}-{1\over2} H_{-+\bar n}=0~,
\eea
The dilaton is invariant only under the isometries generated by $e_+$. The above conditions
imply that
\bea
&&\partial_+\Phi=0~,~~~[e_+, e_1]_n=0~,~~~(de^-)^{2,0}=0~,~~~(de^{\bar 1})^{2,0}=0~,~~~
\cr
&&H_{+1\bar 1}-{i\over2} (de_+)_{ij}\,\omega^{ij}=0~,~~~{\cal N}_{ijk}=0~,~~~
\partial_{\bar1}\Phi+{i\over4}  (de_{\bar1})_{ij}\omega^{ij}-{1\over2} H_{-+\bar 1}=0~,
\cr
&&\partial_{\bar n}\Phi-{1\over2} (\theta_\omega)_{\bar n}+{1\over2} [e_1, e_{\bar 1}]_{\bar n}
 +{1\over2} [e_-, e_+]_{\bar n} =0~.
\eea
The difference between $N=1$ and $N=2$ is the vanishing of ${\cal N}$ and the restriction on $de^{\bar1}$ to be a (2,0)-form.
Again ${\cal L}_a\chi\not=0$. In particular, $W_1=W_2=0$ for these backgrounds. If $[\mathfrak{h}, \mathfrak{h}]\subseteq \mathfrak{h}$,
the last condition is modified to
\bea
\partial_{\bar n}\Phi-{1\over2} \theta_{\bar n} =0~.
\eea

\subsubsection{{${\cal A}(1+e_{1234})={\cal A}(e_{15}+e_{2345})=0$}}

The solution of the dilatino Killing spinor equation in this case is
\bea
&& \partial_+\Phi=\partial_-\Phi=\partial_{\underline 1}\Phi=0~,~~~H_{+1\bar1}+H_{+n}{}^n=0~,~~~H_{-1\bar1}-H_{-n}{}^n=0~,~~~
\cr &&
H_{-+\bar1}+H_{\bar 1n}{}^n=-{1\over6} H_{npq}\epsilon^{npq}-{1\over6} H_{\bar n\bar p\bar q}\epsilon^{\bar n\bar p\bar q}~,~~~
\cr
&&H_{+\bar1\bar n}={1\over2} H_{+pq}\epsilon^{pq}{}_{\bar n}~,~~~ H_{-\bar1\bar n}=-{1\over2} H_{-pq}\epsilon^{pq}{}_{\bar n}~,~~~
\cr
&&
H_{-+\bar n}+H_{1\bar 1\bar n}={1\over2} H_{1pq}\epsilon^{pq}{}_{\bar n}+{1\over2} H_{\bar 1pq} \epsilon^{pq}{}_{\bar n}~,
\cr
&&
\partial_{\bar 1}\Phi-{1\over12}H_{npq}\epsilon^{npq}+{1\over12} H_{\bar n\bar p\bar q}\epsilon^{\bar n\bar p\bar q}=0~,
\cr
&&
\partial_{\bar n}\Phi-{1\over2} H_{\bar n p}{}^p-{1\over4} H_{\bar 1 pq} \epsilon^{pq}{}_{\bar n}+{1\over4} H_{1pq} \epsilon^{pq}{}_{\bar n}=0~.
\eea
 The  conditions can be rewritten as
\bea
&&\partial_+\Phi=\partial_-\Phi=\partial_{\underline 1}\Phi=0~,~~~{1\over3!} \e_a{}^{bcd} H_{bcd}-{1\over2}(de_a)_{ij}\omega^{ij}=0~,~~~a=+,-
\cr
&&{1\over3!} \e_{\bar 1}{}^{bcd} H_{bcd}-{1\over2}(de_{\bar 1})_{ij}\omega^{ij}=-{\sqrt{2}\over 6} {\cal N}_{ijk}\, {\rm Re}\chi^{ijk}
\cr
&&[e_+, e_{\bar 1}]_{\bar n}=-{1\over2} (de_+)_{pq} \epsilon^{pq}{}_{\bar n}~,~~~
[e_-, e_{\bar 1}]_{\bar n}={1\over2} (de_-)_{pq} \epsilon^{pq}{}_{\bar n}~,
\cr
&&[e_-, e_+]_{\bar n}+[e_1, e_{\bar 1}]_{\bar n}=-{1\over2} (de_1+de_{\bar1})_{pq}\,\epsilon^{pq}{}_{\bar n}~,
\cr
&&\partial_6\Phi+{1\over 92} {\cal N}_{ijk}\, {\rm Im}\chi^{ijk}=0~,~~~
\partial_{\bar n}\Phi-{1\over2}(\theta_\omega)_{\bar n}+{1\over4} (de_1-de_{\bar 1})_{pq}\, \epsilon^{pq}{}_{\bar n}=0~.
\eea
The dilaton is invariant under the three out of four isometries of the background.

\subsection{N=3}

The dilatino Killing spinor equation is
\bea
{\cal A}(e_{15}+e_{2345})=0~,~~~{\cal A}1=0~.
\eea
The solution to the dilatino Killing spinor equations is
\bea
&& \partial_+\Phi=\partial_-\Phi=\partial_{ 1}\Phi=0~,~~~H_{+1\bar1}+H_{+n}{}^n=0~,~~~
H_{-1\bar1}-H_{-n}{}^n=0~,~~~H_{-+\bar1}+H_{\bar 1n}{}^n=0~,~~~
\cr
&& H_{-\bar1\bar n}=-{1\over2} H_{-pq}\epsilon^{pq}{}_{\bar n}~,~~~
H_{+1n}= H_{+pq}=0~,~~~H_{pqn}=H_{1pq}=0~,
\cr
&&
H_{-+\bar n}={1\over2} H_{\bar 1pq} \epsilon^{pq}{}_{\bar n}-H_{\bar n1\bar 1}~,~~~
\partial_{\bar n}\Phi-{1\over2} H_{\bar n p}{}^p-{1\over4} H_{\bar 1 pq} \epsilon^{pq}{}_{\bar n}=0
\eea
These conditions can be rewritten as
\bea
&&\partial_a\Phi=0~,~~~{1\over3!} \e_a{}^{bcd} H_{bcd}-{1\over2}(de_{a})_{ij}\omega^{ij}=0~,~~~(de^-)^{2,0}=(de^{\bar 1})^{2,0}=0~,~~~
\cr
&&{\cal N}_{ijk}=0~,~~~[e_-, e_{\bar 1}]_{\bar n}=
{1\over2} (de_-)_{pq} \epsilon^{pq}{}_{\bar n}~,~~~[e_-, e_+]_{\bar n}+[e_1, e_{\bar 1}]_{\bar n}=-{1\over2} (de_{\bar 1})_{pq} \e^{pq}{}_{\bar n}~,~~~
\cr
&&[e_+,e_{\bar 1}]_{\bar n}=0~,~~~ \partial_{\bar n}\Phi-
{1\over2} (\theta_\omega)_{\bar n}-{1\over4} (de_{\bar 1})_{pq} \epsilon^{pq}{}_{\bar n}=0~.
\eea
Observe that  if $(de^+)^{2,0}=(de^1)^{2,0}=0$, then the $N=3$ backgrounds admit an additional supersymmetry and so
they admit four supersymmetries. We can  show this  by comparing the conditions above
with those of $N=4$ backgrounds stated below. The same conclusion holds if $[\mathfrak{h}, \mathfrak{h}]\subseteq \mathfrak{h}$.

\subsection{N=4}

The solution to the dilatino Killing spinor equation given in \cite{het} can be summarized as
\bea
&&\partial_a\Phi=0~,~~~{1\over3!} \e_a{}^{bcd} H_{bcd}-{1\over2}H_{aij}\omega^{ij}=0~,~~~~(de^a)^{2,0}=0~,
\cr
&&{\cal N}_{ijk}=0~,~~~{1\over2} \e_{ab}{}^{cd} H_{cdi}-H_{abj} I^j{}_i=0~,~~~\partial_i\Phi-{1\over2} \theta_i=0~.
\eea
In turn, these can be rewritten as
\bea
&&\partial_a\Phi=0~,~~~{1\over3!} \e_a{}^{bcd} H_{bcd}-{1\over2}(de_a)_{ij}\omega^{ij}=0~,~~~~(de^a)^{2,0}=0~,
\cr
&&{\cal N}_{ijk}=0~,~~~{1\over2} \e_{ab}{}^{cd}\, [e_c, e_d]_i-[e_a, e_b]_j  I^j{}_i=0~,~~~\partial_i\Phi-{1\over2} \theta_i=0~.
\eea
The case that has been investigated in detail in \cite{het} is that for which the algebra of four isometries closes. We shall
not expand on this further here.

\subsection{ Reduction of holonomy}

As in the $G_2$ case we have already investigated, we take that $dH=0$, ${\rm hol}(\hat\nabla)\subseteq SU(3)$ and use the field equations
to identify  the additional $\hat\nabla$-parallel
forms. As   has been shown in \cite{het}, and elaborated on in the $G_2$ case,
  either $[\mathfrak{h}, \mathfrak{h}]\subseteq \mathfrak{h}$, $\mathfrak{h}=\bR<e_a>$,  or the holonomy of ${\rm hol}(\hat\nabla)\subset SU(3)$.
This is because the commutator of two parallel $\hat\nabla$-vectors is $\hat\nabla$-parallel.
Similarly, one can show that $H_{abc}$ are constant and they can be identified with the structure constants  of $\mathfrak{h}$.

Applying the Bianchi identity (\ref{bbb}) for  $B=a, C=p, D=\bar q$,  contracting it with $\delta^{p\bar q}$
and using the condition that ${\rm hol}(\hat\nabla)\subseteq SU(3)$, i.e.
\bea
\hat R_{AB, aC} =0~,~~~\hat R_{AB, p}{}^p =0~,
\eea
one can also show that
\bea
\tau_1=iH_{aq}{}^q\, e^a
\eea
are $\hat\nabla$-parallel. Since $e^a$ are also $\hat\nabla$-parallel, $iH_{aq}{}^q=u_a$ are constants.
Similarly one can show that
\bea
\tau_2^a={1\over2} H^a{}_{pq} e^p\wedge e^q~,
\eea
are also $\hat\nabla$-parallel. In this case, either $\tau^a_2=0$ or ${\rm hol}(\hat\nabla)\subset SU(3)$.

Next applying the Bianchi identity (\ref{bbb}) for  $B=p, C=q, D=n$, one can show that
\bea
\tau_3={1\over3!} {\cal N}_{pqn}\,e^p\wedge e^q\wedge e^n=  {4\over3!} H_{pqn} \,e^p\wedge e^q\wedge e^n~,
\eea
is $\hat\nabla$-parallel, see \cite{wit, howegpf, stefang2} for the properties  of the Nijenhuis tensor of
Riemannian
almost complex manifolds with compatible $\hat\nabla$-connection.
 Since ${\rm Re}\chi$ and ${\rm Im} \chi$ are also (3,0) and (0,3) and
$\hat\nabla$-parallel,
\bea
\tau_3={\cal N}= a \, {\rm Re}\chi+ b\, {\rm Im}\chi~,~~~
\eea
for some constants $a,b\in\bR$.

Using similar arguments to those we have made for the $G_2$ case, one can also show that
\bea
\tau_4=\partial_a\Phi\, e^a
\eea
and
\bea
\tau_5=(2\partial_i\Phi-(\theta_\omega)_i) \, e^i
\eea
are $\hat\nabla$-parallel. Since $e^a$ are also  $\hat\nabla$-parallel,  $\partial_a\Phi=v_a$ are constants.
Similarly, either $\tau_5=0$ or ${\rm hol}(\hat\nabla)\subset SU(3)$.

The implication that these additional parallel forms have on the
$N=L=4$ supersymmetric backgrounds is as follows. It is clear that
in this case the conditions $dH=0$, ${\rm hol}(\hat\nabla)= SU(3)$
and  the field equations are not sufficient to imply the dilatino
Killing spinor equations from the gravitino ones. For this to be
the case, one has to impose in addition $\tau_3=\tau_4=0$ and
relate $\tau_1$ to the structure constants of $\mathfrak {h}$.

The condition  ${\rm hol}(\hat\nabla)= SU(3)$ imposed on the $N=3$ descendant implies enhancement of supersymmetry to $N=4$.
Backgrounds with $N=3$ supersymmetry may exist but these require reduction of the holonomy.
On the other hand backgrounds with $N=1$ and $N=2$ supersymmetry may exist even if ${\rm hol}(\hat\nabla)= SU(3)$.
For example, these can be linear dilaton backgrounds.

\newsection{The descendants of $SU(2)$}

A (complex) basis in the space of parallel spinors with  ${\rm Stab}(\e_1,\dots,\e_8)=SU(2)$ is
\bea
1~,~~~e_{12}~,
~~~e_{15}~,~~~e_{25}~.
\la{basissu2}
\eea
The subspace ${\cal P}$ in $S^+$ spanned by the above spinors can be identified  with the
 positive chirality symplectic Majorana-Weyl representation $\Delta_{\bf 8}^{+s}$ of $Spin(5,1)$.  To see this, first
 observe that ${\rm Stab}({\cal P})=Spin(5,1)\times Spin(4)$, where the Lie algebra of $Spin(5,1)$ is spanned
 by Clifford algebra directions $0,5,1,6,2,7$ and the Lie algebra of $Spin(4)=SU(2)\times SU(2)$ is spanned by the Clifford algebra
 directions $3,4,8,9$. In addition we have  ${\rm Stab}(\e_1,\dots,\e_8)=SU(2)\subset Spin(4)\subset {\rm Stab}({\cal P})$, so
 $\Sigma({\cal P})={\rm Stab}({\cal P})/{\rm Stab}(\e_1,\dots,\e_8)=Spin(5,1)\times SU(2)$. It is known that $Spin(5,1)=SL(2, \bH)$
 and that $Spin(5,1)$  does not admit
 Majorana-Weyl representations. However, it admits symplectic Majorana-Weyl representation after twisting with $SU(2)$, i.e.~taking two copies of the positive chirality (complex) Weyl representation and imposing a symplectic reality condition. This reality
 condition is precisely that inherited from the reality condition of the Majorana-Weyl spinors $S^+$ of $Spin(9,1)$. In the explicit basis (\ref{basissu2}) of ${\cal P}$, one can show that
 the Lie algebra $\mathfrak{su}(2)$ of the $SU(2)$ subgroup of
 $\Sigma({\cal P})$ can be identified as $\mathfrak{su}(2)=\bR<\Gamma^{34},
 \Gamma^{\bar3\bar4}, {i\over2}(\Gamma^{3\bar3}+\Gamma^{4\bar4})>$.

One can similarly examine ${\cal Q}$ which is required for investigating the normal spinors to the Killing spinors.
In particular, one can show that ${\cal Q}=\Delta^{-s}_{\bf 8}$, where $\Delta^{-s}_{\bf 8}$ is
 the negative chirality Majorana-Weyl symplectic representation of $Spin(5,1)$. Furthermore,   $\Sigma({\cal Q})=Spin(5,1)\times SU(2)$.
The $N=8$ supersymmetric backgrounds have already been investigated in \cite{het}. These are the backgrounds
for which all parallel spinors are Killing.  So it remains to investigate the backgrounds with $N<8$. For
$4< N<8$, we shall use $\Sigma({\cal Q})$ to choose directions for the normal spinors while for $1\leq N\leq 4$, we shall use
$\Sigma({\cal P})$ to choose directions in the space of parallel spinors. We shall not elaborate on the choice of normals
to the Killing spinors
for $N>4$ because it follows directly from the choice of Killing spinors for $N\leq 4$. So we shall simply state the dilatino
Killing spinor equations in each case. The stability groups of the Killing spinors are summarized in table 15.
\begin{table}[ht]
 \begin{center}
\begin{tabular}{|c|c|}\hline
$N$&${\rm Stab}_{\Sigma}$\\ \hline \hline
$1$&$(SU(2)\times SU(2))\ltimes \bR^4$ \\ \hline
$2$&$(U(1)\times SU(2))\ltimes \bR^4, \, U(1) \times U(1)$ \\ \hline
$3$&$SU(2)\ltimes \bR^4, \, U(1), \, \{ 1 \}$ \\ \hline
$4$&$SU(2)\ltimes \bR^4, \, U(1), \, \{ 1 \}$ \\ \hline
\end{tabular}
\end{center}
\caption{The first column denotes the number of supersymmetries and the second column the stability
subgroup of  Killing spinor  in $\Sigma({\cal P})=Spin(5,1)\times SU(2)$.}
\end{table}

\subsection{Geometry of the gravitino Killing spinor equation}

The condition that ${\rm hol}(\hat\nabla)\subseteq SU(2)$ is equivalent to requiring that the
forms
\bea
e^a~,~~~\omega_I=-(e^3\wedge e^8+e^4\wedge e^9)~,~~~\omega_J+i\omega_K=(e^3+ie^8)\wedge (e^4+ie^9)
\eea
are $\hat\nabla$-parallel, where $a=+,-,1,\bar 1, 2,\bar 2$. As in previous cases, $i_aH=(de_a)$ and
in addition one has that
\bea
(de_a)^{2,0+0,2}_{ij}&=&-{1\over2}[i_I(\nabla_a\omega)]_{ij}~,
\cr
(de_a)_{ij}\, \omega^{ij}_I&=&(\nabla_a \omega_J)_{ij}\,  \omega_K^{ij}~,
\eea
where $i,j=3,4,8,9$.
Furthermore, one finds that
\bea
H^{\rm rest}&=&- i_I\tilde d\omega_I=- i_J\tilde d\omega_J~,
\cr
{\cal N}(I)_{ijk}&=&{\cal N}(J)_{ijk}=0~.
\eea
The geometry of Riemannian four-dimensional manifolds with an $SU(2)$- structure and compatible
connection with skew-symmetric torsion has been extensively investigated, see \cite{hull, strominger, howegp, howegpc}.
As in previous cases, we define $\mathfrak{h}=\bR<e_a>$, and for the descendants $[\mathfrak{h}, \mathfrak{h}]\nsubseteq \mathfrak{h}$.
However if we demand $[\mathfrak{h}, \mathfrak{h}]\subseteq \mathfrak{h}$, $\mathfrak{h}$ is a Lorentzian (5+1)-dimensional Lie algebra.
These have been classified and the have been found to be
\bea
&&\bR\oplus^5 \mathfrak{u}(1)~,~~~\bR\oplus^2 \mathfrak{u}(1)\oplus \mathfrak{su}(2)~,~~~\mathfrak{sl}(2,\bR)\oplus^3\mathfrak{u}(1)~,~~~
\cr
&&\mathfrak{sl}(2,\bR)\oplus^3\mathfrak{su}(2)~,~~~\mathfrak{cw}_4\oplus^2\mathfrak{u}(1)~,~~~\mathfrak{cw}_6~,
\eea
where $\mathfrak{cw}_n$ denote pp-wave algebras of dimension $(n-1)+1$.
For $N=8$ backgrounds, the dilatino Killing spinor equation implies that
$[\mathfrak{h}, \mathfrak{h}]\subseteq \mathfrak{h}$ and $\mathfrak{h}$ is self-dual. These   have been shown in \cite{cham} to
be isomorphic to
$\bR\oplus^5 \mathfrak{u}(1)$, $\mathfrak{sl}(2,\bR)\oplus \mathfrak{su}(2)$ and $\mathfrak{cw}_6$.

\subsection{N=1}

The group action of $\Sigma({\cal P})=Spin(5,1)\times SU(2)$ on ${\cal P}$ can be most easily described in terms
of quaternions. First identify ${\cal P}=\bH^2$. Then $Spin(5,1)=SL(2,\bH)$ acts on ${\cal P}$ from the left with quaternionic
matrix multiplication while $SU(2)$ acts with quaternionic multiplication from the right, i.e.
\bea
{\underline x}\,\longrightarrow L\,{\underline x}\, \bar a~,~~~~{\underline x}\in {\cal P}~,~~~L\in SL(2,\bH)~,~~~a\in SU(2)=Sp(1)
\eea
where $\bar a$ is the quaternionic conjugate of $a$. It is easy then to see that $\Sigma({\cal P})$ has
a single orbit in ${\cal P}$ of codimension zero. The stability subgroup is $(SU(2)_L\times SU(2)_R)\ltimes \bR^4.$
 So a representative  can be chosen
as $1+e_{1234}$.
In turn, the dilatino Killing spinor equation is
\bea
{\cal A}(1+e_{1234})=0~.
\eea
The solution has been given in (\ref{dn=1sp2}), and the conditions can be interpreted in a similar way.
A different way of organizing the conditions is in terms of $SU(2)$ representations.
This allows to compare the results with the $N>1$ cases. In particular, we find that
\bea
&&\partial_+\Phi=0~,~~~H_{+1\bar1}+H_{+2\bar2}+(de_+)_{n}{}^n=0~,~~~-H_{+\bar1\bar2}+{1\over2} (de_+)_{mn} \e^{mn}=0~,~~~
\cr
&&[e_+,e_{\bar 1}]_{\bar n}-
[e_+, e_2]_m{}\e^m{}_{\bar n}=0~,~~~
\cr
&&\partial_{\bar 1}\Phi-{1\over2} (de_2)_{mn}\e^{mn}-{1\over2}(de_{\bar 1})_n{}^n-{1\over2} H_{\bar 1 2\bar 2}-{1\over2} H_{-+\bar 1}=0~,
\cr
&&
\partial_{\bar 2}\Phi+{1\over2}(de_1)_{mn} \e^{mn}-{1\over2} (de_{\bar 2})_m{}^m-{1\over2} H_{\bar21\bar1}-{1\over2} H_{-+\bar2}=0~,
\cr
&&
\partial_{\bar n}\Phi-[e_1, e_2]_m \e^m{}_{\bar n}+{1\over2} [e_2, e_{\bar 2}]_{\bar n}+
{1\over2} [e_1, e_{\bar 1}]_{\bar n}+{1\over2} [e_-,e_+]_{\bar n}-{1\over2} (\theta_{\omega_I})_{\bar n }=0~,
\eea
where $m,n=3,4$ are Hermitian indices.
It is clear that even if we take $[\mathfrak{h}, \mathfrak{h}]\subseteq
\mathfrak{h}$, $\mathfrak{h}$ is not necessarily
a self-dual  Lie algebra. In fact there is not a condition on the structure constants of $\mathfrak{h}$.
The rotation $\tilde de^a$ of the $\hat\nabla$-vector fields is also not  restricted an a priori  to lie in some subalgebra
of $\mathfrak{so}(4)$. The dilaton is invariant only under $e_+$.

\subsection{N=2}

\subsubsection{Killing spinors}

The first  Killing spinor $\e_1$ can be chosen as in the $N=1$ case above, $\e_1=\e=1+e_{1234}$.  To continue, we shall first explain how the
${\rm Stab}(\e)=(SU(2)_L\times SU(2)_R)\ltimes \bR^4$
acts on ${\cal P}/{\cal K}$, where ${\cal K}=\bR<\e>$.
First identify ${\cal K}$ with the real axis in one of the quaternionic subspaces of ${\cal P}=\bH^2$ and write
 ${\cal P}=\bR<1+e_{1234}>\oplus {\rm Im} \bH\oplus \bH$.
 Thus we can set ${\cal P}/{\cal K}= {\rm Im} \bH\oplus \bH$.
 Then $SU(2)_L\times SU(2)_R$ acts as
 \bea
 ({\bf x},y)\rightarrow (a {\bf x}\bar a, b y \bar a)~,~~~{\bf x}\in {\rm Im} \bH~,~~~y\in\bH~, ~~~a\in SU(2)_L~,~~~b\in SU(2)_R~.
 \eea
In addition the $\bR^4$ subgroup acts with null boosts on ${\cal P}$ with fixed point set $\bR<1+e_{1234}>\oplus {\rm Im} \bH$.
In the explicit basis for ${\cal P}$ in
(\ref{basissu2}),  $\mathfrak{su}(2)_L=\bR<\Gamma^{\bar1 \bar2}+\Gamma^{34}, \Gamma^{1 2}+\Gamma^{\bar3\bar 4},
{i\over2} (\Gamma^{1\bar1}+
\Gamma^{2\bar2}-\Gamma^{3\bar3}-\Gamma^{4\bar4})>$, $\mathfrak{su}(2)_R=
\bR<\Gamma^{1\bar2}, \Gamma^{\bar12}, {i\over2}(\Gamma^{1\bar1}-\Gamma^{2\bar2})>$
and $\mathfrak{\bR}^4=\bR<\Gamma^{-1},
\Gamma^{-\bar 1}, \Gamma^{-2}, \Gamma^{-\bar2}>$. In addition,
  ${\rm Im} \bH=\bR<i(1-e_{1234}), (e_{12}-e_{34}), i(e_{12}+e_{34})>$ and $\bH$ is spanned by the
  rest of the basis.

To continue first observe that $\Sigma({\cal K})=(Spin(1,1)\times SU(2)_L\times SU(2)_R)\ltimes \bR^4$, where $Spin(1,1)$
is generated by $\Gamma^{-+}$. $\Sigma({\cal K})$ has two types of orbits on ${\cal P}/{\cal K}$. One
has codimension zero in ${\rm Im}\bH$ and the other has codimension zero in ${\cal P}/{\cal K}$. To see this, consider the
orbits of $SU(2)_L\times SU(2)_R$ in ${\rm Im}\bH\oplus \bH$. There are three type of orbits. One orbit is an $S^2$  contained
in  ${\rm Im}\bH$
with stability subgroup $U(1)_L\times SU(2)_R$, another is an  $S^3$ contained  in $\bH$ with stability subgroup
$(SU(2)_L\times SU(2)_R)/SU(2)=SU(2)$ and the third is a codimension two $(SU(2)_L\times SU(2)_R)/U(1)$
orbit in ${\rm Im}\,\bH\oplus \bH$. The latter orbit has representatives which have non-vanishing components in both
${\rm Im}\,\bH$ and $\bH$ subspaces. However one can show that such a representative lies in the same orbit of $\Sigma({\cal K})$
as that of the $S^3$ using an $\bR^4$ transformation. This can be easily seen by choosing the representative
of the third orbit as
\bea
i \l_1 (1-e_{1234})+\l_2 (e_{15}+e_{2345})~,~~~\l_1,\l_2\not=0~.
\eea
Clearly an $\bR^4$ transformation along the $\Gamma^{-6}$ direction will transform  a representative along $e_{15}+e_{2345}$
to the representative above. Thus $\Sigma({\cal K})$ has only two orbits, one with stability subgroup
$(U(1)_L\times SU(2)_R)\ltimes \bR^4$ and the other stability subgroup $SU(2)$ in $\Sigma({\cal K})$.
Therefore there are two choices for the second normal spinor each associated with the two orbits. Thus
we can choose
either
\bea
\e_2=i(1-e_{1234})
\eea
which lies in ${\rm Im}\bH$, or
\bea
\e_2=(e_{15}+e_{1235})
\eea
which lies in $\bH$. So there are two dilatino Killing spinor equations to consider.

\subsubsection{${\cal A}1=0$}

The solution of this Killing spinor equation  has been given in (\ref{dn=2sp2}). Decomposing the solution in $SU(2)$ representations
as in the $N=1$ case, one finds,
\bea
&&\partial_+\Phi=0~,~~H_{+12}=0~,~~~[e_+, e_1]_n=[e_+,e_2]_n=[e_1, e_2]_n=0~,
\cr
&&
(de_+)_{mn}=(de_1)_{mn}=(de_2)_{mn}=0~,~~~H_{+1\bar1}+H_{+2\bar2}+(de_+)_n{}^n=0~.
\cr
&&
\partial_{\bar1}\Phi-{1\over2} (de_{\bar 1})_n{}^n-{1\over2} H_{\bar 1 2\bar2}-{1\over2} H_{-+\bar 1}=0~,~~~
\partial_{\bar 2}\Phi-{1\over2} (de_{\bar 2})_p{}^p-{1\over2} H_{\bar 21\bar1}-{1\over2} H_{-+\bar 2}=0~,
\cr
&&
\partial_{\bar n}\Phi+{1\over2} [e_2, e_{\bar 2}]_{\bar n}+{1\over2} [e_1, e_{\bar 1}]_{\bar n}+{1\over2} [e_-, e_+]_{\bar n}-{1\over2}
(\theta_{\omega_I})_{\bar n}=0~.
\eea
In general $[\mathfrak{h}, \mathfrak{h}]\nsubseteq \mathfrak{h}$. Moreover $\tilde de^- \in \mathfrak{u}(2)$ and
$\tilde de^{\bar 1}, \tilde de^{\bar 2} \in \mathfrak{u}(2)\oplus_s \Lambda^{0,2}(\bC^2)$ and $\tilde de^+\in \mathfrak{so}(4)$.
The rotations can be restricted further if for example $\mathfrak{h}$ is abelian. The dilaton is invariant only under $e_+$.

\subsubsection {${\cal A} (1+e_{1234})={\cal A} (e_{15}+e_{2345})=0$}

The solution of the latter is given in \cite{het}, see also (\ref{dn=2g2}). Expressing it in $SU(2)$ representations, one finds
\bea
&& \partial_+\Phi=\partial_-\Phi=\partial_{\underline 1}\Phi=0~,~~~H_{+1\bar1}+H_{+2\bar2}+(de_+)_{n}{}^n=0~,~~~
H_{-1\bar1}-H_{+2\bar2}-(de_-)_{n}{}^n=0~,~~~
\cr &&
H_{-+\bar1}+H_{\bar1 2\bar2}+(de_{\bar 1})_n{}^n=-{1\over2} (de_2)_{pq}\epsilon^{pq}-{1\over2} (de_{\bar2})_{\bar p\bar q}
\epsilon^{\bar p\bar q}~,~~~
\cr
&&H_{+\bar1\bar 2}={1\over2} (de_+)_{pq}\epsilon^{pq}~,~~~[e_+, e_{\bar 1}]_{\bar n}=[e_+, e_2]_p \epsilon^{p}{}_{\bar n}~,~~~
\cr
&&
H_{-1\bar 2}=-{1\over2} (de_-)_{pq}\epsilon^{pq}~,~~~[e_-, e_{1}]_{\bar n}=-[e_-, e_2]_p \epsilon^{p}{}_{\bar n}~,~~~
\cr
&&
H_{-+\bar 2}+H_{1\bar 1\bar 2}={1\over2} (de_1)_{pq}\epsilon^{pq}+{1\over2} (de_{\bar 1})_{pq} \epsilon^{pq}~,
~~~[e_-, e_+]_{\bar n}+ [e_1, e_{\bar 1}]_{\bar n}=[e_1, e_2]_p\,\epsilon^{p}{}_{\bar n}+ [e_{\bar 1}, e_2]_p\, \epsilon^{p}{}_{\bar n}~,
\cr
&&
\partial_{\bar 1}\Phi-{1\over4} (de_2)_{pq}\,\epsilon^{pq}+{1\over4} (de_{\bar 2})_{\bar p\bar q}\,\epsilon^{\bar p\bar q}=0~,
~~~
\partial_{\bar 2}\Phi-{1\over2} (de_{\bar 2})_p{}^p-{1\over4} (de_{\bar 1})_{pq} \epsilon^{pq}+{1\over4}
(de_1)_{pq}  \epsilon^{pq}=0~,
\cr
&&
\partial_{\bar n}\Phi+{1\over2} [e_{\bar 1}, e_2]_p \epsilon^{p}{}_{\bar n}-{1\over2}
[e_1, e_2]_p \epsilon^{p}{}_{\bar n}+{1\over2}
[e_2, e_{\bar 2} ]_{\bar n}-{1\over2} (\theta_{\omega_I})_{\bar n} =0~.
\eea
There is no apparent restriction on the rotations $\tilde de^a$ unless $\mathfrak{h}$ is abelian in which case
$\tilde d e^+, \tilde de^-, \tilde de^{\underline 1}\in \mathfrak{su}(2)$. The dilaton is invariant under
$e_+, e_-$ and  $e_{\underline 1}$ $\hat\nabla$-parallel vectors.
\subsection{N=3}

\subsubsection{Killing spinors}

There are two cases to investigate depending on the choice of the first two Killing spinors. These lead to different
results, so they will be examined separately.

Suppose that $\e_1=1+e_{1234}$ and $\e_2=i(1-e_{1234})$ and ${\cal K}$ is spanned by these two spinors.
After some computation one can show that ${\rm Stab}({\cal K})=(Spin(1,1) \times Spin(2)_L\times Spin(2)_R\times Sp(1))\ltimes \bR^4$.
To see how this acts, first  write ${\cal P}/{\cal K}=\bR^2\oplus \bH$. Then the action of
 the subgroup
$Spin(2)_L\times Spin(2)_R\times Sp(1)$ is
\bea
({\underline x}, y)\longrightarrow (L {\underline x}, a y R^{-1} )~,~~~L\in Spin(2)_L~,~~~a\in Sp(1)~,~~~R\in Spin(2)_R~.
\eea
 In particular in the basis (\ref{basissu2}),  $\mathfrak{sp}(1)=\bR<{i\over2} (\Gamma^{1\bar1}-\Gamma^{2\bar2}),
 \Gamma^{1\bar2}, \Gamma^{\bar12}>$, $\mathfrak{spin}(2)_L=\bR<{i\over2} (\Gamma^{1\bar1}+\Gamma^{2\bar2})>$,
$\mathfrak{spin}(2)_R=\bR<{i\over2} (\Gamma^{3\bar3}+\Gamma^{4\bar4})>$, $Spin(1,1)$ acts with boosts in the $\Gamma_{-+}$ direction
and  $\mathfrak{\bR}^4=\bR<\Gamma^{-1},
\Gamma^{-\bar 1}, \Gamma^{-2}, \Gamma^{-\bar2}>$. In addition, $\bR^2=\bR<e_{12}-e_{34}, i(e_{12}+e_{34})>$ and $\bH$
spans the rest of the directions. Observe that both $Spin(2)_L$ and $Spin(2)_R$ act on ${\cal K}$. There are
two type of orbits of ${\rm Stab}({\cal K})$ in ${\cal P}/{\cal K}$, one is co-dimension zero in $\bR^2$ and the other
is codimension zero in ${\cal P}/{\cal K}$. To see this, observe that the orbit of $Sp(1)$ in $\bH$ is an $S^3$ sphere
and that $Spin(2)_L\times Spin(2)_R\times Sp(1)$ has three types of orbits in ${\cal P}/{\cal K}$. However, the representatives
of two of the orbits are related by an $\bR^4$ transformation as in the $N=2$ case.
Choosing representatives for the two orbits of ${\rm Stab}({\cal K})$ in ${\cal P}/{\cal K}$ the third Killing spinor
can be chosen either as
\bea
\e_3=i(e_{12}+e_{34})
\eea
or as
\bea
\e_3=e_{15}+e_{2345}~.
\la{su2n22}
\eea
Consequently, the dilatino Killing spinor equation becomes either
\bea
{\cal A}1={\cal A}(e_{12}+e_{34})=0
\eea
or
\bea
{\cal A}1={\cal A}(e_{15}+e_{2345})=0
\eea
respectively.

Next suppose  that
$\e_1=1+e_{1234}$ $\e_2=e_{15}+e_{2345}$ and that ${\cal K}$ is spanned by these two spinors.
It turns out  that ${\rm Stab}({\cal K})=SL(2,\bR)\times SO(3)$, and $SL(2,\bR)=Spin(2,1)$ acts on ${\cal K}$ with the two-dimensional
representation. To see how this group acts
on ${\cal P}/{\cal K}$ write ${\cal P}/{\cal K}=\bR^3\oplus \bR^3$. Then we have
\bea
({\bf x}, {\bf y})\longrightarrow (O {\bf x}, O{\bf y}) L^{-1}~,~~~O\in SO(3)~,~~~L\in SL(2,\bR)~.
\eea
In the basis (\ref{basissu2}), we have that\footnote{$\Gamma^{\underline 1}$ denotes the gamma matrix along the real
direction 1 to distinguish it from that the complex direction 1 used for the generators of $SO(3)$.}  $\mathfrak{sl}(2,\bR)=\bR<\Gamma^{-+}, \Gamma^{+\underline 1}, \Gamma^{-\underline 1}>$,
$\mathfrak{so}(3)=\bR<{i\over2}(\Gamma^{3\bar3}+\Gamma^{4\bar4}-2\Gamma^{2\bar2}), \Gamma^{1\bar2}-\Gamma^{\bar1\bar2}-\Gamma^{34},
\Gamma^{\bar1 2}-\Gamma^{12}-\Gamma^{\bar3\bar4}>$, and one of the $\bR^3$ subspaces is spanned by
$\bR^3=\bR<i(e_5-e_{12345}), e_{125}-e_{345}, i(e_{125}+e_{345})>$ and the other by the rest elements of the basis.

To find the orbits of ${\rm Stab}({\cal K})$ consider the invariant
\bea
I={\bf x}^2  {\bf y}^2-({\bf x}\cdot {\bf y})^2~,
\eea
where ${\bf x}\cdot {\bf y}$ is the Euclidean inner product of ${\bf x}$ and ${\bf y}$,  ${\bf x}^2={\bf x}\cdot {\bf x}$,
and similarly for ${\bf y}$. If $I\not=0$, there is a co-dimension one orbit $(SL(2,\bR)\times SO(3))/SO(2)$ represented
by two non-colinear non-vanishing elements  ${\bf x}$ and ${\bf y}$. If $I=0$, then either ${\bf y}=0$ or ${\bf x}=0$
or ${\bf x}$ is colinear to ${\bf y}$. In the first two cases, the orbits are codimension zero in the first subspace $\bR^3$
or the second subspace $\bR^3$, respectively. The latter case is not independent because there is always an $SL(2,\bR)$
transformation to transform $\pm ({\bf x}, {\bf x})$ to an element in one of the two $\bR^3$ subspaces of ${\cal P}/{\cal K}$.
Therefore there are three types of orbits to consider and the representatives can be chosen as
\bea
\e_3=i(1-e_{1234})~,
\eea
or
\bea
\e_3=i(e_{15}-e_{2345})~,
\la{su2x}
\eea
or
\bea
\e_3=i(1-e_{1234})+(e_{25}-e_{1345})~.
\eea
In the latter case  we have used the
freedom to choose the overall scale of the Killing spinor to be one.

To give the independent Killing spinor equations, observe  that
$\e_1=1+e_{1234}$, $\e_2=e_{15}+e_{2345}$, $\e_3=i(e_{15}-e_{2345})$ and $\e_1=1+e_{1234}$, $\e_2=i(1-e_{1234})$,
$\e_3=e_{15}+e_{2345}$ are related by a $Spin(9,1)$ transformation. Consequently, two of the above three case are related
to the two   cases described before. Thus the only additional independent dilatino Killing spinor equation is
\bea
{\cal A}(1+e_{1234})={\cal A}(e_{15}+e_{2345})={\cal A}[i(1-e_{1234})+e_{25}-e_{1345}]=0~.
\eea

\subsubsection{${\cal A}1={\cal A}(e_{12}+e_{34})=0$}

The solution can be found in \cite{het} and it can be re-expressed in $SU(2)$ representations as
\bea
&&\partial_+\Phi=0~,~~~H_{+12}=H_{+1\bar1}+H_{+2\bar2}=0~,~~~[e_+, e_1]_n=[e_+, e_2]_n=[e_+, e_2]_{\bar n}-[e_+, e_{\bar 1}]_p \e^p{}_{\bar n}=0~,
\cr
&&
[e_{\bar 1}, e_{\bar 2}]_n+[e_1, e_{\bar 1}]_{\bar p} \e^{\bar p}{}_n+[e_2, e_{\bar 2}]_{\bar p} \e^{\bar p}{}_n=0~,
~~~[e_{\bar 1}, e_{\bar 2}]_{\bar n}=0~,
\cr
&&
(de_+)_{pq}=(de_+)_n{}^n=0~,~~~
(de_{\bar 1})_{\bar n\bar m}=(de_{\bar 2})_{\bar n\bar m}=0~,~~~
\cr
&&
{1\over2} (de_2)_{\bar n\bar m} \e^{\bar n\bar m}+(de_{\bar 1})_m{}^m=0~,~~~
-{1\over2} (de_1)_{\bar n\bar m} \e^{\bar n\bar m}+(de_{\bar 2})_m{}^m=0~,
\cr
&&
\partial_{\bar 1}\Phi+{1\over4} (de_2)_{\bar n\bar m} \e^{\bar n\bar m}-{1\over2} H_{2\bar2 \bar1}+{1\over2} H_{+-\bar1}=0~,
\cr
&&
\partial_{\bar 2}\Phi-{1\over4} (de_1)_{\bar n\bar m} \e^{\bar n\bar m}-{1\over2} H_{1\bar1 \bar2}+{1\over2} H_{+-\bar2}=0~,
\cr
&&
\partial_{\bar n}\Phi+{1\over2} [e_{\bar 1}, e_{\bar 2}]_p \e^p{}_{\bar n}
-{1\over2} [e_+, e_-]_{\bar n}-{1\over2} (\theta_\omega)_{\bar n}=0~.
\eea
The dilaton is invariant only under the action of $e_+$, and $\tilde de^-\in \mathfrak{su}(2)$.

\subsubsection{${\cal A}1={\cal A}(e_{15}+e_{2345})=0$}
The solution can be easily found by combining   (\ref{dn=2g2}) with (\ref{dn=2sp2}). Expressing
the conditions in $SU(2)$ representations, one finds
\bea
&& \partial_+\Phi=\partial_-\Phi=\partial_{ 1}\Phi=0~,~~~H_{+1\bar1}+H_{+2\bar2}+(de_+)_{n}{}^n=0~,~~~
\cr
&&
H_{-1\bar1}-H_{-2\bar2}-(de_-)_{n}{}^n=0~,~~~H_{-+\bar1}+H_{\bar1 2\bar2}+(de_{\bar1})_{n}{}^n=0~,~~~
\cr
&& H_{-1\bar 2}=-{1\over2}(de_-)_{pq}\epsilon^{pq}~,~~~[e_-, e_1]_{\bar n}=-[e_-, e_2]_p \epsilon^{p}{}_{\bar n}~,~~~
\cr
&&H_{+12}=0~,~~~[e_+, e_1]_n=[e_+, e_2]_n=[e_1, e_2]_n=0~,~~~(de_+)_{pq}=(de_2)_{pq}=(de_1)_{pq}=0~,
\cr
&&
H_{-+\bar 2}={1\over2} (de_{\bar 1})_{pq}  \epsilon^{pq}-H_{\bar 21\bar 1}~,~~~
[e_-, e_+]_{\bar n}=[e_{\bar 1}, e_2]_p  \epsilon^{p}{}_{\bar n}-[e_1, e_{\bar 1}]_{\bar n}~,~~~
\cr
&&
\partial_{\bar 2}\Phi-{1\over2} (de_{\bar 2})_p{}^p-{1\over4} (de_{\bar 1})_{pq} \epsilon^{pq}=0~,
~~~
\cr
&&
\partial_{\bar n}\Phi+{1\over2} [e_2, e_{\bar 2}]_n+{1\over2} [e_{\bar 1},e_2]_p \e^p{}_{\bar n}
-{1\over2}(\theta_{\omega})_{\bar n} =0~, \label{N=3-A2}
\eea
The anti-self dual part of $\tilde de^a$, for $a=-,+, 1$ i.e.~the (2,0)+(0,2) and hermitian trace, is entirely
expressed in terms of the structure constants of $\mathfrak{h}$. Therefore  if $\mathfrak{h}$ is abelian
the $\tilde de^-, \tilde de^+, \tilde de^1$ take values in $\mathfrak{su}(2)$.
The dilaton is invariant under four of the six parallel vectors.

\subsubsection{${\cal A}(1+e_{1234})={\cal A}(e_{15}+e_{2345})={\cal A}[i(1-e_{1234})+e_{25}-e_{1345}]=0$}

The solution of the first two conditions can be found in (\ref{dn=2sp2}). The third condition is
new. The solution of the dilatino Killing spinor equations expressed in $SU(2)$ representations is
 \bea
  && \partial_- \Phi = \partial_+ \Phi = \partial_1 \Phi = \partial_2 \Phi = 0 \,,
  \cr
 && ([e_1, e_2]_m + [e_{\bar 1}, e_2]_m )\e^m{}_{\bar n} = [e_-,
 e_+]_{\bar n} + [e_1, e_{\bar 1}]_{\bar n} \,, \cr
  &&  ([e_1, e_{\bar 2}]_p - [e_{\bar 1}, e_2]_p)\e^p{}_{\bar n} = -
    [e_1 , e_{\bar 1}]_{\bar n} + [e_2, e_{\bar 2}]_{\bar n} + 2 i
    [e_+, e_{\bar 2}]_{\bar n} \,, \cr
    && i ([e_-, e_2]_{\bar m} - [e_-, e_1]_p \e^p{}_{\bar m} ) = 2
    [e_1 , e_2]_p \e^p{}_{\bar m} \,, \cr
   && [e_+, e_{\bar 1}]_{\bar m} - [e_+, e_2]_n \e^n{}_{\bar m} =
    [e_-, e_1]_{\bar m} + [e_-, e_2]_n \e^n{}_{\bar m} = 0 \,, \cr
 && \e^{mn} (de_+)_{mn} = 2 H_{+ \bar 1 \bar 2} \,, \quad
    (de_+) \cont{n} = - H_{+ 1 \bar 1} - H_{+ 2 \bar 2} \,, \cr
  && \e^{mn} (de_-)_{mn} = -2 H_{- 1 \bar 2} \,, \quad
     (de_-) \cont{n} =  H_{- 1 \bar 1} - H_{- 2 \bar 2} \,, \cr
  && (de_{\bar 1}) \cont{n} = - H_{-+ \bar 1} - H_{\bar 1 2 \bar 2} + i H_{- 1 \bar 2} - i H_{- \bar 1
  2} \,, \cr
  && (de_1)_{mn} \e^{mn} = -i H_{- 1 \bar 1} +i H_{- 2 \bar 2} \,,
  \quad
  (de_{\bar 1})_{mn} \e^{mn} = 2 H_{- + \bar 2} + 2 H_{\bar 2 1 \bar 1} + i H_{- 1 \bar 1} - i H_{- 2 \bar 2}
  \,, \cr
  && \e^{mn} (de_2)_{mn} = - i (H_{- 1 \bar 2} - H_{- \bar 1 2}) \,, \quad
  (de_{\bar 2}) \cont{n} = - H_{- + \bar 2} - H_{\bar 2 1 \bar 1} - i H_{- 1 \bar 1} + i H_{- 2 \bar 2} \,, \cr
  &&
  \e^{mn} (de_{\bar 2})_{mn} = - 2 H_{-+ \bar 1} - 2 H_{\bar 1 2 \bar 2} + i (H_{- 1 \bar 2} - H_{- \bar 1 2}) + 4 i H_{+12}
  \,, \cr
  && \partial_{\bar n}\Phi+{1\over2} [e_2, e_{\bar 2}]_{\bar n}+{1\over2} [e_1, e_{\bar 1}]_{\bar n}+{1\over2} [e_-, e_+]_{\bar n}-{1\over2}
(\theta_{\omega_I})_{\bar n} - [e_1 , e_2]_m \e^m{}_{\bar n}  = 0~. \label{N=3-B}
 \eea
The anti-self dual part of $\tilde de^a$ is entirely
expressed in terms of the structure constants of $\mathfrak{h}$. The dilaton is invariant under all
parallel vector fields. Observe that if $\mathfrak{h}$ is abelian, then the above conditions
are the same as those that one can derive from the dilatino Killing spinor equation of
$N=8$ backgrounds \cite{het}, see also (\ref{dn=8su2}). So the supersymmetry
enhances to\footnote{Note that this case, together with the subsequent cases
  where requiring $\mathfrak{h}$ to be abelian implies $N=8$, are exactly
  those which have Stab$_\Sigma = 1$.} $N=8$.

\subsection{N=4}

\subsubsection{Killing spinors}
To begin, suppose that $\e_1=1+e_{1234}$, $\e_2=i(1-e_{1234})$, $\e_3=i(e_{12}+e_{34})$ and that ${\cal K}$
is spanned by these three spinors. Then, ${\rm Stab}({\cal K})=(Sp(1)_L\times Sp(1)_R\times Spin(1,1))\ltimes \bR^4$.
Writing ${\cal P}/{\cal K}=\bR\oplus \bH$, the subgroup $Sp(1)_L\times Sp(1)_R$ acts only on $\bH$ as
$y\rightarrow a y\bar b$, where $a\in Sp(1)_L$ and $b\in Sp(1)_R$. In the basis (\ref{basissu2}),
$\mathfrak{sp}(1)_L=\bR<{i\over2} (\Gamma^{1\bar1}-\Gamma^{2\bar2}), \Gamma^{1\bar2}, \Gamma^{\bar1 2}>$,
$\mathfrak{sp}(1)_R=\bR<{i\over2} (\Gamma^{1\bar1}+\Gamma^{2\bar2}-\Gamma^{3\bar3}-\Gamma^{4\bar4}),
\Gamma^{\bar1\bar2}-\Gamma^{34}, \Gamma^{1 2}-\Gamma^{\bar3\bar4}>$, $Spin(1,1)$ is generated by boosts along $\Gamma^{+-}$
and the Lie algebra of $\bR^4$ is generated by $\mathfrak{\bR}^4=\bR<\Gamma^{-1}, \Gamma^{-\bar1}, \Gamma^{-2}, \Gamma^{-\bar2}>$.
In addition if the subspace $\bR$ of ${\cal P}/{\cal K}$ is chosen along the  $i(e_{12}+e_{34})$ direction, $\bH$
spans the rest of the directions. Using a similar argument as in previous cases, it is easy to see that ${\rm Stab}({\cal K})$
has two types of orbits in ${\cal P}/{\cal K}$ one has codimension zero in $\bR$ and the other has codimension zero
in ${\cal P}/{\cal K}$. So the forth Killing spinor can be chosen either as
\bea
\e_4=e_{12}- e_{34}~,
\eea
or as
\bea
\e_4=e_{15}+e_{2345}~.
\eea
So the  dilatino Killing spinor equation is either
\bea
{\cal A}1={\cal A}e_{12}=0
\eea
or
\bea
{\cal A}1={\cal A}(e_{12}+e_{34})={\cal A}(e_{15}+e_{2345})=0
\eea

Next suppose that $\e_1=1+e_{1234}$, $\e_2=i(1-e_{1234})$, $\e_3=e_{15}+e_{2345}$ and that ${\cal K}$ is spanned
by these three spinors. It turns out that ${\rm Stab}({\cal K})=(U(1)\times U(1)\times Spin(1,1))\ltimes \bR^2$.
In the basis (\ref{basissu2}),
$\mathfrak{u}(1)\oplus \mathfrak{u}(1)=\bR<{i\over4} (\Gamma^{1\bar1}-\Gamma^{2\bar2}+\Gamma^{3\bar3}+\Gamma^{4\bar4}), {i\over2}
(\Gamma^{1\bar1}+\Gamma^{2\bar2})>$, $Spin(1,1)$ is generated
by the boosts  $\Gamma^{+-}$ and $\bR^2$ is generated by $\Gamma^{-{\underline 1}},  \Gamma^{-{\underline 6}}$.
Writing
${\cal P}/{\cal K}=\bR^2\oplus \bR^2\oplus \bR$, where the former $\bR^2$ is spanned by $e_{12}-e_{34}, i(e_{12}+e_{34})$, and the latter
is spanned by $e_{25}-e_{1345}, i(e_{25}+e_{1345})$  and  $\bR$ by the
remaining direction. Each $U(1)$ acts on a $\bR^2$ with the two-dimensional representation.
There are several types of orbits which can be represented by
\bea
\e_4=i(e_{12}+e_{34})~,
\eea
\bea
\e_4=i(e_{15}-e_{2345})~,
\eea
\bea
\e_4=i(e_{12}+e_{34})+i(e_{15}-e_{2345})
\eea
and
\bea
\e_4=\cos\varphi (e_{25}-e_{1345})+i \sin\varphi (e_{15}-e_{2345})
\eea
Only the latter three choices give  independent new cases. The dilatino Killing spinor equations are
\bea
{\cal A}1={\cal A}e_{15}=0~,
\eea
\bea
{\cal A}1={\cal A}(e_{15}+e_{2345})={\cal A}(e_{12}+e_{34}+(e_{15}-e_{2345}))=0~,
\eea
\bea
{\cal A}1={\cal A}(e_{15}+e_{2345})={\cal A}(\cos\varphi (e_{25}-e_{1345})+i \sin\varphi (e_{15}-e_{2345}))=0~,
\eea

Suppose that  $\e_1=1+e_{1234}$, $\e_2=e_{15}+e_{2345}$, $\e_3=i(1-e_{1234})+ e_{25}-e_{1345}$ and
that ${\cal K}$ is spanned by these three spinors. One can show that
${\rm Stab}({\cal K})=SO(3)$ and acts on ${\cal P}/{\cal K}$ with the symmetric   traceless product of the
vector representation.
In the basis (\ref{basissu2}), $\mathfrak{so}(3)=\bR<t_1, t_2, [t_1,t_2]>$, where
\bea
&&t_1={1\over4} [3\Gamma^{1\bar2}+3 \Gamma^{\bar 1 2}-\Gamma^{\bar1\bar2}-\Gamma^{12}-\Gamma^{34}-\Gamma^{\bar3\bar4}]+{1\over \sqrt 2}
\Gamma^{-{\underline 6}}
\cr
&&t_2={i\over4}[2\Gamma^{1\bar1}+4\Gamma^{2\bar2}-\Gamma^{3\bar3}-\Gamma^{4\bar4}]-{1\over\sqrt 2} \Gamma^{+{\underline 2}}~.
\eea
{}From this one can easily show that the above generators satisfy the Lie algebra relations of $SO(3)$.
To identify ${\cal P}/{\cal K}$
with the traceless symmetric representation, $S_0^2(\bR^3)$,
 first observe that $SU(2)$ has real representations of dimensions three, four and five,
and all the rest are of higher dimension.
In the first two cases ${\cal P}/{\cal K}$ would have been the sum of an irreducible
and trivial representations. This
means that ${\cal P}/{\cal K}$  would have a one-dimensional invariant subspace under the $SO(3)$ action. However,
one can easily show that such a subspace does not exit. Thus the only other option available is to identify
${\cal P}/{\cal K}=S_0^2(\bR^3)$. A direct computation in appendix C has confirmed this.
There are two types of orbits of $SO(3)$ in $S_0^2(\bR^3)$. One is a generic
orbit of co-dimension two isomorphic to $SO(3)$ and the other is a special $S^2$ orbit.  This can be easily seen
by observing that any $3\times 3$ symmetric traceless matrix can be diagonalized and has
two eigenvalues. If the two eigenvalues are distinct,
then the symmetric matrix represent the generic orbit. If either one of the eigenvalues vanishes or their sum vanishes, then
the symmetric matrix represents the special $S^2$ orbit.
A representative
of the special orbit can be identified with a spinor that is invariant under one of the generators of $SO(3)$.
Therefore, we can choose as a fourth Killing spinor either
\bea
\e_4=i\cos \varphi  (e_{15}-e_{2345})+i\sin\varphi (e_{12}+e_{34})
\eea
where $\varphi$ is a constant angle, or
\bea
\e_4=i(e_{12}+e_{34})~.
\eea
Thus the dilatino Killing spinor equation is either
\bea
&&{\cal A}(1+e_{1234})={\cal A}(e_{15}+e_{2345})={\cal A}(i(1-e_{1234})+ e_{25}-e_{1345})
\cr
&&={\cal A}[\cos \varphi  (e_{15}-e_{2345})+\sin\varphi (e_{12}+e_{34})]=0~,
\eea
or
\bea
{\cal A}(1+e_{1234})={\cal A}(e_{15}+e_{2345})={\cal A}(i(1-e_{1234})+ e_{25}-e_{1345})={\cal A}(e_{12}+e_{34})=0~.
\eea
The latter case is a special case of the former for $\sin\varphi=1$.

\subsubsection{${\cal A}1={\cal A}e_{12}=0$}

The solution of the dilatino Killing spinor equation expressed in $SU(2)$ representations is
\bea
&&\partial_+\Phi=0~,~~~H_{+\a\b}=H_{+\a}{}^\a=0~,~~~[e_+,e_\a]_i=0~,~~~[e_\a, e_\b]_i=[e_\a, e^\a]_i=0~,
\cr
&&(de_+)_{np}=(de_+)_n{}^n=0~,~~~(de_\a)_{np}=(de_\a)_{\bar n\bar p}=(de_\a)_n{}^n=0~,~~~
\cr
&&\partial_{\bar \a}\Phi-{1\over2} H_{\bar \a \b}{}^\b-{1\over2} H_{-+\bar \a}=0~,~~~
\cr
&&\partial_{\bar n} \Phi-{1\over2}\theta_{\bar n} +{1\over2}[e_-, e_+]_{\bar n} =0~,~~~\a=1,2~,~~~n,p=3,4~,~~~i=3,4, \bar 3, \bar 4.
\eea
If $\mathfrak{h}$ is not abelian, the dilaton is invariant under $e_+$,  $\tilde de^-, \tilde de^\a\in \mathfrak{su}(2)$
but $\tilde de^-$ is not restricted.
For abelian $\mathfrak{h}$, the dilaton is invariant under $e_+, e_1, e_2$.

\subsubsection{${\cal A}1={\cal A}(e_{12}+e_{34})={\cal A}(e_{15}+e_{2345})=0$}

The solution of the dilatino Killing spinor equation is
 \bea
  && \partial_- \Phi = \partial_+ \Phi =
  \partial_1 \Phi =
  (\partial_2 - \partial_{\bar 2}) \Phi = 0 \,, \quad
  H_{+12} = H_{+ 1 \bar 1} + H_{+ 2 \bar 2} = 0 \,, \cr
  && [e_1, e_2 ]_{n} = [e_+, e_1 ]_n = [e_+, e_2 ]_n =
  [e_1 , e_{\bar 1} ]_{\bar n} + [e_- , e_+ ]_{\bar n} - [e_{\bar 1} , e_{2} ]_p \e^p{}_{\bar n}
  = 0 \,, \cr
  &&
  [e_1 , e_{\bar 1} ]_{\bar n} + [e_2 , e_{\bar 2} ]_{\bar n} - [e_{\bar 1} , e_{\bar 2} ]_p \e^p{}_{\bar n} = [e_+, e_2 ]_{\bar n} - [e_+, e_{\bar 1} ]_p \e^p{}_{\bar n} = [e_-, e_1 ]_{\bar n} + [e_-, e_2 ]_p \e^p{}_{\bar n} = 0 \,, \cr
  && (de_+) \cont{n} = (de_+)_{pq} = 0 \,, \quad
  H_{- 1 \bar 1} - H_{- 2 \bar 2} - (de_-) \cont{n} = 0 \,, \quad
  H_{- 1 \bar 2} + \frac{1}{2} (de_-)_{pq} \e^{pq} = 0 \,, \cr
  && H_{2 1 \bar 1} + H_{+ - 2} + \frac{1}{2} \e^{\bar m \bar n} (de_1)_{\bar m \bar n} = (de_1)_{mn} =
  (de_{\bar 1}) \cont{n} + H_{- + \bar 1} + H_{\bar 1 2 \bar 2} = 0 \,, \cr
  && H_{+- \bar 1} - H_{\bar 1 2 \bar 2} + \frac{1}{2} \e^{\bar m \bar n} (de_2)_{\bar m \bar n} =
  H_{2 1 \bar 1} + H_{+ - 2} + (de_{\bar 2}) \cont{p} = (de_2)_{mn} = 0 \,,  \,, \cr
  && 2 \partial_{\bar 2} \Phi + H_{+-2} + H_{+- \bar 2} + H_{2 1 \bar 1} - H_{\bar 2 1 \bar 1} =0 \,,
  \cr
  &&  2 \partial_{\bar n} \Phi + [e_1, e_{\bar 1}]_{\bar n} + [e_2, e_{\bar 2}]_{\bar n} - [e_+, e_-]_{\bar n}
  - (\theta_\omega)_{\bar n} = 0 \,.
 \eea
The dilaton is invariant under five parallel vector field. Moreover $\tilde de^-\in \mathfrak{su}(2)$ and the
anti-self dual part of $\tilde de^+$, $\tilde de^1$ and $\tilde de^2$ are determined in terms of the
structure constants of $\mathfrak{h}$. So if  $\mathfrak{h}$ is abelian, all rotations are in $\mathfrak{su}(2)$. In addition
$\Phi$ is invariant under all parallel vectors fields. As a consequence, there is supersymmetry enhancement to $N=8$.

\subsubsection{${\cal A}1={\cal A}e_{15}=0$}
The solution  in $SU(2)$ representations is
\bea
&&\partial_+\Phi=\partial_-\Phi=\partial_1\Phi=0~,~~~H_{-1\bar 2}=H_{+12}=H_{-+\bar2}+H_{\bar2 1\bar1}=0~,
\cr
&&(de_+)_{np}=0~,~~~(de_+)_n{}^n+H_{+2\bar2}+H_{+1\bar1}=0~,~~~
\cr
&&(de_-)_{np}=0~,~~~H_{-1\bar1}-H_{-2\bar2}-(de_-)_n{}^n=0~,
\cr
&&(de_{\bar 1})_{np}=(de_{ 1})_{np}=0~,~~~(de_{\bar 1})_n{}^n+H_{\bar 12 \bar2}+H_{-+\bar 1}=0~,
\cr
&&(de_2)_{np}=0~,~~~[e_-, e_+]_{\bar n}+[e_1, e_{\bar 1}]_{\bar n}=0~,
\cr
&&[e_+, e_2]_n=[e_+, e_1]_n=[e_-, e_2]_n=[e_-, e_{\bar 1}]_n=[e_{\bar 1}, e_2]_n=[e_{1}, e_2]_n=0
\cr
&&
\partial_{\bar 2}\Phi-{1\over2} (de_{\bar 2})_n{}^n=0~,~~~\partial_{\bar n}\Phi-
{1\over2} \theta_{\bar n}+{1\over2} [e_2, e_{\bar 2}]_{\bar n}=0~. \label{N=4-B1}
\eea
The dilaton is invariant under $e_+, e_-$ and $e_1$. Moreover $\tilde d e^a\in \mathfrak{u}(2)$, $a=-,+,1,2$.
For the first three rotations, the hermitian trace depends on the structure constants of $\mathfrak{h}$.
Consequently even if $\mathfrak{h}$ is abelian, the hermitian trace of $\tilde d e^2$ does not vanish.
There is no supersymmetry enhancement to $N=8$.

\subsubsection{${\cal A}1={\cal A}(e_{15}+e_{2345})={\cal A}(e_{12}+e_{34}+(e_{15}-e_{2345}))=0$}
The solution for the first three Killing spinor equations  is given in  \eqref{N=3-A2}. The forth Killing spinor equation
 gives the additional constraints
 \bea
  &&  (de_+)\cont{m} = - 2 H_{\bar 2 1 \bar 1} - 2 H_{- + \bar 2} = 2 (de_{\bar 2}) \cont m \,, \cr
  && \e^{\bar m \bar n} (de_2)_{\bar m \bar n} = 4 H_{- \bar 1  2} + 2 H_{- + \bar 1} + 2 H_{\bar 1 2 \bar 2} \,, \cr
  &&[e_{\bar 1}, e_{\bar 2}]_p \e^p{}_{\bar n} = [e_1, e_{\bar 1}]_{\bar n} + [e_2, e_{\bar 2}]_{\bar n} + 2 [e_-, e_{\bar 2}]_{\bar n} \,, \cr
  && [e_+, e_2]_{\bar m} + \e_{\bar m}{}^p [e_+, e_{\bar 1}]_p = 2 [e_-, e_+]_{\bar n} + 2 [e_1, e_{\bar 1}]_{\bar n}
  \,.
 \eea
Note that these conditions  together with  \eqref{N=3-A2} imply that $\partial_2 \Phi = 0$. Therefore the dilaton
is invariant under all parallel vector fields.
The anti-self dual part of $\tilde de^a$, for $a=-,+, 1, 2$  is entirely
expressed in terms of the structure constants of $\mathfrak{h}$. Hence, if $\mathfrak{h}$ is abelian
then $\tilde de^a$ takes values in $\mathfrak{su}(2)$ and supersymmetry enhances to $N=8$.

\subsubsection{${\cal A}1={\cal A}(e_{15}+e_{2345})={\cal A}(\cos\varphi (e_{25}-e_{1345})+i \sin\varphi (e_{15}-e_{2345}))=0$}
The solution for the first three Killing spinor equations  is given in \eqref{N=3-A2} while the forth implies the additional constraints
 \bea
  && H_{-1 \bar 2} = H_{- \bar 1 2} \,, \quad
 \cos \varphi (de_-)\cont{n} - i \sin \varphi (H_{-1 \bar 2} + H_{- \bar 1 2})
 = 0
 \,, \cr
  && (de_{\bar 2})\cont{m} + H_{\bar 2 -+} + H_{\bar 2 1 \bar 1} = 0 \,, \cr
 && \cos \varphi (\tfrac12 (de_{\bar 2})_{mn} \e^{mn} + H_{\bar 1 - +} + H_{\bar
    1 2 \bar 2}) - 2 i \sin \varphi (H_{-+ \bar 2} + H_{\bar 2 1 \bar 1}) = 0
  \,, \cr
  && \cos \varphi ([e_2, e_{\bar 2}]_{\bar m} + [e_-,e_+]_{\bar m} - [e_1,
  e_{\bar 2}]_p \e^p{}_{\bar m} ) - 2 i \sin \varphi ([e_-,e_+]_{n} -
  [e_1,e_{\bar 1}]_{n}) \e^n{}_{\bar m} = 0 \,, \cr
 &&  \cos \varphi (-[e_-, e_2]_{\bar m} + [e_-, e_1]_p \e^p{}_{\bar m}) - 2 i \sin
  \varphi [e_-,e_1]_{\bar m} = 0 \,,
 \eea
where we have assumed that $\cos \varphi$ and $\sin \varphi$ do not
vanish. In the special case in
which $\sin \varphi = 0$, the fourth Killing spinor equation  gives the additional constraints
 \bea
  && H_{-1 \bar 2} = H_{- \bar 1 2} \,, \quad (de_-)\cont{n} = 0
  \,, \cr
  && (de_{\bar 2})\cont{m} + H_{\bar 2 -+} + H_{\bar 2 1 \bar 1} = \tfrac12 (de_{\bar 2})_{mn} \e^{mn} + H_{\bar 1 - +} + H_{\bar
    1 2 \bar 2} = 0 \,, \cr
  && [e_2, e_{\bar 2}]_{\bar m} + [e_-,e_+]_{\bar m} - [e_1,
  e_{\bar 2}]_p \e^p{}_{\bar m} = -[e_-, e_2]_{\bar m} + [e_-, e_1]_p
  \e^p{}_{\bar m} = 0 \,.
 \eea
The additional constraints from the fourth Killing spinor equation imply both in the
generic and in the special case that $\partial_2 \Phi = 0$. Therefore we conclude that
in both cases the dilation is invariant under all parallel vectors $e_a$. The anti-self dual part of
all rotations $\tilde d e^a$ depends on the structure constants of $\mathfrak{h}$. Therefore if $\mathfrak{h}$
is abelian, then $\tilde d e^a$ is self-dual, i.e.~takes values in $\mathfrak{su}(2)$.
In such a case, supersymmetry enhances to $N=8$.

\subsubsection{${\cal A}(1+e_{1234})={\cal A}(e_{15}+e_{2345})={\cal A}(i(1-e_{1234})+ e_{25}-e_{1345})={\cal A}[\cos \varphi
 (e_{15}-e_{2345})+\sin\varphi (e_{12}+e_{34})]=0$}

The solution  for the first three Killing spinor equations has been given in \eqref{N=3-B}.
In addition, the (generic) fourth Killing spinor equation, assuming both $\sin \varphi$ and $\cos \varphi$ do not vanish, gives
the additional constraints
 \bea
  && H_{-1 \bar 2} - H_{- \bar 1 2} = H_{+ 12} + H_{+ \bar 1 \bar 2} = 0 \,, \cr
  && H_{-+ 2} + H_{- + \bar 2} + H_{\bar 2 1 \bar 1} - H_{2 1 \bar 1} + i (H_{- 1 \bar 1} - H_{- 2 \bar 2}) =  0 \,, \cr
  && \sin \varphi (H_{+ 1 \bar 1} + H_{+ 2 \bar 2}) - \cos \varphi (H_{- + \bar 2} - H_{-+2} + H_{\bar 2 1 \bar 1} + H_{2 1 \bar 1}) = 0 \,, \cr
  && \cos \varphi (H_{- 1 \bar 2} + H_{- \bar 1 2}) + \sin \varphi (- H_{1 2 \bar 2} +H_{\bar 1 2 \bar 2} + H_{-+ 1} + H_{- + \bar 1}) = 0 \,,\cr
  && \sin \varphi ([e_{\bar 1}, e_{\bar 2}]_m + [e_1, e_2]_m - \e_m{}^{\bar n} ([e_1, e_{\bar 1}]_{\bar n} + [e_2, e_{\bar 2}]_{\bar n}
  ))
  - 2 \cos \varphi [e_-, e_{\bar 1}]_m = 0 \,, \cr
  && \sin \varphi ([e_+, e_2]_{\bar m} + \e_{\bar m}{}^p [e_+,e_{\bar 1}]_p ) - 2 \cos \varphi [e_{\bar 1}, e_2]_p \e^p{}_{\bar m} = 0 \,.
 \eea
Moreover for the special orbit that corresponds to  $\sin\varphi=1$, we find that the solution to
fourth Killing spinor equations is
 \bea
  && H_{+ 1 \bar 1} + H_{+ 2 \bar 2} = H_{+ 12} + H_{+ \bar 1 \bar 2} = 0 \,, \cr
  && H_{-+ 2} + H_{- + \bar 2} + H_{\bar 2 1 \bar 1} - H_{2 1 \bar 1} + i (H_{- 1 \bar 1} - H_{- 2 \bar 2}) =  0 \,, \cr
  && - H_{1 2 \bar 2} +H_{\bar 1 2 \bar 2} + H_{-+ 1} + H_{- + \bar 1} - i (H_{- 1 \bar 2} - H_{- \bar 1 2}) = 0 \,, \cr
  && [e_{\bar 1}, e_{\bar 2}]_m + [e_1, e_2]_m - \e_m{}^{\bar n} ([e_1, e_{\bar 1}]_{\bar n} + [e_2, e_{\bar 2}]_{\bar n} ) = 0 \,, \cr
  && [e_+, e_2]_{\bar m} + \e_{\bar m}{}^p [e_+,e_{\bar 1}]_p = 0 \,.
 \eea
It is easy to see that the conditions in both cases restrict the
commutators of the vectors fields $e_a$ and the structure constants
of $\mathfrak{h}$. As in the \eqref{N=3-B}, the anti-self dual part
of $\tilde de^a$ is entirely expressed in terms of the structure
constants of $\mathfrak{h}$, and  the dilaton is invariant under all
parallel vector fields. As a consequence, if $\mathfrak{h}$ is
abelian, supersymmetry enhances to $N=8$.

\subsection{N=5}
\subsubsection{Killing spinors}

In this case, it is more convenient to use the gauge symmetry to determine the normals
to the Killing spinors. The correspondence $N\leftrightarrow 8-N$ suggests that the normals can be chosen in a way
similar to the Killing spinors for $N=3$ backgrounds. In turn these can be used to find the
Killing spinors of the theory.  In particular as for $N=3$ supersymmetric backgrounds, there are three cases to consider.

\subsubsection{${\cal A}1={\cal A}(e_{15}+e_{2345})={\cal A}e_{12}=0$}

The solution of the Killing spinor equations is
 \bea
 && \partial_+ \Phi = \partial_- \Phi = \partial_1 \Phi = \partial_2 \Phi = 0
 \,, \quad
 \partial_{\bar n} \Phi - \tfrac 12 (\theta_{\omega_I})_{\bar n} + \tfrac 12 [e_-,
 e_+]_{\bar n} = 0
 \,, \cr
 && H_{+12} = H_{+ 1 \bar 1} + H_{+ 2 \bar 2} = H_{-+ \bar 1} + H_{\bar 1 2 \bar
 2} = H_{-+ \bar 2} + H_{\bar 2 1 \bar
 1} = 0 \,, \cr
 && [e_+,e_1]_m = [e_+,e_2]_m = [e_+,e_1]_{\bar m} = [e_+,e_2]_{\bar m} =
 [e_-,e_1]_{\bar n} + [e_-,e_2]_m \e^m{}_{\bar n} = 0 \,, \cr
 &&[e_1, e_2]_{\bar m} = [e_1, e_{\bar 1}]_m + [e_2, e_{\bar 2}]_m =  [e_-, e_+]_{\bar m} + [e_1, e_{\bar 1}]_{\bar m} - [e_{\bar 1}, e_2]_p
 \e^p{}_{\bar m} = 0 \,, \cr
 && (de_+)_{mn} = (de_+)\cont{n} = 0 \,, \quad
 (de_-)\cont{n} - H_{-1\bar 1} + H_{- 2 \bar 2} =
 \tfrac12 \e^{mn} (de_-)_{mn} + H_{- 1 \bar 2} = 0 \,, \cr
 && (de_1)_{mn} = (de_{\bar 1})_{mn} = (de_1)\cont n = 0 \,, \quad
(de_2)_{mn} = (de_{\bar 2})_{mn} = (de_2)\cont n = 0 \,.
 \eea
The dilaton is invariant under all parallel vector field $e_a$. Moreover $\tilde d e^a$, $a=-,+,1$, is self-dual, i.e.~takes values in $\mathfrak{su}(2)$, while $\tilde d e^2$ takes values in $\mathfrak{u}(2)$. The hermitian trace
of the latter depends on the structure constants of $\mathfrak{h}$.
Therefore if $\mathfrak{h}$ is abelian,  there is supersymmetry enhancement to $N=8$.

\subsubsection{${\cal A}1={\cal A}(e_{12}+e_{34})={\cal A}e_{15}=0$}

The solution to the Killing spinor equations is
 \bea
 && \partial_+ \Phi = \partial_- \Phi = \partial_1 \Phi = \partial_2 \Phi = 0
 \,, \quad
 \partial_{\bar n} \Phi - \tfrac 12 (\theta_{\omega_I})_{\bar n} + \tfrac 12 [e_2,
 e_{\bar 2}]_{\bar n} = 0
 \,, \cr
 && H_{+12} = H_{+ 1 \bar 1} + H_{+ 2 \bar 2} = H_{- 1 \bar 2} = H_{-+ \bar 2} + H_{\bar 2 1 \bar 1} = 0 \,, \cr
 && [e_+,e_1]_m = [e_+,e_2]_m = [e_+,e_2]_{\bar m} + \e_{\bar m}{}^n
 [e_+,e_{\bar 1}]_{n} =
 [e_-,e_1]_{\bar n} = [e_-,e_2]_m = 0 \,, \cr
 && [e_1, e_2]_m = [e_{\bar 1}, e_2]_m = [e_1, e_{\bar 1}]_{\bar m} + [e_2,
 e_{\bar 2}]_{\bar m} + \e_{\bar m}{}^p [e_{\bar 1}, e_{\bar 2}]_p =  [e_-, e_+]_{\bar m} + [e_1, e_{\bar 1}]_{\bar m} = 0 \,, \cr
 && (de_+)_{mn} = (de_+)\cont{n} = 0 \,, \quad
 (de_-)\cont{n} - H_{-1\bar 1} + H_{- 2 \bar 2} = (de_-)_{mn}= 0 \,, \cr
 && (de_1)_{mn} = (de_{\bar 1})_{mn} = (de_{\bar 1})\cont n + H_{-+ \bar 1} +
 H_{\bar 1 2 \bar 2} = 0 \,, \cr
 && (de_2)_{mn} = \tfrac 12 (de_{2})_{\bar m \bar n} \e^{\bar m \bar n} - H_{-+
 \bar 1} - H_{\bar 1 2 \bar 2} = (de_2)\cont n = 0 \,. \label{N=5-A}
 \eea
The dilaton is invariant under all parallel vector field $e_a$, and $\tilde d e^-$ is self-dual. Moreover
 $\tilde d e^a$, $a=+,1$ takes values in $\mathfrak{u}(2)$, where the hermitian trace
 depends on the structure constants of $\mathfrak{h}$. Similarly all the anti-self-dual components $\tilde d e^2$
 depend on the structure constants of $\mathfrak{h}$.
Therefore if $\mathfrak{h}$ is abelian,  there is supersymmetry enhancement to $N=8$.

\subsubsection{${\cal A}1={\cal A}e_{15}= {\cal A}(e_{25}-e_{1345}+i(e_{12}+e_{34}))=0$}

The first four Killing spinor equations give the conditions  \eqref{N=4-B1}, while the  fifth   implies the
additional constraints
 \bea
  && (de_-) \cont n - (de_+) \cont n = (de_1) \cont n - (de_{\bar 1}) \cont n =
  (de_2) \cont n = 0 \,, \cr
  && \tfrac 12 (de_2)_{\bar m \bar n} \e^{\bar m \bar n} + (de_1) \cont n - i
  (de_-) \cont n = 0 \,, \cr
  && - [e_-, e_{\bar 2}]_m + [e_-, e_{\bar 1}]_{\bar p} \e^{\bar p}{}_m - i
  ([e_{\bar 1}, e_{\bar 2}]_m - \e_m{}^{\bar n} ([e_1, e_{\bar 1}]_{\bar n} +
  [e_2, e_{\bar 2}]_{\bar n}) = 0 \,, \cr
  && -[e_+ , e_1]_{\bar m} + \e_{\bar m}{}^p [e_+ , e_{\bar 2}]_p + i (- [e_1, e_{\bar 1}]_{\bar n} +
  [e_2, e_{\bar 2}]_{\bar n} - [e_1, e_{\bar 2}]_p \e^p{}_{\bar m}) = 0 \,. \label{N=5-B}
 \eea
 Note that the above conditions together with those in \eqref{N=4-B1} imply that  $\partial_2 \Phi = 0$.
 Therefore the dilaton is invariant under all parallel vector fields $e_a$. Moreover $\tilde d e^a$, $a=-,+,1$,
 take values in $\mathfrak{u}(2)$, and the hermitian traces depend on the structure constants of $\mathfrak{h}$.
 Similarly the anti-self dual part of $\tilde d e^2$ depends on the structure constants of $\mathfrak{h}$.
 So again there is supersymmetry enhancement to $N=8$, if $\mathfrak{h}$ is abelian.

\subsection{N=6}

\subsubsection{Killing spinors}

As in the $N=5$ case, we use the gauge symmetry to determine the normals to the Killing spinors.
Comparing with the $N=2$ case, we conclude that there are two different possibilities.

\subsubsection{${\cal A}1={\cal A}e_{15}={\cal A}e_{12}=0$}

The solution of the dilatino Killing spinor equation is
 \bea
 && \partial_- \Phi = \partial_+ \Phi = \partial_1 \Phi = \partial_2 \Phi = 0 \,, \cr
 && H_{+- \bar 1} - H_{\bar 1 2 \bar 2} = H_{+- \bar 2} - H_{\bar 2 1 \bar 1} = H_{+ 1 \bar 1} + H_{+ 2 \bar 2}
 = H_{+ 1 2 } = H_{- 1 \bar 2} = 0 \,, \cr
 && [e_1,e_2]_{\bar m} = [e_1,e_2]_{m} = [e_1, e_{\bar 2}]_{\bar m} = [e_+, e_-]_{\bar n} - [e_1, e_{\bar 1}]_{\bar m}
 = [e_1, e_{\bar 1}]_{\bar m} + [e_2, e_{\bar 2}]_{\bar m} = 0 \,, \cr
 && [e_+, e_1]_m = [e_+ , e_2]_m = [e_+, e_1]_{\bar m} = [e_+ , e_2]_{\bar m} = 0 \,, \quad
  [e_-, e_1]_{\bar m} = [e_-,e_2]_m = 0 \,, \cr
 && (de_+) \cont{n} = (de_+)_{mn} = (de_-)\cont{n} - H_{- 1 \bar 1} + H_{- 2 \bar 2} = (de_-)_{mn} = 0 \,, \cr
 && (de_1)\cont{n} = (de_1)_{mn} = (de_1)_{\bar m \bar n} = 0 \,, \quad
 (de_2)\cont{n} = (de_2)_{mn} = (de_2)_{\bar m \bar n} = 0 \,, \cr
 && \partial_{\bar n} \Phi - \tfrac{1}{2} (\theta_{\omega_I})_{\bar n} + [e_-, e_+]_{\bar n} = 0 \,.
  \label{N=6-SU(2)}
 \eea
Clearly the dilaton is invariant under all parallel vectors $e_a$. Moreover $\tilde d e^a$, $a=-,1,2$ take
values in $\mathfrak{su}(2)$, while $\tilde de^+$ takes values in $\mathfrak{u}(2)$. The hermitian trace
of the latter is determined by the structure constant of $\mathfrak{h}$. So there is supersymmetry
enhancement to $N=8$, if $\mathfrak{h}$ is abelian.

 \subsubsection{${\cal A}1={\cal A} e_{15}={\cal A}(e_{25}-e_{1345})={\cal A} (e_{12}+e_{34})=0$}
The solution of this Killing spinor equation is given in \eqref{N=5-A} and supplemented with the
conditions
 \bea
   && (de_-) \cont n = (de_1) \cont n - (de_{\bar 1}) \cont n = 0 \,, \cr
   &&  - [e_-, e_{\bar 2}]_m + [e_-, e_{\bar 1}]_{\bar p} \e^{\bar p}{}_m = - [e_1, e_{\bar 1}]_{\bar n} +
  [e_2, e_{\bar 2}]_{\bar n} - [e_1, e_{\bar 2}]_p \e^p{}_{\bar m} = 0 \,.
 \eea
 The dilaton is invariant under all parallel vector fields $e_a$. Moreover $\tilde d e^a$, $a=-,+$,
 take values in $\mathfrak{su}(2)$, and $\tilde d e^1$ takes values in $\mathfrak{u}(2)$ with
 the hermitian trace to depend on the structure constants of $\mathfrak{h}$.
 Similarly the anti-self dual part of $\tilde d e^2$ depends on the structure constants of $\mathfrak{h}$.
 So again there is supersymmetry enhancement to $N=8$, if $\mathfrak{h}$ is abelian.

\subsection{N=7}

\subsubsection{${\cal A} 1={\cal A}e_{15}={\cal A}e_{12}={\cal A}(e_{25}-e_{1345})=0$}

The solution is given by \eqref{N=6-SU(2)} with the additional constraints
 \bea
  (de_-)\cont{m}=0 \,, \quad
  [e_-, e_2]_{\bar m} - [e_-,e_1]_p \e^p{}_{\bar m} =
  [e_-, e_+]_{\bar m} - \tfrac{1}{2} [e_1, e_{\bar 2}]_p \e^p{}_{\bar m} = 0
 \,, \label{N=7-constraints}
 \eea
It is straightforward to see that the difference between the solution of the dilatino
Killing spinor equation for $N=7$ backgrounds and that of   $N=8$ backgrounds, see \cite{het} and (\ref{dn=8su2}) below,
is that in the former case the commutators $[e_-, e_2]_i$ and $[e_-, e_+]_i$ do not vanish.
 Therefore if $[\mathfrak{h}, \mathfrak{h}]\subseteq \mathfrak{h}$,
then the $N=7$ backgrounds admit eight supersymmetries. We will discuss the
 case $[\mathfrak{h}, \mathfrak{h}] \not\subseteq \mathfrak{h}$ in section
 \ref{red-hol-SU(2)}.

\subsection{Comparison with N=8}

The solutions to the dilatino Killing spinor equation are \cite{het}
\bea
&&\partial_a\Phi=0~,~~~(de_a)_n{}^n=0~,~~~(de_a)_{mn}=0~,
\cr
&&[e_a, e_b]_i=0~,~~~H_{a_1a_2a_3}+{1\over3!} \epsilon_{a_1a_2a_3}{}^{b_1b_2b_3} H_{b_1b_2b_3}=0~,
\cr
&&2\partial_{\bar n} \Phi-(\theta_{\omega_I})_{\bar n} =0~,
\la{dn=8su2}
\eea
where $\e_{+-1\bar12\bar2}=-1$.  In particular $\mathfrak{h}=\bR<e_a>$, $[\mathfrak{h}, \mathfrak{h}]\subseteq \mathfrak{h}$ and spans a
self-dual Lorentzian Lie algebra. As has been already mentioned, these
have been classified in \cite{cham}.

It is clear that the differences between $N=8$ and $1\leq N<8$ supersymmetric backgrounds lie in the invariance
properties of the dilaton under the action of the parallel vector field, the properties of the
commutator $[\mathfrak{h}, \mathfrak{h}]$ and the values of the rotations  $\tilde d e^a$ in $\mathfrak{su}(2)^\perp \subset \Lambda^2(\bR^4)$.
All the different cases can be characterized in terms of these three criteria. However unlike the
cases with parallel spinors that admit a non-compact isotropy  group, the comparison is much more involved.
In the case by case analysis we have presented,  we have not used the fact that there is a classification
of Lorentzian metric Lie groups. In particular, this may impose some additional conditions on the structure
constants of $\mathfrak{h}$ that arise from the Jacobi identities. In turn, this may lead to some additional
simplifications to the solutions of the dilatino Killing spinor equations. We shall investigate this aspect elsewhere.

\subsection{ Reduction of holonomy } \label{red-hol-SU(2)}

As in the investigation of the holonomy reduction in previous cases, we assume that $dH=0$, ${\rm hol}(\hat\nabla)\subseteq SU(2)$ and use the field equations
to identify  the additional $\hat\nabla$-parallel
forms. As  before
  either $[\mathfrak{h}, \mathfrak{h}]\subseteq \mathfrak{h}$ or ${\rm hol}(\hat\nabla)\subset SU(2)$.
Similarly, one can show that the $H_{abc}$ are constant and they can be identified with the structure constants  of $\mathfrak{h}$.

In fact, in this case the constraints that arise from  $[\mathfrak{h}, \mathfrak{h}] \not
\subseteq \mathfrak{h}$ are particularly strong. Suppose there is some
component $H_{abi} = - [X_a, X_b]_i$ non-vanishing. Since it is parallel with
respect to $\hat \nabla$ the supercurvature has to satisfy the integrability condition
 \bea
   {\hat R}_{AB,i}{}^j H_{abj} = 0 \,, \label{SU(2)-intcond}
 \eea
One can show that this implies the vanishing of ${\hat R}$ in the following
way. First suppose that either $H_{ab3}$ or $H_{ab4}$ vanishes. Then the above
constraint readily implies the vanishing of the supercurvature. If both
components of $H_{abi}$ are non-vanishing, one can show that the
supercurvature has to satisfy
 \bea
  - {\hat R}_{AB, 3 \bar 4} {\hat R}_{AB, \bar 3 4} = - {\hat R}_{AB, 3 \bar
    3} {\hat R}_{AB, \bar 4 4} \,.
 \eea
The left hand side is non-positive while the right hand side is non-negative
(using the fact that ${\hat R}$  takes values in $SU(2)$) and hence the supercurvature
has to vanish. We conclude that in the $SU(2)$ case either $[\mathfrak{h},
\mathfrak{h}]\subseteq \mathfrak{h}$ or ${\rm hol}(\hat\nabla) = 1$.

Furthermore,
\bea
\tau_1=iH_{ap}{}^p\, e^a
\eea
is $\hat\nabla$-parallel. Since $e^a$ are also $\hat\nabla$-parallel  $iH_{ap}{}^p=u_a$ are constants.
Similarly one can show that
\bea
\tau_2^a={1\over2} H^a{}_{pq} e^p\wedge e^q~,
\eea
are also $\hat\nabla$-parallel. In this case, one can set
\bea
 \tau^a_2=\l^a\,  \omega_J^{2,0}~,~~~\l^a\in\bC~.
 \eea

Using similar arguments to those we have made for the $G_2$ case, one can also show that
\bea
\tau_3=\partial_a\Phi\, e^a
\eea
and
\bea
\tau_4=(2\partial_i\Phi-(\theta_{\omega_I})_i) \, e^i
\eea
are $\hat\nabla$-parallel. Since $e^a$ are also  $\hat\nabla$-parallel,  $\partial_a\Phi=v_a$ are constants.
Similarly, either $\tau_4=0$ or ${\rm hol}(\hat\nabla)\subset SU(2)$.

Let us now turn to investigate some of the implications that the above parallel forms
have for supersymmetric backgrounds. As can be seen from the conditions for
$N=8$ backgrounds,  $dH=0$, ${\rm hol}(\hat\nabla)= SU(2)$ and  the field equations
are not sufficient to imply the dilatino Killing spinor equations from the gravitino ones.
In particular, one has to  impose in addition $\tau_1=\tau_2=\tau_3=0$.
Moreover, a direct inspection of the conditions for the descendants with $N<7$ reveals that they may
be solutions with  ${\rm hol}(\hat\nabla)=SU(2)$. All such solutions are principal bundles over
a four-dimensional manifold. However the base manifold may not admit an
$SU(2)$-structure.

For $N=7$ we found that if $[\mathfrak{h}, \mathfrak{h}]\subseteq \mathfrak{h}$ this would reduce to
the $N=8$ case. However, due to the above integrability condition
\eqref{SU(2)-intcond}, if $[\mathfrak{h}, \mathfrak{h}] \not \subseteq
\mathfrak{h}$ the holonomy of $\hat \nabla$  reduces to the identity. As
mentioned before and as will be discussed in section \ref{flat}, such backgrounds preserve at least 8
 supersymmetries.
This arises as a consequence of the conditions $dH=\hat R=0$ and the dilatino
 Killing spinor equation \cite{kawano, jfofhet}.
This case is reminiscent of  type II backgrounds with 31 supersymmetries \cite{iibpreons,bandos, Mpreons}.


\newsection{The descendants of $1$}\la{flat}

The  Killing spinor equation implies that $\hat R=0$ and so the spacetime is parallelizable. In addition one can argue that
$dH=0$. This is certainly the case in the lowest order in $\a'$. The Bianchi identity of $H$ receives  anomaly  contributions from
the gravitational sector and the gauge sector. The gravitational contribution can be expressed in terms of $\check R$
where $\check R$ can be found from $\hat R$ after setting $H$ to $-H$.
 Now if $dH=0$, $\hat R_{AB, CD}=\check R_{CD,AB}$ and since $\hat R=0$ for these backgrounds,
 the gravitational contribution to the anomaly vanishes. The gauge contribution also vanishes if we assume that
all parallel spinors also solve the gaugino Killing spinor equation. Parallelizable supersymmetric
backgrounds with $dH\not=0$ have been investigated in \cite{jfofhet}.

The dilatino Killing spinor equation imposes additional conditions on the spacetime. There are two cases
to consider depending on whether or not the one-form $d\Phi$ is null. Suppose that $d\Phi$ is not null
and $|d\Phi|^2\not=0$. In this case, one can show that the dilatino Killing spinor equation \cite{kawano, jfofhet} implies that
\bea
  \Pi = {1\over2} + \frac{\partial_M \Phi H_{NPQ} \Gamma^{MNPQ}}{24 |d\Phi|^2
 }   \,,
 \eea
is a projector, $\Pi^2=\Pi$. Since ${\rm tr}\, \Pi=8$, backgrounds with $dH=\hat R=0$ and $|d\Phi|\not=0$
preserve at least half of the supersymmetry. Moreover one can also show that
$d\Phi$ is $\nabla$-parallel, spacelike and $i_{d\Phi} H=0$, see e.g \cite{kawano, het}. All these are linear dilaton backgrounds. Moreover
$d\Phi$ spans a flat direction orthogonal to the rest of spacetime.

On the other hand if $|d\Phi|=0$, i.e.~either $d\Phi$  is null or $d\Phi=0$, then $H$ is null. If $d\Phi\neq 0$, then using the condition
$i_{d\Phi} H=0$ one can show  that these backgrounds preserve at least eight  supersymmetries. In the following we confirm the
results of \cite{jfofhet}.

\subsection{N=8}

The solutions for which $|d\Phi|\not=0$ have been classified and have been found to be isometric to
\bea
&&AdS_3\times S^3\times S^3\times \bR~,~~~AdS_3\times S^3\times\bR^4~,~~~\bR^{1,1}\times SU(3)~,~~~\bR^{3,1}\times S^3\times S^3~,~~~
\cr
&&\bR^{6,1}\times S^3~,~~~CW_4\times S^3\times \bR^3~,~~~CW_6\times S^3\times \bR~,
\eea
where $CW$ stands for Cahen-Wallach spaces.
In fact it turns out that the full content of the dilatino Killing spinor equation is the
projection $\Pi\e=\e$. So these backgrounds preserve precisely 8 supersymmetries.

On the other hand if $|d\Phi|=0$, $d\Phi\not=0$, it has been shown that the spacetime is isometric to
\bea
CW_{10}~,~~~CW_{8}\times\bR^2~,~~~CW_6\times \bR^4~,~~~CW_4\times \bR^6~,~~~\bR^{9,1}~.
\eea
The eight Killing spinors can be chosen, up to a gauge transformation, to satisfy $\Gamma^+\e=0$.
A basis in the space of these Killing spinors is $(e_{\a 5}, e_{\a \b \g 5})$, i.e.~these Killing spinors
span the $\Delta_{\bf 8}^-$ representation of $Spin(8)$. All these backgrounds can be
thought of as degenerations of $CW_{10}$. The only non-vanishing component of the flux is
\bea
H=e^+\wedge \b~,~~~\b={1\over2} \b_{ij} e^i\wedge e^j~,
\eea
and the dilaton is linear.
The form $\b$ is a generic element in $\Lambda^2(\bR^8)=\mathfrak{spin}(8)$. The Maurer-Cartan
structure
equations of the Cahen-Wallach group manifolds are
\bea
de^+=0~,~~de^-={1\over2}\b_{ij} e^i\wedge e^j~,~~~de^i=-\b^i{}_j e^+\wedge e^j~.
\eea
The backgrounds with $N>8$ supersymmetries are special cases of such backgrounds
with constant dilaton and appropriate restrictions on $\b$.

\subsection{N= 10}

To begin observe that $\Sigma({\cal K})=Spin(1,1)\times Spin(8)$. The additional
Killing spinor lies in $\Delta_{\bf 8}^+$. Using a similar argument to that we have applied
to determine the descendants in the $\bR^8$ case, the Killing spinor can be chosen as
\bea
\e_9= 1+e_{1234}~.
\eea
The dilatino Killing spinor equation gives
 \bea
  \partial_+ \Phi = 0, \qquad H_{+} \cont{\a} = 0, \qquad H_{+ \a \b} -
  \tfrac{1}{2} \epsilon_{\a \b}{}^{\bar \g \bar \d} H_{+ \bar \g \bar \d} = 0~.
  \,,
 \eea
Therefore for such backgrounds, we have found that
\bea
\Phi={\rm const}~,~~~\b \in \mathfrak{spin}(7)~.
\eea
Since $\b$ is constant and $Spin(7)$ acts on the remaining spinors with the vector representation, the
dilatino Killing spinor equation $\b_{ij} \Gamma^{ij}\e=0$ has another solution. This is because an element of $SO(2k+1)$
acting of $\bR^{2k+1}$ leaves
an axis invariant. The stability subgroup in this case is $SU(4)=Spin(6)$. Therefore,
there is supersymmetry enhancement to $N=10$. The second spinor can be chosen as
\bea
\e_{10}=i(1-e_{1234})
\eea
and $\b\in \mathfrak{su}(4)$. These backgrounds are the special cases of Cahen-Wallach
space-times that have two additional supersymmetries \cite{jfofhet}.
Therefore there are no {\it isolated} backgrounds with $N=9$ supersymmetry.
However deformations families of $N=10$ backgrounds which can be constructed by allowing
$\b$ to take values in $\mathfrak{spin}(7)$  have $N=9$ supersymmetries.

\subsection{N=12}

Applying the same arguments as in the $\bR^8$ case, we can choose an additional solution
of the dilatino Killing spinor equation $\b_{ij}\Gamma^{ij}\e=0$ as
$\e_{11}=i(e_{12} + e_{34})$.  The stability subgroup is $Sp(2)=Spin(5)$. Again
$Sp(2)$ acts with the vector representation on the remaining spinors and so
there is an additional Killing spinor which can be chosen as $\e_{12}=e_{12}-e_{34}$
with stability subgroup $Spin(4)$ and so
\bea
\b\in \mathfrak{spin}(4)=\mathfrak{su}(2)\oplus \mathfrak{su}(2)~.
\eea
Again there are no backgrounds with $N=11$ supersymmetries. Moreover the components of $\b$ for $N>8$
are those of $H_{+ij}$ found for the corresponding  descendants of the $\bR^8$.

\subsection{N=14}

It can be arranged such that the next solution of the dilatino
Killing spinor equation is $\e_{13}=e_{13}+e_{24}$ with stability subgroup $Spin(3)$ which
again acts on the remaining three spinors with the vector representation. Therefore
there is an addition Killing spinor which can be chosen $\e_{14}=i(e_{13}-e_{24})$. Moreover
\bea
\b\in \mathfrak{su}(2)~.
\eea

\subsection{N=16}

There no backgrounds with $N=15$ supersymmetries, see e.g \cite{Mpreons}. The backgrounds
with $N=16$ supersymmetries are isometric to Minkowski spacetime \cite{jfgpa}.

\newsection{Concluding remarks}

We have solved, using the spinorial geometry technique of \cite{ggp},
 the Killing spinor equations for all supersymmetric type I backgrounds.
In particular, we  utilized the gauge symmetry of the Killing spinor
equations of the theory to construct representatives for the Killing spinors in all cases.
We have approached the problem by first solving the gravitino Killing spinor equation whose
solutions are parallel spinors with respect to a metric connection,  $\hat\nabla$, with torsion a three-form $H$.
The solutions have been characterized by the isotropy group of the spinors in $Spin(9,1)$.
Then  in each case, we have used as gauge symmetry the subgroup $\Sigma({\cal P})$ of $Spin(9,1)$ that leaves invariant
the space of parallel spinors ${\cal P }$ to find representatives for the
solutions of the dilatino Killing spinor equation for the descendant backgrounds.   The Killing spinors are characterized by the
isotropy group of the associated parallel spinors and their stability subgroup in $\Sigma({\cal P})$.

There are two  classes of supersymmetric backgrounds depending on whether the isotropy group
of the parallel spinors is compact $K$ or non-compact $K\ltimes \bR^8$. In the latter case,
all backgrounds admit a null $\hat\nabla$-parallel vector field. Moreover, their geometries
can be characterized in terms of the properties of the rotation of the parallel vector field and
those of endomorphisms of the tangent bundle that behave as almost complex structures
on the transverse directions to the light-cone. In particular, the geometries depend on the integrability
of these endomorphisms and on a relation between their Lee forms. In addition the Gray-Hervella class $W_2$ vanishes
for all the endomorphisms.
We have also shown that if one imposes
$dH=0$ and the field equations, then the holonomy of $\hat\nabla$ reduces for all the descendant
backgrounds. This is because the spacetime admits more parallel forms than those
allowed by the holonomy of $\hat\nabla$. Moreover, under the same assumptions, if one insists that the holonomy of $\hat\nabla$
is precisely the isotropy group of the parallel spinors, then the gravitino Killing spinor
equation implies the dilatino one and all parallel spinors are Killing. These are the backgrounds
explored in \cite{het}.

On the other hand if the isotropy group of the parallel spinors is compact, i.e.~$G_2$, $SU(3)$, $SU(2)$ and $\{1\}$, then the spacetime
admits, 3, 4,  6  and 10 $\hat\nabla$-parallel vector fields, respectively.
In addition all the invariant forms associated with these groups are also $\hat\nabla$-parallel.
The geometry of the backgrounds depends on the properties of the rotation of the
parallel vector fields and their commutators, the integrability conditions of the endomorphisms invariant under the above groups,
and the relation between the Lee forms of the remaining $\hat\nabla$-parallel forms. In addition $W_2=0$ vanishes
for all endomorphisms associated with the invariant Hermitian forms. The pattern of relations between the various tensors that characterize the
geometry is more involved in this case.
We have also shown that if $dH=0$ and the field equations are satisfied, then in many cases there are additional
parallel forms on the spacetime than those allowed by the holonomy groups. Therefore if these
additional forms do not vanish, the holonomy reduces. Hence, if one insists that the
holonomy of $\hat\nabla$
is precisely the isotropy group of the parallel spinors, this imposes additional conditions on the existence
of descendants. In particular, this would imply that the vector space spanned by $\hat\nabla$-parallel vector fields
closes under Lie brackets and many terms in the solution of dilatino Killing spinor equation
for the descendants would vanish. However unlike the non-compact case, there are descendants
with holonomy precisely the isotropy group of the parallel spinors.

As we have already mentioned,  the assumption that the holonomy of $\hat\nabla$ is precisely the isotropy group of
the parallel spinors puts strong conditions on the existence of most descendants. However, one may allow the
holonomy of $\hat\nabla$ to be reduced. In some of the cases, this will imply enhancement of supersymmetry but not
always. We have shown that the $N=7$ descendant of $SU(2)$ always admits an additional supersymmetry and so it can be identified with the $N=8$
backgrounds. The full pattern or web of reductions is rather involved and it may be worth a systematic investigation.
There are additional conditions on the geometry that we have not investigated. For example, there is a classification
of Lorentzian Lie groups \cite{medina} and so a priori there are additional conditions on the structure constants
of the Lie algebra of the parallel vectors. These have not been implemented in the analysis of the descendants. This
mostly affects the descendants of the $SU(2)$ case and the results will be reported elsewhere \cite{prep}.

So far we have investigated the geometry of supersymmetric backgrounds. A natural question arises whether all solutions
can be classified. If a background admits Killing spinors with a non-compact isotropy group, then from the results
of \cite{het}, $dH=0$, the Killing spinor equations,  and the vanishing of $E_{--}$ and $LH_{-+}$ components
of the Einstein and two-form gauge potential field equations, respectively,  imply that all field equations are satisfied.
Similarly, if a background admits Killing spinors with a compact isotropy group, then the Killing spinor equations
and $dH=0$ imply all field equations. However despite  these simplifications, it is unlikely that all the solutions
can be classified in full generality in the near future. This is because such a task is related to
other classic classification problems like for example those of $G_2$ and $Spin(7)$ manifolds that remain unresolved.
Nevertheless some classes of solutions can be understood better. One such class is that of compactifications of type I
supergravities with fluxes. It is clear that some backgrounds with $N$, $N=1,2,3,4,5,6,8$,  parallel spinors which have a non-compact
isotropy group can serve as the vacuum configurations of compactification of type I to $1+1$ dimensions.
This is confirmed by the property of $\Sigma({\cal P})$ to be isomorphic to $Spin(1,1)\times R$ where $R$ and be thought of
as an $R$-symmetry group of the $1+1$ supergravity. A similar observation can be made for backgrounds with parallel spinors
which have compact isotropy groups. In particular backgrounds with parallel spinors that have $G_2$, $SU(3)$ and $SU(2)$  isotropy groups
can be used for compactifications to $2+1$-, $3+1$- and $5+1$-dimensions. The $\Sigma({\cal P})$ group
has the appropriate structure.
It is also possible to go beyond the vacuum configurations and compare supersymmetric solutions
of type I supergravity with those of lower dimensional supergravities that are related via a compactification. This will give an insight
into how supersymmetric solutions are related in a compactification scenario.

One may also wonder whether the classification of geometries of all supersymmetric
backgrounds in type I supergravity can be extended to those of type II supergravities.
The nature of the problem in type II is different. This is because the gauge group
of the Killing spinor equations in type II supergravities is a proper subgroup of the holonomy group of the
supercovariant connection. This, and its consequences, have been explained in detail in the conclusions of \cite{rev} and we shall not
 repeat the analysis here.
Nevertheless the results of this paper can be adapted to solve the algebraic Killing spinor equations of
type II supergravities provided that a solution of the gravitino Killing spinor equation is known. In particular
the group $\Sigma({\cal P})$ that preserves the space of parallel spinors can again be introduced and then it can be used
to find representatives for the solutions of the algebraic Killing spinor equations. Clearly, this can be
applied in type IIA and IIB supergravities and well as in other supergravities in lower dimensions.

\vskip 0.5cm \noindent{\bf Acknowledgements} \vskip 0.1cm Part of
this project was done while U.G.~was a post-doc at K.U.~Leuven,
Belgium, where he was funded by the Research Foundation K.U.~Leuven.
In addition, he is presently funded by the Swedish Research Council.
D.R.~is supported by the European EC-RTN project
MRTN-CT-2004-005104, MCYT FPA 2004-04582-C02-01 and CIRIT GC
2005SGR-00564. P.S.~is supported by a
PPARC studentship.

\vskip 0.5cm

\setcounter{section}{0}

\appendix{Geometric structures}

\subsection{Compact stability subgroup}

To determine the geometry of supersymmetric backgrounds, one has to understand the different geometric
structures that can occur.
For parallel spinors with compact stability subgroups $K$ in $Spin(9,1)$, and so ${\rm hol}(\hat\nabla)\subseteq K$,  the spacetime admits one time-like
 and $n=2$ or $3$ or $5$ or $9$ spacelike $\hat\nabla$-parallel
vectors fields denoted by $X_a$ and
some  $K$-invariant $\hat\nabla$-parallel forms, which we denote collectively by $\tau$. It is always possible to choose
a basis in the ring of invariant forms  such that
\bea
i_{a}\tau=0~,
\la{aaaa}
\eea
where $i_a$ denotes inner derivation with respect to the vector field $X_a$. Since $X_a$ are parallel they are nowhere zero and so span
a topologically trivial subbundle $\Xi$ of the tangent vector bundle $TM$ of the spacetime $M$. Thus we have
\bea
0\rightarrow \Xi\rightarrow TM \rightarrow \Pi\rightarrow 0~,
\eea
such that $\Pi$ is the orthogonal complement of $\Xi$ in $TM$ with respect to the spacetime metric,  $TM=\Xi\oplus \Pi$.
Since $\Xi$ is trivial, the topological structure group of $M$ reduces to $K\subset Spin(9-n)\subset Spin(9,1)$.
In general $[\Xi, \Xi]\nsubseteq \Xi$, and so $M$ is {\it not} always foliated.  Nevertheless, the above decomposition
of $TM$ and its dual can be used to decompose the various tensors of $M$ along the directions of $\Xi$ and $\Pi$.
In particular,  introduce the
dual one forms $e^a$ of $X_b$, i.e.~$e^a(X_b)=\delta^a{}_b$. Since $\hat\nabla$ is a metric connection $g(X_a, X_b)$ is constant
and so one can always choose $g(X_a, X_b)=\eta_{ab}$, where $\eta_{ab}$ is the standard Lorentz metric. Therefore, we can set
for the metric and $H$,
\bea
&&ds^2=\eta_{ab} e^a e^b+\delta_{ij} e^i e^j~,
\cr
&&H={1\over3!}H_{abc} e^a\wedge e^b\wedge e^c+{1\over2} H_{abi} e^a\wedge e^b\wedge e^i+{1\over2} H_{aij} e^a\wedge e^i\wedge e^j
\cr
&&~~~~~~~~~~~~~~~~+{1\over3!}H_{ijk} e^i\wedge e^j\wedge e^k~,
\eea
where $e^i$ is a local basis of one-forms spanning the fibers of the dual of $\Pi$, i.e.~the spacetime frame index
decomposes as $A=(a,i)$ and $H|_{\Pi}={1\over3!}H_{ijk} e^i\wedge e^j\wedge e^k$.
In this basis, the remaining $\hat\nabla$-parallel forms can be written as
\bea
\tau={1\over k!}\tau_{i_1,\dots, i_k} e^{i_1}\wedge\dots \wedge e^{i_k}~,
\eea
i.e.~$\tau=\tau|_{\Pi}$.
This follows from (\ref{aaaa}).

To find the conditions imposed on the geometry by $\hat\nabla X_a=\hat\nabla\tau=0$, we observe that
\bea
\hat\nabla_A(X_a)_B=0&\Longleftrightarrow& i_a H=\eta_{ab}de^b~,~~~~{\cal L}_ag=0~,
\cr
\hat\nabla_A\tau_{B_1\dots B_k}=0&\Longleftrightarrow& \nabla_a\tau_{j_1\dots j_k}
=
{k\over2}(-1)^k H_a{}^i{}_{[j_1} \tau_{j_2\dots j_k]i}~,~~~
\hat\nabla_i\tau_{j_1\dots j_k}=0~.
\la{parcon}
\eea
Therefore $X_a$ is Killing and its rotation is given in terms of $H$. In turn this implies that all
components of $H$ of the type $H_{aAB}$ are determined. In particular, we have that
\bea
&&H_{aij}=(i_aH)_{ij}=(de_a)_{ij}~,~~~~H_{abi}=(i_b i_aH)_i=(de_a)_{bi}=-[X_a, X_b]_i~,~~~~
\cr
&&H_{abc}=i_ci_bi_aH=(de_a)_{bc}= -g([X_a,X_b], X_c)~,
\eea
where $de_a=\eta_{ab} (de^b)$.
If  $[\Xi, \Xi]\subseteq \Xi$, i.e.~$X_a$ span a Lie algebra,  then $H_{abi}=0$. We shall examine this case in more detail later.

Next focus on the conditions in (\ref{parcon}) involving $\tau$.
It is clear from the above equations that some of the components  $H_{aij}$ are also determined  in terms  the
the covariant derivative of $\tau$. Compatibility requires a restriction on the geometry, i.e.~a relation between the exterior
derivative of $X_a$, which also determines $H_{aij}$,
and the covariant derivative of $\tau$. To find such geometric conditions, we begin
with the first pair of the above equations, and decompose $\Lambda^2(\bR^{9-n})=\mathfrak{k}\oplus \mathfrak{k}^\perp$, where
$\mathfrak{k}$ is the Lie algebra of $K$. This induces a decomposition of the two-form $i_a H|_{\Pi}$ as
$i_a H|_{\Pi}=i_a H^{\mathfrak{k}}+i_aH^{\mathfrak{k}^\perp}$. It is clear that $i_aH^{\mathfrak{k}}$ is not determined
by the first equation because the forms $\tau$ are invariant under the action of $K$. However,
$i_aH^{\mathfrak{k}^\perp}$ can be expressed in terms of both the covariant derivative of
$\tau$ and the rotation of $X_a$. As a result, the $\mathfrak{k}^\perp$ component of the rotation of $X_a$ can be expressed
in terms of the $\nabla_a$ covariant derivative of $\tau$, i.e schematically we have
\bea
(de_a)^{\mathfrak{k}^\perp}=(\nabla_a\tau)^{\mathfrak{k}^\perp}~.
\la{geomcon}
\eea
It remains to investigate the condition $\hat\nabla_i\tau_{j_1\dots j_k}=0$. This condition can be used to
investigate the $H|_{\Pi}$ component of $H$. The analysis is similar to that which one does
 in the context of $(9-n)$-dimensional manifolds with $K$-structure compatible with a connection with skew-symmetric torsion.
The end results depends on
the $K$ structure, it may or may not give additional conditions on the geometry. In all cases, $H|_{\Pi}$ is entirely
determined in terms of the geometry. We shall not give further details here but we describe the end result
in each case separately.

Using $\hat\nabla X_a=\hat\nabla\tau=0$, one can also compute the  Lie derivative
of $\tau$ along $X_a$ to find
\bea
 {\cal L}_a\tau_{A_1A_2\dots A_k}=k(-1)^k H_a{}^B{}_{[A_1} \tau_{A_2\dots A_k]B}&\Longleftrightarrow&
 {\cal L}_a\tau_{i_1i_2\dots i_k}=k(-1)^k H_a{}^j{}_{[i_1} \tau_{i_2\dots i_k]j}~,~~~~
 \cr
 &&{\cal L}_a\tau_{bi_1\dots i_{k-1}}=(-1)^k H_a{}^j{}_{b} \tau_{i_1\dots i_{k-1}j}~.
\eea
Thus if $i_aH^{\mathfrak{k}^\perp}$ vanishes and $[\Xi, \Xi]\subseteq \Xi$, then ${\cal L}_a\tau=0$.
Moreover observe that if $dH=0$, then ${\cal L}_aH=0$.

One can utilize the relation of $H$ to the rotation of $X_a$ to write
 $H$  in terms of $X_a$ in various ways. For example, one can write
\bea
H&=&\eta_{ab}\, e^a\wedge de^b+{1\over3} g([X_a,X_b], X_c) e^a\wedge e^b\wedge e^c+{1\over2} [X_a, X_b]_i e^a\wedge e^b\wedge e^i
\cr
&&+{1\over 3!}
H_{ijk} e^i\wedge e^j\wedge e^k~,
\eea
where as we have mentioned the expression for $H|_{\Pi}$ depends on the $K$-structure.

As has been observed in \cite{het}, there is an alternative way to write $H$ in the case that $[\Xi, \Xi]\subseteq \Xi$. In particular,
one has that  $H_{abi}=0$, the spacetime is a principal bundle, $\l^a=e^a$ is identified with a principal bundle
connection, and $H_{abc}$ are the structure constants of the Lie algebra spanned by $X_a$. In this case, it is more convenient to write
\bea
H={1\over3} \eta_{ab} \l^a\wedge  d\l^b+{2\over3} \eta_{ab} \l^a \wedge {\cal F}^b+{1\over 3!}
H_{ijk} e^i\wedge e^j\wedge e^k~,
\eea
where
\bea
{\cal F}={1\over2} H^a{}_{ij} e^i\wedge e^j=d\l^a-{1\over2} H^a{}_{bc} \l^b\wedge \l^c~,
\eea
is the curvature of the principal bundle. Sometimes we write $H^{\rm rest}=H|_{\Pi}$.

The dilatino Killing spinor equation will impose additional conditions on $H$ and on the geometry.
These are determined on a case by case basis from the solutions of the dilatino Killing spinor equations
and depend on the choice of Killing spinors up to  Lorentz transformations. This
is unlike the conditions we have described above which depend on the
$\hat\nabla$-parallel spinors that the spacetime admits, i.e.~the solutions of the gravitino Killing spinor equation.

\subsection{Non-compact stability subgroup}

If the stability subgroup of the parallel spinors is not compact, $K\ltimes\bR^8$, the spacetime admits
a $\hat\nabla$-parallel null vector field $X$ and null $\hat\nabla$-parallel forms which we collectively denote with $\tau$ such that
\bea
i_X\tau=0~.
\eea
Since the null vector field is nowhere vanishing, the tangent bundle of the spacetime admits a trivial
rank one subbundle $\Xi$ and so
\bea
0\rightarrow \Xi\rightarrow TM\rightarrow L\rightarrow 0~.
\eea
Choosing $X=e_+$, and so the associated $\hat\nabla$-parallel one-form is $e^-$, the spacetime metric can be written
as
\bea
ds^2=2 e^- e^++\delta_{ij} e^i e^j~,
\eea
where $e^+, e^i$ is a local basis in $L$. The structure group of $TM$,
 which is a subgroup of the holonomy group $K\ltimes \bR^8\subset Spin(8)\ltimes\bR^8$, acts as
\bea
e^-\rightarrow e^-~,~~~e^+\rightarrow e^+-O_{ij} q^i \,e^j-O_{ij} q^i q^j\, e^-~,~~~e^i\rightarrow  O^i{}_j\, e^j+ q^i\, e^-~,
\eea
where $O$ is an element of the vector representation of $Spin(8)$ and $q\in \bR^8$. There is no natural definition
of the $e^+$ light-cone direction or of the ``transverse'' $e^i$ directions to the lightcone.
  Next observe the bundle of $(k+1)$-forms
of $M$, $\Lambda^{k+1}(M)$, contains a subbundle
\bea
N^{k+1}=\{\a\in\Lambda^{k+1}(M)~,{\rm s.t.},~ i_X\a=0~,~~~e^-\wedge \a=0\}~.
\eea
The  $\hat\nabla$-parallel forms $\tau$ are sections of this bundle. The
transition functions of $N^{k+1}$ are those associated with the k-fold skew-symmetric
product of the vector representation of $SO(8)$, i.e.~the transition functions of $N^{k+1}$ are those of a k-form bundle of a ``transverse space'' to the light-cone.
In particular, one can define ``transverse'' $(k+1)$-forms on the spacetime $M$ up to sections of $N^{k+1}$. This can be seen
from the sequence
\bea
0\rightarrow N^{k+1}\rightarrow M^{k+1}\rightarrow  \Omega^{k+1}\rightarrow 0
\eea
where
\bea
M^{k+1}=\{\a\in\Lambda^{k+1}(M)~,{\rm s.t.},~ i_X\a=0~\}~,
\eea
and the ``transverse forms'' are sections of $\Omega^{k+1}$. Moreover observe that the map
$e^-\wedge: \Omega^k\rightarrow N^{k+1}$ is an isomorphism. In addition $\Omega^{k+1}$ is equipped
with a fiber metric induced from the spacetime metric.

The $\hat\nabla$-parallel forms $\tau$  can be written as $\tau=e^-\wedge \phi$, where $\phi$ are $K$-invariant forms
which can be thought of as sections of $\Omega^k$.
 The condition that $X$ and $\tau$ are $\hat\nabla$-parallel can be written as
\bea
\hat\nabla_A X_B = 0 &\Longleftrightarrow& de^-=i_X H~,~~~~{\cal L}_Xg=0
\cr
\hat\nabla_A\tau_{B_1\dots B_{k+1}}=0&\Longleftrightarrow&
\nabla_+\phi_{j_1\dots j_k}={k\over2}(-1)^k H_+{}^i{}_{[j_1}\phi_{j_2\dots j_k]i}~,
\cr
&&
\hat\nabla_-\phi_{j_1\dots j_k}=\nabla_-\phi_{j_1\dots j_k}+(-1)^{k}{k\over2} H^i{}_{-[j_1}\phi_{j_2\dots j_k]i}=0~
\cr
&&\hat\nabla_i\phi_{j_1\dots j_k}=0~,~~~~
\la{lalala}
\eea
So the $i_XH_{AB}=H_{+AB}$ components of $H$ are determined in terms of $e^-$, and $X$ is a  Killing vector field. Next let us focus on
\bea
\nabla_+\phi_{j_1\dots j_k}={k\over2}(-1)^k H_+{}^i{}_{[j_1}\phi_{j_2\dots j_k]i}~.
\la{mamama}
\eea
This can be viewed as conditions on $H_{+ij}$. Since $i_Xi_XH=0$, $i_X H$ is a section of $M^2$. The above condition
imposes a restriction on the ``transverse'' components of $i_XH$. In particular,  decomposing $\Lambda^2(\bR^8)=\mathfrak{k}\oplus \mathfrak{k}^\perp$,
(\ref{mamama})  is independent of $i_XH^{\mathfrak{k}}$, and expresses $i_XH^{\mathfrak{k}^\perp}$ in terms
of the covariant derivative of $\tau$ along the $X$ direction. In turn this is related to the $\mathfrak{k}^\perp$ component
of the rotation $de^-$. This is a condition on the geometry as that of (\ref{geomcon}) for compact stability subgroups mentioned above.
Similarly, one can see from the remaining conditions in (\ref{lalala}) that the $H_-^{\mathfrak{k}}$ is not determined
by the parallel transport equation while the $H_-^{\mathfrak{k}^\perp}$
is expressed in terms of the $\nabla_-$ covariant derivative of $\tau$.

It remains to investigate the condition $\hat\nabla_i\phi_{j_1\dots j_k}=0$. This condition can be analyzed as though
it is examined in the context of $8$-dimensional manifolds with $K$-structure compatible with a connection with skew-symmetric torsion.
This is because as we have mentioned $\Omega^{k}$ has the properties of the bundle of k-forms of ``transverse space''
to the light-cone.
The end result depends on
the $K$ structure, it may or may not give additional conditions on the geometry. In all cases, $H_{ijk}$ is entirely
determined in terms of the geometry.

The Lorentzian structures we have presented above are reminiscent
of the Cauchy-Riemann (CR) structures. This is not a surprise
since the CR structures also arise in the context of null Maxwell
fields in General Relativity, and they can be associated with a
$U(n)\ltimes \bR^{2n}$ type of structures, for a recent review see
\cite{traut}. One can give various generalizations of the CR
structures by using a Gray-Hervella type of classification for the
$K\ltimes\bR^L$-structures.

The Lie derivative of a k-form along the $\hat\nabla$-parallel vector field $X$ is
\bea
 {\cal L}_X   \tau_{A_1A_2\dots A_{k+1}}
 &=&(k+1)(-1)^{k+1} i_XH{}^B{}_{[A_1}\tau_{A_2\dots A_{k+1}]B}
 \cr
 &\Longleftrightarrow&
{\cal L}_X\tau_{-i_1\dots i_k}=k i_XH{}^j{}_{[i_1}\tau_{i_2\dots i_k]j-}~.~~~
\eea
Thus ${\cal L}_X\tau=0$ for all $\tau$, iff $i_XH^{\mathfrak{k}^\perp}=0$.

The geometry and fluxes can be written as
\bea
&&ds^2=2 e^- e^++\delta_{ij} e^i e^j
\cr
&& H=e^+ \wedge de^-+ {1\over2} (H^{\mathfrak{k}}+H^{\mathfrak{k}^\perp})_{-ij} e^-\wedge e^i \wedge e^j
+{1\over3!} H_{ijk} e^i\wedge e^j\wedge e^k
\la{nullgeom}
\eea
where $H_-^{\mathfrak{k}}$ is not determined by the Killing spinor equations.

{}For pp-wave backgrounds $de^-=0$. In such a case, one can write $e^-=dv$ for some coordinate $v$ and $X$ is parallel with respect to the
Levi-Civita connection. The transverse space $B$ to the pp-wave can then be defined as $u,v={\rm const.}$, where $u$ is the affine
parameter of the of the null geodesics. In all cases $B$ admits a $K$-structure, see \cite{het}
for more details.

\subsection{Integrability conditions, field equations and holonomy}

To investigate the existence of certain supersymmetric backgrounds, it is useful to incorporate the Bianchi identities
and the field equations in the conditions for supersymmetry. The derivation of the field equations from the integrability conditions
of the Killing spinor equations can be found in \cite{biran, het}. Some additional useful formulae are the Bianchi identities
\bea
&&\hat R_{[AB, CD]}=-{1\over4} (dH)_{ABCD}+ {1\over2} H_{E[AB} H^E{}_{CD]}~,
\cr
&& \hat R_{A[B,CD]}=-{1\over3} \hat\nabla_A H_{BCD}-{1\over6} (dH)_{ABCD}~,
\cr
&& \hat R_{[AB,C]D}=-{1\over3} dH_{ABCD}-{1\over3}\hat\nabla_D H_{ABC}- H^E{}_{D[A} H_{BC]E}~.
\eea
of $\hat R$. In particular, the second identity will be used to investigate the reduction of the holonomy
of $\hat\nabla$
for the descendants.

\appendix{Revisiting the singlets}

In the introduction, we have listed the Lie subgroups of $Spin(9,1)$ that leave some Majorana-Weyl spinors invariant.
Here we shall provide an argument to show that the list in the introduction is complete.
This is essentially a Lie algebra computation. There are two additional cases that occur in the type I backgrounds
in addition to those that have not been investigated in \cite{het}.

There is a single type of orbit of $Spin(9,1)$ in Majorana-Weyl representation $S^+$
of co-dimension zero and a representative is $1+e_{1234}$, see \cite{figueroab, het}.
Two  spinors are invariant either under the subgroup
$SU(4)\ltimes\bR^8$ or $G_2$. The representative of the second spinor \cite{het} can be chosen as $i(1-e_{1234})$ or $e_{15}+e_{2345}$,
respectively.

To proceed, we decompose $S^+$ under the action of $SU(4)$, as\footnote{With $V_{\bR}$ we denote the associated
real representation of a complex representation, i.e.~$(\bC<1>)_{\bR}=\bR<(1+e_{1234}, i(1-e_{1234})>$.}  $S^+=(\bC<1>)_{\bR}\oplus {\rm Re}\Lambda^2(\bC^4)\oplus  (\Lambda^1(\bC^4))_{\bR}$, where we
have chosen a $1$ as representative for the first two invariant spinors to make the analysis more transparent.
There is an orbit of co-dimension one of $SU(4)$ in ${\rm Re}\Lambda^2(\bC^4)$ with stability stability subgroup
$Sp(2)$. In addition under the action of $Sp(2)$, $S^+$ decomposes as $S^+=(\bC<1>)_{\bR}\oplus\bR<i(e_{12}+e_{34})>\oplus \Lambda^1(\bR^5)\oplus \bH^2$,
thus
there are only three $Sp(2)\ltimes\bR^8$-invariant spinors. $SU(4)$ has an orbit of co-dimension 2 in $(\Lambda^1(\bC^4))_{\bR}$ with stability
subgroup $SU(3)$. However this case can be thought of  descending from $G_2$ and so it will be investigated later.

To investigate the case of four invariant spinors,  $Sp(2)=Spin(5)$ acts with the vector representation on
$\Lambda^1(\bR^5)\subset S^+$. So there is a single orbit with stability
subgroup $SU(2)\times SU(2)$. In addition under $SU(2)\times SU(2)$, $S^+$ decomposes as
$S^+=(\bC<1, e_{12}>)_{\bR}\oplus \bH\oplus (\bC^2\oplus \bC^2)_{\bR}$, thus
there are only four $(SU(2)\times SU(2))\ltimes\bR^8$-invariant spinors. A key point is that $SU(2)\times SU(2)$ acts on $\bH$
with the left and right multiplication by unit quaternions, i.e.
\bea
x\rightarrow ax\bar b~,~~~x\in \bH~,~~~a\in SU(2)~,~~b\in SU(2)~.
\la{ayb}
\eea
$Sp(2)$ has also an orbit in $\bH^2$ with stability subgroup $Sp(1)$
but this can also be thought of as the descending from the $G_2$ case and it will be investigated later.

Next $SU(2)\times SU(2)$ has a single orbit in $\bH$ with stability subgroup $SU(2)$,
$SU(2)\subset SU(2)\times SU(2)$
is the diagonal subgroup.
This case has not been consider in \cite{het}. Moreover under this $SU(2)$, $S^+$ decomposes as
$S^+=(\bC<1,e_{12}>)_{\bR}\oplus \bR<e_{13}+e_{24}>\oplus{\rm Im}\bH\oplus (\bC^2)_{\bR}\oplus (\bC^2)_{\bR}$, where $SU(2)$ acts on both
copies of $\bC^2$ with the fundamental representation.
Thus $SU(2)\ltimes\bR^8$ leaves  invariant five spinors in $S^+$.  Moreover there three types of  orbits of $SU(2)\times SU(2)$  in
$(\bC^2\oplus \bC^2)_{\bR}$. Two of those have $SU(2)$ stability subgroup. These two cases can be thought of descending from
the $G_2$ case and they will be investigated later. The third type has trivial stability subgroup and so there are no more invariant spinors.

To proceed, observe that $SU(2)\subset SU(2)\times SU(2)$ acting as (\ref{ayb}), for $a=b$, on ${\rm Im}\bH$ has a orbit
of codimension one which has stability subgroup $U(1)$.
In addition $S^+$ decomposes under $U(1)$ as
$S=(\bC<1, e_{12}, e_{13}>)_{\bR}\oplus^4 (\bC)_{\bR}$, where $U(1)$ acts on $\bC$ with the fundamental representation.
Thus there are six $U(1)\ltimes\bR^8$-invariant spinors.
In addition, the orbits of $SU(2)$ in $\oplus^2(\bC^2)_{\bR}$ have stability subgroup $\{1\}$ in $Spin(9,1)$.  Thus there are no
more invariant spinors.
This concludes the cases with non-compact stability subgroups.

Next let us consider the descendants of $G_2$. The $G_2$ decomposition of $S^+$ is
$S^+=\bR<1+e_{1234}>\oplus \bR<e_{15}+e_{2345}>\oplus \Lambda^1(\bR^7)\oplus \Lambda^1(\bR^7)$.
In addition $G_2$ has a single orbit in $\Lambda^1(\bR^7)$ of co-dimension one which has stability subgroup
$SU(3)$. Moreover $S^+$ under $SU(3)$ decomposes as $S^+=(\bC<1, e_{15}>)_{\bR}\oplus (\Lambda^2(\bC^3))_{\bR}
\oplus (\Lambda^1(\bC^3))_{\bR}$, thus there are four $SU(3)$-invariant spinors.
In either $(\Lambda^2(\bC^3))_{\bR}$ or $(\Lambda^1(\bC^3))_{\bR}$, $SU(3)$
acts with stability subgroup $SU(2)$. In addition, $S^+$ decomposes under $SU(2)$ as
$S^+=(\bC<1, e_{15}, e_{12}, e_{25}>)_{\bR}\oplus^2 (\bC^2)_{\bR}$. Thus there are eight $SU(2)$-invariant spinors.
Moreover, the orbits of $SU(2)$ in
$\oplus^2(\bC^2)_{\bR}$ have stability subgroup $\{1\}$. So there are no other cases to investigate. This concludes the analysis.

\appendix{$SO(3)$ transformations}

The first three $SU(2)$-invariant Killing spinors are given by $1 + e_{1234}$, $e_{15} +
e_{2345}$ and $i (1-e_{1234}) + e_{25} - e_{1345}$. Therefore the fourth
Killing spinor is spanned by the following basis elements:
 \begin{align}
   &  \l_1 = i (1-e_{1234}) - e_{25} + e_{1345} \,, \quad \l_2 = i (e_{15} - e_{2345}
     \,, \quad \l_3 = i (e_{25} + e_{1345}) \,, \notag \\
& \l_4 = e_{12} - e_{34}
     \,, \quad \l_5 = i ( e_{12} + e_{34} ) \,.
 \end{align}
The action of the generators $t_i$ of $SO(3)$ on these is given by (omitting
terms proportional to the first three Killing spinors, i.e.~restricting to
${\cal P} / {\cal K}$ as discussed in the text)
 \begin{align}
 & t_1 (\l_1) = - 2 \l_5 \,, \quad t_2 (\l_1) = 2 \l_3 \,, \quad t_3 (\l_1) = 2
 (\l_2 + \l_4) \,, \notag \\
 & t_1 (\l_2) = - 2 \l_3 \,, \quad t_2 (\l_2) = 0 \,, \quad t_3 (\l_2) = - \l_1 \,, \notag \\
& t_1 (\l_3) =  2 \l_2 - \l_4 \,, \quad t_2 (\l_3) = - \tfrac{1}{2} \l_1 \,,
 \quad t_3 (\l_3) = - \l_5 \,, \notag \\
& t_1 (\l_4) = 0 \,, \quad t_2 (\l_4) = - 2 \l_5 \,, \quad t_3 (\l_4) = - \l_1 \,, \notag \\
& t_1 (\l_5) = \tfrac{1}{2} \l_1 \,, \quad t_2 (\l_5) = - \l_2 + 2 \l_4 \,,
 \quad t_3 (\l_5) = \l_3 \,.
 \label{SO(3)-transf}
 \end{align}
One can also see explicitly that the $\l$'s constitute the symmetric
traceless representation of $SO(3)$. Define the matrix
 \begin{align}
  M = \left( \begin{array}{ccc} 2 \l_4 & \l_1 & 2 \l_5 \\ \l_1 & -2 \l_2 & - 2
  \l_3 \\ 2 \l_5 & - 2 \l_3 & 2 \l_2 - 2 \l_4 \end{array} \right) \,.
 \end{align}
The transformation \eqref{SO(3)-transf} corresponds to
 \begin{align}
  t_i (M) = M t_i  -  t_i M \,,
 \end{align}
with the generators given by
 \begin{align}
 t_1 = \left( \begin{array}{ccc} 0 & 0 & 0 \\ 0 & 0 & 1 \\ 0 & -1 & 0
 \end{array} \right) \,, \quad
 t_2 = \left( \begin{array}{ccc} 0 & 0 & 1 \\ 0 & 0 & 0 \\ -1 & 0 & 0
 \end{array} \right) \,, \quad
 t_3 = \left( \begin{array}{ccc} 0 & 1 & 0 \\ -1 & 0 & 0 \\ 0 & 0 & 0
 \end{array} \right) \,.
 \end{align}
{}From these formulae it is clear that while the first three Killing spinors
transform in the fundamental representation of $SO(3)$, the remaining five
basis elements form the symmetric traceless representation.

\appendix{Null spinor bilinears}

The spinor bilinear vectors for the extra two basis
elements $e_{13}$ and $e_{24}$ are given by
 \begin{align}
  \kappa(e_{13}, e_{24}) = (e^0 - e^5) \,,
 \end{align}
and other combinations vanishing. Similarly, the non-vanishing
bilinear three-forms are given by
 \begin{align}
  & \xi(1,e_{13}) = - (e^0 - e^5) \wedge (e^2 + i e^7) \wedge (e^4 + i
  e^9) \,, \notag \\
  & \xi(e_{1234},e_{13}) = -(e^0 - e^5) \wedge (e^1 - i e^6) \wedge (e^3 - i
  e^8) \,, \notag \\
  & \xi(e_{12},e_{13}) = -(e^0 - e^5) \wedge (e^1 - i e^6) \wedge (e^4 + i
  e^9) \,, \notag \\
  & \xi(e_{24},e_{13}) = -i (e^0 - e^5) \wedge (\omega_1 - \omega_2)\,, \notag \\
  & \xi(e_{34},e_{13}) = -(e^0 - e^5) \wedge (e^2 + i e^7) \wedge (e^3
  - i e^8) \,, \notag \\
  & \xi(1,e_{24}) = -(e^0 - e^5) \wedge (e^1 + i e^6) \wedge (e^3
  + i e^8) \,, \notag \\
  & \xi(e_{1234},e_{24}) = -(e^0 - e^5) \wedge (e^2 - i e^7) \wedge
  (e^4 - i e^9) \,, \notag \\
  & \xi(e_{12},e_{24}) = (e^0 - e^5) \wedge (e^2 - i e^7) \wedge (e^3
  + i e^8) \,, \notag \\
  & \xi(e_{34},e_{24}) = (e^0 - e^5) \wedge (e^1 + i e^6) \wedge (e^4
  - i e^9) \,.
 \end{align}
Finally, the non-vanishing bilinear five-forms are
 \begin{align}
  & \tau(1,e_{13}) = i (e^0 - e^5) \wedge (e^2 + i e^7) \wedge (e^4 + i
  e^9) \wedge \omega_1 \,, \notag \\
  & \tau(e_{1234},e_{13}) = - i (e^0 - e^5) \wedge (e^1 - i e^6) \wedge (e^3 - i
  e^8)  \wedge \omega_2 \,, \notag \\
  & \tau(e_{12},e_{13}) = - i (e^0 - e^5) \wedge (e^1 - i e^6) \wedge (e^4 + i
  e^9)  \wedge (e^2 \wedge e^7 - e^3 \wedge e^8)\,, \notag \\
  & \tau(e_{13},e_{13}) = (e^0 - e^5) \wedge (e^1 - i e^6) \wedge (e^2 + i
  e^7) \wedge (e^3 - i e^8) \wedge (e^4 + i e^9) \,, \notag \\
  & \tau(e_{24},e_{13}) = - \tfrac{1}{2} (e^0 - e^5) \wedge
  (\omega_1 - \omega_2) \wedge (\omega_1 - \omega_2) \,, \notag \\
  & \tau(e_{34},e_{13}) = i (e^0 - e^5) \wedge (e^2 + i e^7) \wedge (e^3
  - i e^8) \wedge (e^1 \wedge e^6 - e^4 \wedge e^9) \,, \notag \\
  & \tau(1,e_{24}) = i (e^0 - e^5) \wedge (e^1 + i e^6) \wedge (e^3
  + i e^8) \wedge \omega_2 \,, \notag \\
  & \tau(e_{1234},e_{24}) = - i (e^0 - e^5) \wedge (e^2 - i e^7) \wedge
  (e^4 - i e^9) \wedge \omega_1 \,, \notag \\
  & \tau(e_{12},e_{24}) = i (e^0 - e^5) \wedge (e^2 - i e^7) \wedge
  (e^3 +  i e^8) \wedge (e^1 \wedge e^6 - e^4 \wedge e^9) \,, \notag \\
  & \tau(e_{24},e_{24}) = (e^0 - e^5) \wedge (e^1 + i e^6) \wedge (e^2 - i
  e^7) \wedge (e^3 + i e^8) \wedge (e^4 - i e^9) \,, \notag \\
  & \tau(e_{34},e_{24}) = - i (e^0 - e^5) \wedge (e^1 + i e^6) \wedge (e^4
  - i e^9) \wedge (e^2 \wedge e^7 - e^3 \wedge e^8) \,.
 \end{align}
Here we have used the following definitions
 \begin{align}
   \omega_1 = e^1 \wedge e^6 + e^3 \wedge e^8 \,, \quad \omega_2 = e^2 \wedge
   e^7 + e^4 \wedge e^9 \,,
 \end{align}
for the two-forms.

{}From these expressions one can derive the inner products for the null Majorana
spinors, which are
 \begin{align}
  \e_1 = 1 + e_{1234} \,, \quad \e_2 = i (1 - e_{1234}) \,, \notag \\
  \e_3 = e_{12} - e_{34} \,, \quad \e_4 = i (e_{12} + e_{34}) \,, \notag \\
  \e_5 = e_{13} + e_{24} \,, \quad \e_6 = i (e_{13} - e_{24}) \,.
 \end{align}
In the Majorana basis of spinors, the bilinear vectors read
 \begin{align}
  \k (\e_5,\e_5) = 2 (e^0 - e^5) \,, \notag \\
  \k (\e_6,\e_6) = 2 (e^0 - e^5) \,.
 \end{align}
The bilinear three-forms are given by
 \begin{align}
  & \xi (\e_1, \e_5) = - 2 (e^0 - e^5) \wedge (e^2 \wedge e^4 - e^7 \wedge e^9 +
  e^1 \wedge e^3 - e^6 \wedge e^8 ) \,, \notag \\
  & \xi (\e_1, \e_6) = - 2 (e^0 - e^5) \wedge (-e^2 \wedge e^9 + e^4 \wedge e^7 +
  e^1 \wedge e^8 - e^3 \wedge e^6 ) \,, \notag \\
  & \xi (\e_2, \e_5) = - 2 (e^0 - e^5) \wedge (- e^2 \wedge e^9 + e^4 \wedge e^7 -
  e^1 \wedge e^8 + e^3 \wedge e^6 ) \,, \notag \\
  & \xi (\e_2, \e_6) = 2 (e^0 - e^5) \wedge (e^2 \wedge e^4 - e^7 \wedge e^9 -
  e^1 \wedge e^3 + e^6 \wedge e^8 ) \,, \notag \\
  &\xi (\e_3, \e_5) = - 2 (e^0 - e^5) \wedge (e^1 \wedge e^4 + e^6 \wedge e^9 -
  e^2 \wedge e^3 - e^7 \wedge e^8 ) \,, \notag \\
  & \xi (\e_3, \e_6) = - 2 (e^0 - e^5) \wedge (-e^1 \wedge e^9 - e^4 \wedge e^6 -
  e^2 \wedge e^8 - e^3 \wedge e^7 ) \,, \notag \\
  & \xi (\e_4, \e_5) = - 2 (e^0 - e^5) \wedge (-e^1 \wedge e^9 - e^4 \wedge e^6 +
  e^2 \wedge e^8 + e^3 \wedge e^7 ) \,, \notag \\
  & \xi (\e_4, \e_6) = 2 (e^0 - e^5) \wedge (e^1 \wedge e^4 + e^6 \wedge e^9 +
  e^2 \wedge e^3 + e^7 \wedge e^8 ) \,, \notag \\
  & \xi (\e_5, \e_6) = 2 (e^0 - e^5) \wedge (\omega_1 - \omega_2) \,.
 \end{align}
Finally, the five-forms are
 \begin{align}
  & \tau(\e_i, \e_5) = - \xi(\e_i,\e_6) \wedge (\omega_1 - \omega_2) \,, \quad i=1,\ldots,4, \notag \\
  & \tau(\e_i, \e_6) = \xi(\e_i,\e_5) \wedge (\omega_1 - \omega_2) \,, \quad i=1,\ldots,4, \notag \\
  & \tau(\e_5, \e_5) = - (e^0 - e^5) \wedge (\omega_1 - \omega_2) \wedge
  (\omega_1 - \omega_2) + 2  (e^0 - e^5) \wedge \text{Re}(\chi)  \,, \notag \\
  & \tau(\e_5, \e_6) = - 2 (e^0 - e^5) \wedge \text{Im}(\chi)  \,, \notag \\
  & \tau(\e_6, \e_6) = - (e^0 - e^5) \wedge (\omega_1 - \omega_2) \wedge
  (\omega_1 - \omega_2) - 2  (e^0 - e^5) \wedge \text{Re}(\chi) \,,
 \end{align}
where we have used
 \begin{align}
 \chi = (e^1 - i e^6) \wedge (e^2 +
  i e^7) \wedge (e^3 - i e^8) \wedge (e^4 + i e^9) \,.
 \end{align}

\end{document}